\documentclass[12pt]{article}
\usepackage{tikz}
\usetikzlibrary{arrows.meta,positioning}
\usepackage{amsfonts}
\usepackage{amsmath, amsthm, amssymb}
\usepackage{indentfirst}
\usepackage[left=1.5in,right=1.5in,top=1.5in,bottom=1.5in]{geometry}
\usepackage[natbibapa]{apacite}
\usepackage{multirow}
\usepackage{verbatim}
\usepackage{bbm}
\usepackage{scrextend}
\usepackage{lipsum}% for demo only!

\usepackage[bookmarksopen,pdfstartview=FitH,pdfpagemode=UseNone,urlcolor=blue,citecolor=black,linkcolor=black,colorlinks]{hyperref}

\usepackage{bbm}
\usepackage{graphicx}
\usepackage{ragged2e}
\usepackage{booktabs}
\usepackage{mathtools}

\usepackage{float}
\usepackage{mathrsfs}
\usepackage{pdfsync}
\usepackage{tikz}
\newcommand{\defeq}{\mathrel{\mathop:}=}

\usepackage{pdfsync}
\usepackage{tikz}
\usetikzlibrary{shapes,arrows}
\usepackage{graphics,color}
\usepackage{graphicx}
\usepackage{inputenc}
\usepackage[natbibapa]{apacite}

\usetikzlibrary{shapes,arrows}
\usepackage{graphics,color}
\newtheorem{lemma}{Lemma}

\newtheorem{theorem}{Theorem}
\newtheorem{definition}{Definition}

\newtheorem{proposition}{Proposition}

\DeclareMathOperator*{\argmax}{arg\,max}
\definecolor{ThorBlau}{RGB}{31, 129, 192}
\definecolor{darkviolet}{rgb}{0.58, 0.0, 0.83}

\begin{document}

\title{A Theory of Auditability for Allocation Mechanisms\thanks{We thank seminar participants at Bonn, Boston College, Caltech, Duke, Stanford, UC Berkeley, UC San Diego, Upenn, and conference participants at the Iowa State University Market Design workshop, Stony Brook Game Theory conference, Southwest Economic Theory conference, Conference on Economic Design, ACM FAccT'23, and EC'23 for helpful discussions and comments.}}

\author{Aram Grigoryan\thanks{Department of Economics, University of California - San Diego, ag@ucsd.edu} \and Markus M\"{o}ller\thanks{Department of Economics, University of Bonn, mmoelle2@uni-bonn.de}} 
\date{March 2024}

\maketitle

\begin{abstract}
In centralized mechanisms and platforms, participants do not fully observe each others' type reports. Hence, if there is a deviation from the promised mechanism, participants may be unable to detect it. We formalize a notion of auditabilty that captures how easy or hard it is to detect deviations from a mechanism. We find a stark contrast between the auditabilities of prominent mechanisms. We also provide tight characterizations of maximally auditable classes of allocation mechanisms. 
\end{abstract}

\section{Introduction}

Scarce resources are often allocated in a centralized clearinghouse; the designer collects participants' type reports and chooses an allocation through some publicly announced mechanism. Some major examples include allocation of public school seats or subsidized housing, auctions for online ads or spectrum licences, and presidential elections. In these problems, participants know and observe their own type reports and outcomes, but not necessarily those of others. Hence, if there is a deviation from the announced mechanism, participants or an auditing entity may be unable to detect it because of limited information. In this paper, we develop a general theory of auditability to compare mechanisms in terms of how easy or hard it is for participants or some third-party auditing entity to detect deviations. 

As a main application, we study auditability in allocation problems, addressing practical concerns about errors, misconduct, and fraud in school choice and college admissions. For example, there have been multiple controversies during the implementation of admissions to Chicago Public Schools (CPS).\footnote{More generally, the need for transparency and auditability in school choice has been highlighted in \cite{pathak:17} and \cite{benner/boser:18}. The latter argues that an outside entity should audit the assignment mechanism to guarantee the implementation is consistent with enrollment priorities. Such auditing practices have been adopted by school districts in New Orleans and Chicago \citep{benner/boser:18}.} The Office of Inspector-General (OIG) of Chicago Board of Education \citep{schuler:18} states that in the 2016-2017 admission year: 

\begin{addmargin}[1em]{2em}%
\textit{``\dots almost every kind of CPS
elementary school imaginable improperly admitted students last school year. \dots{} Of more than 18,200 elementary-grade admissions audited, nearly 6,900 failed the audit. That’s nearly two of every five."}
\end{addmargin}

The report further states that:

\begin{addmargin}[1em]{2em}% 
\textit{``OIG interviews with principals of 30 audited schools that held more than 500 combined audit failures revealed that many didn’t know all the admissions rules, which are scattered across several locations. Others knew the rules and broke them. In some cases, audit failures may have been caused by documentation errors. \dots{} Several principals weeded out applicants, based on a variety of factors, including attendance concerns. \dots{} Some principals clearly played favorites. Many schools bypassed [admission rules] to give preference to the children of CPS employees, the siblings of existing students or lottery winners, \dots{} One principal owner improperly admitted her four children, her niece and nephew." }
\end{addmargin}

Similarly, in Boston, numerous students were prevented from applying to the city's prestigious exam schools in 2020. As BPS officials acknowledged, the error was attributed to internal communication breakdowns within the BPS.\footnote{https://www.bostonglobe.com/2020/08/31/metro/boston-public-schools-announces-error-exam-school-admissions-that-kept-dozens-out-recent-years/} A student's tutor and not BPS's internal audit uncovered the deviation.\footnote{Also in 2023, the BPS erroneously excluded some students from applying to some exam schools. See, for instance, https://www.bostonglobe.com/2023/04/12/metro/bps-miscalculated-student-gpas-wrongly-informing-students-they-were-eligible-apply-exam-schools/.}

Transparency and auditability considerations are also salient in other environments. In online ads auctions, the lack of bid-observability has led to numerous allegations of misconduct, including the recent antitrust lawsuit against Google.\footnote{https://www.wired.com/story/google-antitrust-ad-market-lawsuit/} Concerns of misconduct have also been raised for transplant organ allocation \citep{mcmichael:22}

In our model, there are $N$ individuals and a commonly known mechanism that specifies an outcome for each realization of individuals' type reports. Given a profile of type reports, a deviation is an outcome different from the one specified by the mechanism. A group of individuals detects a deviation, if the group's outcome could not have resulted from the mechanism for any type report profile consistent with the group members' type reports. We define a straightforward index-based measure to evaluate mechanisms' auditability properties. For a given type report profile, a mechanism's \textbf{auditability index} is the smallest integer such that for any deviation, there is a group of individuals whose size does not exceed this integer, and who detects the deviation. The \textbf{worst-case auditability index} of a mechanism is its largest auditability index across all type report profiles.  In essence, a mechanism with a large auditability index is more vulnerable to undetected errors or fraud. A more detailed interpretation of our auditability notion is given in Section \ref{sec:allocation_problems}. 

In our main application, we study the auditability properties of mechanisms for allocation problems without money. We unveil sharp contrasts in some prominent mechanisms' auditability properties. More specifically, we say a mechanism is \textit{maximally auditable} if its worst-case auditability index is 2, that is, any deviation can be detected by some two individuals. We say a mechanism is \textit{minimally auditable} if its worst-case auditability index is $N$, that is, some deviations cannot be detected by any proper subset of individuals. We find that many prominent mechanisms fall into one of these two extremes. 

In the context of priority-based allocation, we investigate the auditability properties of a large class of mechanisms, which we call Deferred-Acceptance (DA)-representable. 
Among many other well-known classes of mechanisms, we study the entire family of \textit{application-rejection} mechanisms \citep{chen/kesten:17}---a class that includes the prominent Deferred Acceptance and Immediate Acceptance (aka the `Boston' mechanism) as special cases. We provide a tight characterization of maximally auditable DA-representable mechanisms  (Theorem \ref{thm:DA-representable-full-characterization-two}). The result reveals a close connection between a mechanism's auditability and the structure of \textit{stable} outcomes. As a corollary of our characterization result, we establish that Immediate Acceptance is maximally auditable (Proposition \ref{cor:IA}). Immediate Acceptance is the unique maximally auditable mechanism among all application-rejection mechanisms, and every other application-rejection mechanism, including Deferred Acceptance, is minimally auditable (Proposition \ref{prop:AR}).

These findings may contribute to the ongoing discussions and evaluations of popular mechanisms for practical application. Traditionally, many school districts in the US have been using Immediate Acceptance for centralized assignment. Starting from the early 2000s, many major school districts transitioned to the Deferred Acceptance mechanism, potentially because the latter mechanism is \textit{strategy-proof} \citep{roth:82,dubins/freedman:81}. That means, in theory, parents do not have to worry about finding the best strategy for ranking schools. Currently, Deferred Acceptance is used for student assignment in Boston, New York City, Denver, Chicago, New Orleans, Newark, and Indianapolis. Despite these transitions, Immediate Acceptance remains one of the most common school assignment mechanisms. The mechanism is used in Charlotte-Mecklenburg, Miami-Dade, Minneapolis, Seattle, Tampa-St. Petersburg and many other school districts in the US and worldwide. In this paper, we formalize a notion of auditability. To the best of our knowledge, we provide the first theoretical comparison of Immediate Acceptance and Deferred Acceptance concerning this dimension. Our findings potentially highlight the importance of addressing trust and transparency considerations for real-life assignment problems.

Next, we study a house allocation setup (e.g., \cite{hylland/zeckhauser:79}, \cite{svensson:99}), and we focus on the large class of \textit{hierarchical exchange mechanisms} of \cite{papai:00}. Hierarchical exchange mechanisms characterize Pareto efficient, strategy-proof, non-bossy, and reallocation-proof mechanisms \citep{papai:00}. At each step of implementation, each unmatched object is owned by a still unmatched individual, and the ownership determines objects' pointing for the Top-Trading Cycle (TTC) algorithm \citep{shapley/scarf:74}. 
 
 We characterize all maximally auditable hierarchical exchange mechanisms as \textit{vice ownership mechanisms} (Theorem \ref{theorem:TTCcharacterization}). Essentially, in this new class of mechanisms, ownership distribution over all objects follows a strict ordering over levels of individuals: Each level contains at most two individuals, and each individual is in at most one level. Basically, within each level and for each object, one of the level's members owns it unless an individual from a strictly lower level has already claimed it. Vice ownership mechanisms are a strict subclass of hierarchical exchange mechanisms that are obviously strategy-proof (OSP) \citep{li2017obviously} and thus offer high standards for strategic simplicity. 
 
 More recently, \cite{pycia/troyan:23} established that only a few mechanisms achieve even higher standards for strategic simplicity than OSP and that these mechanisms closely resemble the class of sequential dictatorships \citep{ehlers/klaus:03, papai:01, papai:00}. We find that even sequential dictatorships can be hard to audit (Proposition \ref{proposition:seq}) and that \textit{almost} serial dictatorship emerges as the unique maximally auditable mechanism (Theorem \ref{theorem:vicedictatorship}) in the entire class of sequential dictatorships.

For general object allocation problems, we give a simple characterization of the entire class of mechanisms with an auditability index of one (Theorem \ref{characterization:pr-sp index_one}). We establish that these mechanisms are highly restrictive in a sense that they violate very basic notions of efficiency (Proposition \ref{prop:full-range}). 

In Appendix \ref{appendix:other-applications}, we describe a general model of auditability that captures other important social choice applications. In a single-item auction setting, we observe that the first-price price auction is maximally auditable, while the second price auction is minimally auditable. We give a simple characterization of the majority voting rule through auditability for voting with a binary outcome. 
We also study the auditability properties of different reserves rules for choice with affirmative action \citep{hafalir/yebmez/yildirim:13}.

The remainder of this work is organized as follows. We discuss the related literature in section \ref{sec:literature}. Section \ref{sec:allocation_problems} introduces the model and our notion of auditability. Sections \ref{section:priority-based} and \ref{section:house-allocation} contain our results for priority-based and housing allocation settings, respectively. Section \ref{sec:index_one} discusses mechanisms with an auditability index of one for general allocation problems. Section \ref{section:concluding_remarks} concludes. The Appendix contains further results and discussions of other applications. 
Longer proofs are in the Online Appendix. 

\section{Related Literature}
\label{sec:literature}

A relatively recent strand of literature studies transparency and credibility of mechanisms. For auction problems, the question has been studied by \cite{akbarpour/li:20} and \cite{woodward:20}. \cite{akbarpour/li:20} introduce a general partial commitment framework with sequential private communication between the operator and individuals. They study \textit{credible implementation} (Bayes-Nash implementation with imperfect information) of various standard auction formats. \cite{akbarpour/li:20} and \cite{woodward:20} require that deviations are optimal for the operator, and a credible mechanism guarantees that a single individual can detect a deviation from it. \cite{hakimov2023improving}  and \cite{moller:24} 
study transparency in the context of allocation problems without money.
Like in our paper, their notions of transparency do not restrict attention to optimal deviations. The authors, however, maintain 
the assumptions of public information or private communication from the designer, and they study deviation detection by a single individual. 

Whereas these recent and concurrent papers study specific settings and communication environments, we take a context-free approach and provide an informational measure of auditability to evaluate mechanisms in general social choice environments. One important conceptual novelty of our analysis is that we consider deviation detection through the information of a group of individuals instead of a single individual, and the size of the group emerges as a natural measure of auditability.  Our auditability notion proves tractable and informative in various applications, and our findings on the stark contrasts of prominent mechanisms' auditability have no counterpart in the literature. 

Our paper also relates to the literature on the role of privacy in market design. Without privacy concerns, the central authority could achieve more transparency through some (credible) communication of all participants' type reports and outcomes. When there are privacy concerns, one would prefer that transparency is achieved with minimal informational leakage. 
One implication of our auditability index is that it measures the smallest amount of private information needed for exposing deviations from a mechanism. Hence, a mechanism with a lower auditability index may achieve trust and transparency with lower privacy (and economic) costs. To the best of our knowledge, the privacy question in object allocation (without money) is relatively underexplored. In a recent paper, \cite{haupt/hitzig:23} examines privacy-preserving implementation (without cryptographic commitments) of prominent object allocation mechanisms and auctions through `sequential elicitation protocols'. Specifically, whereas our auditability notion tells how much privacy can be preserved when exposing a deviation, \cite{haupt/hitzig:23} asks whether privacy can be protected in the sense of not eliciting information that is inconsequential for computing the outcome. \cite{Ollar2021} analyzes the influence of privacy on efficiency in a dynamic setup from a market design perspective.\footnote{For a recent survey on privacy in economic theory, see, for instance, \cite{acquisti2016economics}.}   

There is another parallel strand of literature, mainly in the context of auctions, that studies the possibility of improving transparency through cryptographic protocols without sacrificing privacy (e.g., \citep{brandt:01,ferreira/weinberg:20,parkes2006practical,parkes2009cryptographic}) and blockchain (e.g., \citep{chitra/ferreir/kulkarni:23}).\footnote{In a similar context, \cite{canetti/fiat/gonczarowki:23}) study the possibility of trustworthy mechanisms that preserve the privacy of the central authority (i.e., reveal no information about the mechanism), as opposed to maintaining privacy of market participants.} Our study complements this line of literature: we study a situation where transparency and privacy considerations are essential, yet cryptographic protocols are unavailable or impractical. 

On a higher level, our study of auditability contributes to the theoretical understanding of the structure of matching and allocation mechanisms. Naturally, the notion of auditability is closely related to how a single individual (or a group of individuals) can affect the mechanism's outcome by only changing her own type report. \cite{arnosti:20} has discussed this question and notes that the DA mechanism is `unpredictable' in this sense. \cite{gonczarowski/thomas:23} explores this question and studies a parallel notion of `verifiability' in matching problems. \cite{pycia/unver:23} study a related but conceptually different auditability property in social choice. \cite{gangam/mai/raju/vazirani:23} analyze the notion of `robust' stable matching, which is simultaneously a stable matching for two problems that only differ by one type report. Interestingly, the authors highlight the implications of their theory to some notion of credibility or auditability for stable mechanisms. 

 \section{A Model of Auditability}
 \label{sec:allocation_problems}

\subsection{Preliminaries}
\label{sec:model}

Consider the problem of allocating a set of objects $\mathcal{O}$ to a set of individuals $\mathcal{I}$, with $|\mathcal{I}| = |\mathcal{O}| = N \geq 2$. Each individual $i$ has a type report $\theta_i \in \Theta_i$, and $\Theta \subseteq \times_{i \in \mathcal{I}} \Theta_i$ denotes the set of feasible type report profiles. We refer to an element $\theta \defeq (\theta_i)_{i \in \mathcal{I}} \in \Theta$ as an allocation problem, or simply, a problem.

For a problem $\theta$ and a subset of individuals $I \subseteq \mathcal{I}$, let $\theta_I = (\theta_i)_{i \in I}$ denote the type reports of individuals in $I$, and let $\theta_{-I} = (\theta_i)_{i \in \mathcal{I} \setminus I}$ denote the type reports of individuals not in $I$.

For now, we will keep the setup general and not specify what a type report $\theta_i$ of individual $i$ stands for. In Section \ref{section:priority-based}, an individual's type report is represented by a pair of a preference ranking and a vector of priority scores over objects (priority-based allocation). In Section \ref{section:house-allocation}, type reports are preference rankings only (house allocation). 

An allocation is a bijection $\omega: \mathcal{I} \rightarrow \mathcal{O}$. The space of all allocations is denoted by $\mathcal{M}$.  A mechanism is a mapping $\varphi: \Theta \rightarrow \mathcal{M}$ that gives an allocation $\varphi(\theta)$ for every problem $\theta \in \Theta$.

A deviation from mechanism $\varphi$ at problem $\theta$ is an allocation $\omega \neq \varphi(\theta)$. A non-empty set of individuals $I \subseteq \mathcal{I}$ \textbf{detects the deviation} $\omega$ at problem $\theta$ if for any $\theta_{-I}'$ such that $(\theta_I, \theta_{-I}') \in \Theta$, we have
\begin{equation*}
\omega(i) \neq \varphi(\theta_I, \theta_{-I}')(i) \text{ for some } i \in I.
\end{equation*}

For a given mechanism $\varphi$ and problem $\theta$, consider the smallest integer $n \in \mathbb{N}$, such that for any deviation $\omega \neq \varphi(\theta)$, there is a subset of individuals $I \subseteq \mathcal{I}$ that detects the deviation $\omega$ at problem $\theta$, and $|I| \leq n$. We refer to this number as the \textbf{auditability index} of $\varphi$ at problem $\theta$, and we denote it by $\scalebox{1.1}{$\#$} \varphi^{\theta}$.  That is, 

\begin{definition}
\label{def:auditability-problem-specific}
The auditability index of mechanism $\varphi$ at problem $\theta$ is 
\[
\scalebox{1.1}{$\#$} \varphi^{\theta} = \max_{\omega \neq \varphi(\theta)} \min \Big\{ |I| : I \subseteq \mathcal{I}, I \textit{ detects the deviation} \Big\}. 
\]
\end{definition}

As we explained in our introduction, deviations can be caused by errors, operators' intentional actions, or uncoordinated fraudulent actions by multiple participants. Our notion allows us to address all these examples. Thus, our approach differs from previous mechanism design literature. These works focus solely on operators' intentional deviations in strategic settings \citep{akbarpour/li:20, baliga1997theory, bester2000imperfect}.

We additionally define a \textit{worst-case} measure of auditability, which corresponds to the maximum of the auditability indices across all problems. This brings another layer of robustness to our auditability analysis.

\begin{definition}
\label{def:auditability:worst-case}
The worst-case auditability index of mechanism $\varphi$ is 
\[ \scalebox{1.1}{$\#$} \varphi = \max_{\theta \in \Theta} \varphi^{\theta}. \]
\end{definition}

It is immediate from the definition that $1 \leq \scalebox{1.1}{$\#$} \varphi \leq N$. In Section \ref{sec:index_one}, we will show that only a restrictive class of allocation mechanisms have a worst-case auditability index of one. Hence, a reasonable range for the worst-case auditability index is between $2$ and $N$.

We will mainly be interested in mechanisms in the two extremes of the auditability spectrum: (i) mechanisms with $\scalebox{1.1}{$\#$} \varphi = 2$, which we call \textbf{maximally auditable}, and (ii) mechanisms with $\scalebox{1.1}{$\#$} \varphi = N$, which we call \textbf{minimally auditability}. 

We offer several interpretations for our auditability notion. First, our auditability index indirectly measures the likelihood of market participants detecting deviations in a decentralized fashion. For instance, when focusing on minimally and maximally auditable mechanisms, one can find a stark contrast between their deviation-detecting probabilities for (m)any realistic network formation among individuals. We illustrate this with a simple Erdős–Rényi network, where connected nodes are interpreted as individuals accessing one others' types and outcomes. 
 
Consider an i.i.d. uniform random graph where the nodes are the individuals, and each pair of nodes is connected with some probability $p(N)$ that depends on the problem size $N$. Given a problem and a deviation, the probability of detecting the deviation is at least as large as the probability that the nodes of individuals in some deviation detection group are linked in this random graph. Thus, for a maximally auditable mechanism, for any problem and deviation, the probability of detecting the deviation is at least as large as the probability that a particular pair of nodes are linked. When $p(N)$ is in the order of a function between $\frac{1}{N}$ and $\frac{\ln N}{N}$, this probability is strictly positive \citep{erdos/renyu:60}.\footnote{For this range of $p(N)$, the graph will have a `giant component', i.e., one that contains a positive fraction of the nodes. The probability that the relevant pair belongs to this giant component is strictly positive.} Hence, for this range of $p(N)$, any deviation from a maximally auditable mechanism is detected with a positive probability. In contrast, for the same range of $p(N)$, the probability that the graph is (fully) connected is asymptotically zero. Hence, for a minimally auditable mechanism, there is zero probability of detecting an adversarial deviation at some problem.

Second, our auditability index can be informative in a setting with imperfect internal or third-party audit. For example, suppose the auditor observes only some fraction $p \in (0,1)$ of type reports and outcomes. If a mechanism is maximally auditable, any deviation will be detected if a particular pair is included in the auditing sample. This happens with some positive probability, asymptotically close to $p^2$. In contrast, if the auditability index of a mechanism is $N$, then the probability of detecting an adversarial deviation is zero for any $p \in (0,1)$ even when $p$ is arbitrarily close to $1$. 

Finally, our auditability index can be interpreted as the economic and privacy cost of proving deviations.
 If a mechanism has a low auditability index, then a small amount of information is sufficient to expose a deviation. For example, suppose an auditor wants to make a court case against a centralized platform regarding not implementing the promised allocation mechanism. If the auditability index of a mechanism is two, the court case can be settled by revealing the type reports of just two participants. On the other hand, if the auditability index of a mechanism is $N$, the auditor possibly needs to reveal all the information to prove the deviation.\footnote{\cite{pycia/unver:23} study a related property and offer a similar interpretation for their auditability notion.} 

\section{Priority-Based Allocation}
\label{section:priority-based}

The priority-based allocation model covers real-life assignment problems, including school choice and college admissions. Each individual $i \in \mathcal{I}$ has a strict preference ranking $P_i$ over objects $\mathcal{O}$, and $P= (P_i)_{i \in \mathcal{I}}$ denotes a preference ranking profile of all individuals. The space of all preference ranking profiles is $\mathcal{P}$. Each individual $i \in \mathcal{I}$ has object-specific priority scores $r_i \in \mathbb{R}^N$, and $r= (r_i)_{i \in \mathcal{I}}$ denotes the priority scores profile of all individuals. For individuals $i$ and $j$ and an object $o \in \mathcal{O}$, $r_{io} > r_{jo}$ indicates that $i$ has a higher priority at $o$ than $j$. We assume that the priority score profile is in $\mathcal{R} \subseteq (\mathbb{R}^N)^N$, which is defined as the set of all points in $(\mathbb{R}^N)^N$ that do not share any coordinate. That is, for every $r \in \mathcal{R}$, $i,j \in \mathcal{I}$, and $o \in \mathcal{O}$, we assume that $r_{io} \neq r_{jo}$.
We interpret $r \in \mathcal{R}$ as having strict priorities.\footnote{In some allocation problems, such as school choice, applicants are categorized into a few priority classes, and strict priorities are obtained only after ties are broken using randomly drawn lottery numbers. In our setup, the $r$'s denote these strict priority scores obtained after the tie-breaking, i.e., the priority group plus the random number. We think of the refined priorities as private information in our model. In several school districts, such as New York City, the random numbers are privately known by applicants, which is consistent with our setup.} Individual $i$'s type report is $\theta_i = (P_i,r_i)$. The set of feasible problems is $\Theta = \mathcal{P} \times \mathcal{R}$.

Given a problem $\theta = (P,r)$, we say that $o$ is the $n$-th most preferred object for individual $i$ at preference ranking $P_i$ if $\big| \{ o' \in \mathcal{O} : o' P_i o \} \big| = n-1$. In that case, we also say that $i$ ranks $o$ in the $n$-th position or that $o$ has the $n$-th position in $i$'s preference ranking. We use `better position' to refer to a position corresponding to a smaller $n$. 
 
We start by describing the (individual-proposing) \textit{Deferred Acceptance (DA)} algorithm of \cite{gale/shapley:62}. We use the DA to define a class of mechanisms. For this class, we provide a general characterization result for having a problem-specific and worst-case auditability index of two.

\textit{\textbf{Deferred Acceptance (DA) for input $\theta$:} Start with Step 1.}

\textit{ \textbf{Step $t \geq 1$}. Each $i$ claims her most preferred object according to $P_i$, among those that have not rejected her. Each object $o$ that some individuals claim is tentatively matched to the claimant with the highest priority score at $o$ and rejects the rest. If no rejections exist, the algorithm terminates, and the tentative assignments are finalized. Otherwise, we proceed to Step $t+1$.}

Take any problem $\theta = (P, r)$ and denote the outcome of the DA mechanism by $DA(\theta)$.  We proceed with our new class of mechanisms, which modify the priority scores by a mapping $\tau: \Theta \rightarrow \mathcal{R}$, and then we compute the outcome of the mechanism by applying DA to the modified problem $(P, \tau(\theta))$.

\begin{definition}
\label{def:da-representation}
We say a mechanism $\varphi$ is \textbf{DA-representable} if there is a mapping $\tau: \Theta \rightarrow \mathcal{R}$, such that for any problem $\theta = (P,r)$,
\[ \varphi(\theta) = DA(P, \tau(\theta)),\]    
and $\tau$ satisfies the following conditions for arbitrary problems $\theta = (P,r)$ and $\theta' = (P',r')$, individuals $i,j \in \mathcal{I}$ and object $o \in \mathcal{O}$. 

\begin{itemize}

\item \textbf{Independence of Irrelevant Alternatives.} 
Suppose (i) $o$ has the same position in the preference rankings $P_i$ and $P_i'$, (ii) $o$ has the same position in the preference rankings $P_j$ and $P_j'$, and (iii) $r_{io} > r_{jo} \iff r_{io}' > r_{jo}'$. Then \[ \tau_{io}(\theta) > \tau_{jo}(\theta) \iff \tau_{io}(\theta') > \tau_{jo}(\theta'). \]

    \item \textbf{Monotonicity.} Suppose (i) $o$ has a weakly better position in the preference ranking $P_i'$ compared to $P_i$, 
    (ii) $o$ has a weakly worse position in the preference ranking $P_j'$ compared to $P_j$, and (iii) $r_{io} > r_{jo} \implies r_{io}' > r_{jo}'$. Then 
\[
\tau_{io}(\theta) > \tau_{jo}(\theta) \implies \tau_{io}(\theta') > \tau_{jo}(\theta').
\]

      \item \textbf{Equal Treatment.} Suppose (i) $o$ has the same position in preference rankings $P_i$ and $P_j$, and (ii) $r_{io} > r_{jo}$. Then 
    \[
    \tau_{io}(\theta) > \tau_{jo}(\theta).
    \]

\end{itemize}
\end{definition}

DA-representable mechanisms cover many well-studied mechanisms. The DA mechanism is DA-representable through the identity projection $\tau$ of priority scores (i.e., $\tau(P,r) = r$).  Below, we give examples of how a DA presentation can be constructed for IA and the whole class of Application Rejection Mechanisms \citep{chen/kesten:17}. DA-representable mechanisms also cover the First-Priority-First mechanisms \citep{pathak/sonmez:13}, the Secure Immediate Acceptance mechanism \citep{dur/hammond/morrill:19}, and the French-tie-breaking mechanism \citep{bonkoungou:19}. In Appendix \ref{appendix:da-representable}, we discuss other classes of mechanisms studied in the literature and their relation to the DA-representable mechanisms.

Application-rejection (AR) mechanisms are an essential subclass of the DA-representable mechanisms, and they include both IA and DA as special cases. The standard descriptions of the AR and IA are provided in Appendix \ref{app:algorithms}.

\textbf{DA-representation of IA.} 
Consider the following mapping $\tau$. For any $\theta = (P, r)$, $i,j \in \mathcal{I}$, and $o \in \mathcal{O}$, $\tau_{io}(\theta) > \tau_{jo}(\theta)$
if and only if either $o$ is ranked in a strictly better position at $P_i$ than at $P_{j}$, or $o$ is ranked in the same position at $P_i$ and $P_{j}$, and $r_{io} > r_{jo}$.

\textbf{DA-representation of AR.} 
For a given natural number $e \in \mathbb{N}$, let $AR_e$ denote the AR mechanism corresponding to parameter $e$.  Fix an $e \in \mathbb{N}$. We say that $o$ is a \textit{tier} $t \in \mathbb{N}$ object for $i$ if $e(t-1) \leq \big| \big\{ o' \in \mathcal{O} : o' P_i o \big\} \big| < et$. Given any $e \in \mathbb{N}$, consider the following mapping $\tau^e$. For any $\theta = (P, r)$, $i,j \in \mathcal{I}$ and $o \in \mathcal{O}$, $\tau_{io}^e(\theta) > \tau_{jo}^e(\theta)$
if and only if $i$ ranks $o$ in a better (more preferred) tier than $j$, or $i$ and $j$ rank $o$ in the same tier and $r_{io} > r_{jo}$. We define $AR_e$ as the mechanism that is DA-representable through $\tau^e$. When $e=1$, $AR_e$ is equivalent to $IA$, and for any $e \geq N$, $AR_{e}$ is equivalent to $DA$. When $2 \leq e < N$, $AR_e$ is known as a \textit{Chinese parallel} mechanism \citep{chen/kesten:17}.

Before stating the main result of this section, we introduce a few more definitions.
For a given $\theta = (P,r)$, we say that an outcome $\omega$ is \textbf{stable} at problem $\theta$ if there are no two individuals $i,j \in \mathcal{I}$ and an object $o \in \mathcal{O}$ such that (1) $o P_i \omega(i)$, (2)  $\omega(j) = o$, and (3) $r_{io} > r_{jo}$. Here is an important implication of the DA-representation: if $\varphi$ is DA-representable through a mapping $\tau$, then for any problem $\theta = (P,r)$, the outcome $\varphi(\theta)$ is stable at $(P,\tau(\theta))$.\footnote{That is, in our definition stability can also hold with respect to modified priorities and is thus a generalization of the standard stability notion from \citep{gale/shapley:62} or notion of fairness in school choice \citep{balinski/sonmez:99,abdulkadiroglu/sonmez:03}. For related ideas that build on generalizing stability in this way, see, for instance, \cite{ayoade/papai:23}, \cite{dur2022deduction} or \cite{afacan2013alternative}.} This observation follows directly from the definition of DA-representability and stability of the DA with respect to any input of preferences and scores \citep{gale/shapley:62}. 

Given a problem $\theta = (P,r)$, we say that an allocation $\omega$ is \textbf{sufficiently undesirable} for individuals 
$i$ and $j$, if the following conditions hold:
\begin{itemize}
    \item $i$ and $j$ prefer each other's objects at $\omega$ to their own, that is, $\omega(j) P_i \omega(i)$ and $\omega(i) P_j \omega(j)$, 
    \item there is no object $o$ that both $i$ and $j$ prefer less than their own objects, that is, for any $o \notin \{\omega(i), \omega(j)\}$, $o P_i \omega(i)$ or $o P_j \omega(j)$.
\end{itemize}

\noindent In the following theorem, we fully characterize all problems where a given DA-representable mechanism has an auditability index of two. We also fully characterize all DA-representable mechanisms with a worst-case auditability index of two. Our results reveal a close connection between maximal auditibility of DA-representable mechanisms and uniqueness of (desirable) stable allocations in the modified problems. 

\begin{theorem}
\label{thm:DA-representable-full-characterization-two}
Let $\varphi$ be an arbitrary DA-representable mechanism. Then,

\begin{enumerate}
    \item for a given problem $\theta = (P,r)$, $\scalebox{1.1}{$\#$} \varphi^{\theta} = 2$ if and only if either $\varphi(\theta)$ is the unique stable allocation at problem $(P,\tau(\theta))$, or there are stable allocation other than $\varphi(\theta)$, but any such allocation is sufficiently undesirable for some pair of individuals; 

    \item $\scalebox{1.1}{$\#$} \varphi = 2$ if and only if point 1 holds for all problems. 
\end{enumerate}
\end{theorem}

Theorem \ref{thm:DA-representable-full-characterization-two} has important corollaries. First, we can use the result to compute the worst-case auditability index of IA. 

 \begin{proposition}
 \label{cor:IA}
$\scalebox{1.1}{$\#$} IA = 2$. 
 \end{proposition}
 
 \begin{proof}
By Theorem \ref{thm:DA-representable-full-characterization-two}, it is sufficient to show that for any problem $\theta = (P,r)$ there is a unique stable allocation at problem $(P, \tau(\theta))$, where $\tau$ is the mapping in the DA-representation of IA.

 Consider an arbitrary problem $\theta = (P,r)$, and let $\hat{r} = \tau(\theta)$. To show that there is a unique stable allocation at problem $(P,\hat{r})$, we prove the following equivalent claim.
 
 \textbf{Claim.} At every step $t$ of the implementation of IA, if an individual $i$ receives an object $o$, then $i$ receives $o$ at every stable allocation of problem $(P,\hat{r})$.
 
 We prove the claim by induction. Suppose $i$ receives $o$ at step $1$ of the implementation of IA. Then $o$ is the most preferred object of $i$, and she has the highest original priority score at $o$ among all individuals for whom $o$ is the most preferred object. 
 Hence, by construction of $\hat{r}$, individual $i$ has the highest modified priority score at $o$, i.e., $i = \argmax_{j \in \mathcal{I}} \hat{r}_{jo}$. Thus, $i$ must receive $o$ at every stable allocation of problem $(P,\hat{r})$. 
 
Suppose the claim holds for steps $1,2,\dots,t-1$, and $i$ receives $o$ at step $t$ of the implementation of IA. Then $o$ is the $t$-th most preferred object of $i$ according to $P_i$, and there is no other available individual $j\neq i$ who ranks $o$ in a better position than $t$ on $P_j$. Hence, by the construction of $\hat{r}$, individual $i$ has the highest modified priority at $o$ among all available individuals. Thus, $i$ receives $o$ at every stable allocation of problem $(P,\hat{r})$. This completes the proof of the claim and, therefore, the proof of Proposition \ref{cor:IA}.  \end{proof}

Another direct corollary of Theorem \ref{thm:DA-representable-full-characterization-two} is a complete characterization of all problems where DA has an auditability index of two. 

\begin{proposition}
\label{thm:DA-auditable-index<=2}
For a problem $\theta = (P,r)$,
$\scalebox{1.1}{$\#$} DA^{\theta} = 2$ if and only if either $DA(\theta)$ is the unique stable allocation at problem $(P,\tau(\theta))$, or there are stable allocations other than $DA(\theta)$, but any such allocation is sufficiently undesirable for some pair of individuals. 
\end{proposition}

This gives a relatively simple characterization of all problems for which DA has an auditability index of two. The necessary conditions for having $\scalebox{1.1}{$\#$} DA^{\theta} = 2$ are generally restrictive. One can use the characterization result to compute the proportion of problems for which DA achieves an auditability index of two, either computationally or analytically. In Appendix \ref{appendix:DA-problem-specific}, we give a sufficient condition for which the problem-specific auditability index is larger than $M$ for any $M \in \{1,2,\dots,N\}$. This result shows that the DA tends to have a large auditability index for almost all problems. 

Next, we show that DA is a minimally auditable mechanism. We prove this as a part of a more general proposition.

\begin{proposition}
\label{prop:AR}
$\scalebox{1.1}{$\#$} AR_e = N$ for any $e > 1$. In particular, $\scalebox{1.1}{$\#$} DA = N$.
\end{proposition}

\begin{proof}
Consider an arbitrary $AR_e$ with $e > 1$. To prove that $\scalebox{1.1}{$\#$} AR_e = N$, we need to construct a problem $\theta$ and a deviation $\omega \neq AR_e(\theta)$, such that no proper subset of individuals detects this deviation.

Suppose the individuals and objects are indexed, i.e., $\mathcal{I} = \{i_1,i_2,\dots,i_N\}$ and $\mathcal{O} = \{o_1,o_2,\dots,o_N\}$. Consider the following problem $\theta = (P,r)$:
\begin{itemize}
    \item $o_1 P_{i_1} o_2 P_{i_1} o_n$ and $o_2 P_{i_2} o_1 P_{i_2} o_n$ for all $n \in \{3,\dots,N\}$,
    \item $o_n P_{i_n} o_m$ for all $n \in \{3,\dots,N\}$ and for all $m \neq n$,
    \item $r_{i_2 o_1} > r_{i_n o_1} > r_{i_1 o_1}$ and $r_{i_1 o_2} > r_{i_n o_2} > r_{i_2 o_2}$ for all $n \in \{3,\dots,N\}$,
\item otherwise, the preferences and priority scores are arbitrary.
\end{itemize}

The constructed problem can be illustrated by the table below:

\begin{center}
\begin{tabular}{ccccccccccccc}
$i_1$ & $i_2$ & $i_3$ & $\cdots$ & $i_N$ & & & & $o_1$ & $o_2$ & $o_3$ & $\cdots$ & $o_N$  \tabularnewline
\hline 
$o_1$ & $o_2$ & $o_3$ & $\cdots$ & $o_N$ & & & & $i_2$ & $i_1$ & $\cdots$ & $\cdots$ & $\cdots$  \tabularnewline
$o_2$ & $o_1$ & $\cdots$ & $\cdots$ & $\cdots$ & & &  & $\cdots$  & $\cdots$ & $\cdots$  & $\cdots$ & $\cdots$  \tabularnewline
$\cdots$ & $\cdots$  & $\cdots$  & $\cdots$  & $\cdots$ & & & & $\cdots$ & $\cdots$ & $\cdots$ & $\cdots$ & $\cdots$ \tabularnewline
$\cdots$  & $\cdots$  & $\cdots$  & $\cdots$  & $\cdots$ & & & & $i_1$ & $i_2$ & $\cdots$ & $\cdots$ & $\cdots$ \tabularnewline
\end{tabular}
\par\end{center}

Under $AR_e(\theta)$, all individuals are matched to their most preferred objects with respect to $P$. That is, $AR_e(\theta)(i_n) = o_n$ for all $n \in \{1,2,\dots,N\}$. Now consider the deviation $\omega \neq AR_e(\theta)$ that differs from $AR_e(\theta)$ by that $\omega(i_1) = o_2$ and $\omega(i_2) = o_1$. 

Let $I \subsetneq \mathcal{I}$ be an arbitrary proper subset of individuals. We will show that $I$ does not detect the deviation. Consider the following cases:

(i) $i_1 \notin I$ or $i_2 \notin I$. Without loss of generality, suppose $i_1 \notin I$. 
Consider the problem $\tilde{\theta}$ that differs from $\theta$ by only that $i_1$ ranks $o_2$ as her first choice. Then, $AR_e(\theta_{I}, \tilde{\theta}_{-I})  = \omega$. Thus, $I$ does not detect the deviation $\omega$.

(ii) $i_1, i_2 \in I$.  Consider an arbitrary $i_n \in \mathcal{I} \setminus \big(I \cup \{i_1,i_2\} \big)$. Such an $i_n$ exists because $I \subsetneq \mathcal{I}$ and $i_1,i_2 \in I$. Consider the problem $\tilde{\theta}$ that differs from $\theta$ by that $i_n$ ranks $o_1$ as her first choice, and $o_n$ as her second choice.
Then, $AR_e(\theta_{I}, \tilde{\theta}_{-I})  = \omega$.  Thus, $I$ does not detect the deviation $\omega$. This completes the proof.
\end{proof}

Propositions \ref{cor:IA} and \ref{prop:AR} reveal a strong contrast among the AR mechanisms: namely, $\scalebox{1.1}{$\#$} AR_1 = 2$, and $\scalebox{1.1}{$\#$} AR_e = N$ for all $e > 1$. 

A natural question is whether one can facilitate detection by providing additional data or statistics (e.g., cutoffs, aggregate match data, prices).\footnote{See, for instance, \cite{budish2011combinatorial}, \cite{he2018pseudo}, \cite{echenique2021constrained}, \cite{hylland/zeckhauser:79} for the role of prices in the context of matching.}  For instance, a widely used instrument in school choice and college admission is to publish admission thresholds or cutoffs. Another example is the famous NRMP, which provides aggregate match data of residency matchings.\footnote{https://www.nrmp.org/match-data-analytics/residency-data-reports/ .} 
Proposition \ref{prop:AR} is robust to the joint public disclosure of the entire allocation, all reported preferences, and cutoffs. We show this in Appendix \ref{prop:ARalternative}.

In Appendix \ref{appendix:stable-dominating}, we show that many other well-studied mechanisms are minimally auditable. This includes, for instance, all mechanisms that weakly Pareto dominate a stable one \citep{alva2019stable}, such as any stable mechanism and the \textit{Efficiency Adjusted DA} of \cite{kesten:10}.

\section{House Allocation}
\label{section:house-allocation}

In this section, type reports consist of preference rankings only. This is also known as the house allocation problem, with applications like public housing, dormitory room assignment, and organ allocation. For each $i \in \mathcal{I}$, a type report $\theta_i$ now takes the form of a strict preference ranking $P_i$ over objects $\mathcal{O}$. The set of feasible problems is $\Theta = \mathcal{P}$ (i.e., the space of all preference ranking profiles). 
To simplify the analysis and the exposition, we assume $N\geq 6$.

We investigate the auditability properties of hierarchical exchange mechanisms, which constitute the entire class of Pareto efficient, strategy-proof, non-bossy, and reallocation-proof mechanisms \citep{papai:00}. 

We start with some terminology and definitions. A suballocation for $I \subseteq \mathcal{I}$ and $O \subseteq \mathcal{O}$, with $|I|=|O|$, is an allocation $\sigma : I \rightarrow O$ restricted to individuals $I$ and objects $O$. Let $\mathcal{S}$ be the set of all suballocations. 
Denote with $\sigma_I$ the set of individuals matched under suballocation $\sigma$. Let $\bar{I}(\sigma) \defeq I \setminus \sigma_I$ and $\bar{O}(\sigma)$ be the set of unmatched individuals and objects under $\sigma$, respectively. The empty suballocation is denoted with $\sigma_{\emptyset}$.

\begin{definition}\label{definition: ownership-rights}
    An \textit{ownership structure} is a collection of mappings 
    \begin{displaymath}
    c \defeq \{c_{\sigma}: \bar{O}(\sigma) \rightarrow \bar{I}(\sigma) \}_{\sigma \in \mathcal{S}}
    \end{displaymath}
\end{definition}

That is, for a given suballocation $\sigma$ and unmatched object $o \in \bar{O}(\sigma)$, the mapping $c_{\sigma}$ appoints the unmatched individual $c_{\sigma}(o) \in \bar{I}(\sigma)$ as the unique owner of $o$ at $\sigma$. Let $\mathcal{C}$ be the space of all ownership structures. For any $\sigma,\hat{\sigma} \in \mathcal{S}$, $\sigma \subseteq \hat{\sigma}$ means that each individual-object pair matched under $\sigma$ is also matched under $\hat{\sigma}$. A suballocation $\sigma$ has size $n \in \mathbb{N}$, if $|\sigma_I|=n$. A suballocation $\sigma$ is larger (smaller) than another suballocation $\hat{\sigma}$ if $\sigma$ has a larger (smaller) size. 

An ownership structure $c \in \mathcal{C}$ is \textit{consistent} if for any pair of suballocations $\sigma,\hat{\sigma} \in \mathcal{S}$, with $\sigma \subseteq \hat{\sigma}$, when $i \in \bar{I}(\hat{\sigma})$ owns an object $o \in \bar{O}(\hat{\sigma})$, then $i$ owns $o$ at $\sigma$. 
Any hierarchical exchange mechanism $\varphi$ is induced by running a TTC algorithm on a consistent ownership structure $c \in \mathcal{C}$ \citep{pycia/unver:17}.

\textit{\textbf{TTC for input $\theta$:}}

\textit{Denote $\sigma_{t-1}(\theta)$ as the suballocation among individuals and objects matched before step $t \in \mathbb{N}$. Prior to the first step, the suballocation is empty, i.e., $\sigma_{0}(\theta)=\sigma_{\emptyset}$. In step $t \in \mathbb{N}$ and for some $n' \in \mathbb{N}$, there is a \textit{cycle} at $\sigma_{t-1}(\theta)$}
\begin{displaymath}
o_1 \rightarrow i_1 \rightarrow \dots o_{n'} \rightarrow i_{n'} \rightarrow o_1
\end{displaymath}
\textit{in which individual $i_l \in \bar{I}_{\sigma_{t-1}(\theta)}$ points to $o_{l+1} \in \bar{O}_{\sigma_{t-1}(\theta)}$, and object $o_l$ points to individual $i_l$ for all  $l \in \{1,...,n'\}$ and superscripts are added modulo $n'$.} 

Let $\varphi^c$ be the hierarchical exchange mechanism induced via consistent ownership structure $c \in \mathcal{C}$. For the rest of this section, fix an arbitrary consistent $c \in \mathcal{C}$ and a hierarchical exchange mechanisms $\varphi^c$. Then a suballocation $\sigma \in \mathcal{S}$ is \textit{on-path} on $\varphi^c$ if there exists $\theta \in \Theta$ such that $\sigma = \sigma^{t-1}(\theta)$ for some step $t \in \mathbb{N}$ in the course of the TTC algorithm. Unless mentioned otherwise, whenever we refer to a (generic) suballocation $\sigma \in \mathcal{S}$, then we mean a suballocation $\sigma$ that is on-path under $\varphi^c$. Suballocations that are not on-path will not play a role in the analysis.

\begin{proposition}
    \label{prop:TTC}
If there exists a suballocation that has at least $n \in \mathbb{N}$ owners, then  $\# \varphi^c \geq n$.
\end{proposition}

In other words, whenever there is a problem such that there is a cycle of length $n$ in the TTC algorithm, then there is a problem where the auditability index is $n$.

Next, we introduce a new class of mechanisms central to our main characterization. Take any $i \in \mathcal{I}$. Given any $n \in \{1,\dots,N\}$, $i$ is a \textit{level}-$n$ \textit{owner} if $n-1$ is the size of the smallest suballocation at which $i$ is an owner of an object. If $i$ is in level-$1$, then she is a \textit{principal owner}. We say that $i$ is a \textit{vice owner}, if either $(i)$ there is a single principal owner and $i$ is in level-$2$ or level-$3$, or $(ii)$ there are two principal owners and $i$ is in level-$2$. Individual $i$ is a \textit{residual owner} if she is in level $N-1$ or $N$. If $i$ is no principal-, vice- or residual owner, then $i$ is a \textit{regular owner}.

\begin{definition}\label{def:viceownership}
We say $\varphi^c$ is a \textbf{vice ownership mechanism} if the following five conditions are satisfied:
 
\begin{enumerate}
\item[(1)] For each $n \in \{1,\dots N\}$, there are at most two level-$n$ owners.

\item[(2)] Given any two suballocations $\sigma,\hat{\sigma}$, if $\bar{O}(\sigma)=\bar{O}(\hat{\sigma})$ and $\bar{I}(\sigma)=\bar{I}(\hat{\sigma})$, then $c_{\sigma}=c_{\hat{\sigma}}$. 

\item[(3)] Given any $n \in \{3,\dots N\}$, if there are two level-$n$ owners, then in any $\sigma$ where only one level-$n$ owner is matched, the remaining level-$n$ owner owns all objects at $\sigma$. 

\item[(4)] 

Taky any suballocation $\sigma$. If there are two principal owners, then a principal owner is matched under $\sigma$, if all vice owners are matched under $\sigma$. Moreover, if a vice owner $i$ owns an object $o$ at $\sigma$ that some principal owner $k$ owns at $\sigma_{\emptyset}$, then, at any suballocation $\sigma''$ where $i$ and $o$ are unmatched , $i$ owns $o$ provided that

\begin{itemize}
\item[i.]  $o$ is not owned by a principal owner, and 
\item[ii.] there is no suballocation $\sigma'$ that is weakly smaller than $\sigma$, where $i$ is unmatched, while a vice owner $j\neq i$ is unmatched at $\sigma''$ and $j$ owns some $o'$ at $\sigma'$ that principal owner $k$ owns at $\sigma_{\emptyset}$.
\end{itemize}

\item[(5)] A regular owner owns an object at $\sigma$ if and only if 

\begin{itemize}
    \item[i.] she owns it at some $\sigma'$ where only strictly lower level owners are matched and
    \item[ii.] all individuals with a strictly lower level are matched under $\sigma$.
\end{itemize}

\end{enumerate}
\end{definition}
At its core, vice ownership mechanisms follow a straightforward structure: Principal owners and vice owners take precedence over all regular owners, and each regular owner has precedence over all regular owners from a higher level. Once all principal owners and vice owners are matched, the assignment of ownership rights follows a fixed ordering over levels, each capped at two members. Within each level of regular owners, the division of ownership rights is fixed across problems. Concretely, for each object, one of each level's members owns the object unless an individual from a strictly lower level is still unmatched. Finally, according to Definition \ref{def:viceownership} (2), the ownership for the two residual owners is entirely determined by the pair of objects remaining for them.

We are ready for the main result of this section.

\begin{theorem}\label{theorem:TTCcharacterization}
 $\# \varphi^c = 2$ if and only if it is a vice ownership mechanism.
\end{theorem}

The proof is primarily constructive. For example, consider the scenario where only regular owners are still unmatched and one of the conditions of Definition \ref{def:viceownership} is violated. One of the key arguments to show that there is a deviation that no two individuals can detect is as follows: We identify two suballocations $\sigma^1$ and $\sigma^2$ of minimal size, where in $\sigma^1$ some individual $i$ owns $o$, while in $\sigma^2$ another individual $j$ owns $o$. The deviation will only swap the allocations of $i$ and $j$. Specifically, if $\omega(j)= o$, while $i$ finds $\omega(i)$ less desirable than $o$ and $j$ finds $\omega(i)$  more desirable than $o$, any group of individuals that detects $\omega$ must infer that (1) $j$ received $o$ under $\omega$, (2) $\sigma^1\subsetneq\omega$, and (3) $i$ finds $o$ more desirable than $\omega(i)$. However, if the conditions in Definition  \ref{def:viceownership} are not satisfied, then this is impossible with just two individuals.

Conversely, suppose $\varphi^c$ is a vice ownership mechanism and consider any relative indeterminacy for the ownership right of an object as described in the last paragraph. If a pair of regular individuals cannot resolve the indeterminacy by comparing their levels, this means that one of the two individuals can resolve it with one of the principal or vice owners. By contrast, pairs that contain a principal or vice owner can often infer their respective ownership rights among themselves. More generally, being at the top of the hierarchy eases detection. In fact, it is a crucial driver for the more flexible ownership division among principal owners and vice owners (see Definition \ref{def:viceownership} (4)).

We also study a subset of hierarchical exchange mechanisms known as \textit{sequential dictatorships}. A hierarchical exchange mechanism $\varphi^c$ is a \textbf{sequential dictatorship} if each suballocation has at most one owner. A prominent special case of sequential dictatorships is the well-known \textit{serial dictatorship} \citep{satterthwaite1981strategy}. Under a serial dictatorship, individuals choose according to fixed dictatorial ordering. \cite{papai:01} has characterized sequential dictatorships through strategy-proofness, non-bossiness, and citizen sovereignty. Additionally, \cite{pycia/troyan:23} have demonstrated that sequential dictatorships meet stringent standards for strategic simplicity. More recently, \cite{moller:24} showed that sequential dictatorships are equivalent to the class of transparent, efficient, and strategy-proof mechanisms. It is important to remark that the findings presented here are distinct and not directly related to these previously established results.

We first show that even sequential dictatorships can be `almost' minimally auditable. 

\begin{proposition}\label{proposition:seq}
There exists a sequential dictatorship $\varphi^c$ with $\scalebox{1.1}{$\#$} \varphi^c = N-1$.  
\end{proposition}

Our construction of the deviation relies on violating a condition equivalent to Definition \ref{def:viceownership} (2) and exploits the relative indeterminacy of ownership for objects among two individuals. The mechanism we find works like a \textit{serial dictatorship} except for residual owners $i$ and $j$, which are unique. Moreover, there is a unique suballocation $\sigma$, where $i$ owns an object $o$ and $j$ is still unmatched. When reaching this suballocation through a problem where among the remaining objects $o$ is most desirable for $i$ and $j$, the constructed deviation $\omega$ swaps only the assignments of $i$ and $j$. Intuitively, any group that detects $\omega$ must verify that (i) $\sigma$ was realized at step $N-1$, and (ii) $i$ prefers $o$ to her assignment under the deviation. However, this would require a group of at least $N-1$ individuals. 

Theorem \ref{theorem:TTCcharacterization} implies that sequential dictatorships are maximally auditable if and only if they are a vice ownership mechanism. For our final result in this section, we define this class formally, 

\begin{definition}\label{def:vicedictatorship}
    
A sequential dictatorship $\varphi^c$ is a \textbf{vice dictatorship} if: 
\begin{enumerate}
\item There is a unique principal owner and there are at most two vice owners. 
\item level-$n$ is a singleton for each $n \in \{2,\dots, N-2\}$.
\end{enumerate}

\end{definition}
The result below is an immediate corollary of Theorem \ref{theorem:TTCcharacterization} and Definition \ref{def:vicedictatorship}. 

\begin{theorem}\label{theorem:vicedictatorship}
    A sequential dictatorship $\varphi^c$ is maximally auditable if and only if it is a vice dictatorship.
\end{theorem}

A short inspection of Definition \ref{def:vicedictatorship} reveals that vice dictatorships are vice ownership mechanisms that are `almost' serial dictatorial: The dictatorial ordering is fixed except for vice owners and the residual owners.

\section{Auditability Index of One: A Characterization Result}
\label{sec:index_one}

We provide necessary and sufficient conditions for a mechanism to have a (worst-case) auditability index of one. The characterization describes a restrictive class of mechanisms that violate mild (efficiency) properties. 

For a mechanism $\varphi$ and $i \in \mathcal{I}$, we say object $o$ is \textbf{possible} for individual $i$ at type report $\theta_i$, if $o = \varphi(\theta_i, \theta_{-i})$ for some $\theta_{-i}$. Given a subset of objects $O \subseteq \mathcal{O}$, we say an individual $i$ \textbf{clinches} object $o$ from the set $O$ at type report $\theta_i$, if $o$ is the only possible object in $O$ for $i$ at type report $\theta_i$. That is, there is no $o' \in O \setminus \{o\}$ such that $\varphi( \theta_i, \theta_{-i} )(i) = o'$ for some $\theta_{-i}$.

\begin{definition}
\label{def:clinching}
We say $\varphi$ has a \textbf{sequential clinching implementation} at problem $\theta$ if $\varphi(\theta)$ can be computed as the outcome of the following algorithm: Initially, all individuals and objects are available. Start with Step $1$. 

\textit{ \textbf{Step $t \geq 1$}. An available individual $i$ clinches some object $o$ from the set of available objects at type report $\theta_i$. The individual and the object become unavailable. If there is no available individual, the algorithm terminates. Otherwise, proceed to Step $t + 1$. }  
\end{definition}

Typically, there are multiple possible objects for an individual at a given type report. Hence, being able to clinch an object is a restrictive condition, and (most) mechanisms may not have a sequential clinching implementation.  
Interestingly, mechanisms with a sequential clinching implementation are the unique ones with an auditability index of one. 

\begin{theorem}
\label{characterization:pr-sp index_one}
Let $\varphi$ be an arbitrary mechanism. Then,
\begin{enumerate}
    \item for any $\theta \in \Theta$, $\scalebox{1.1}{$\#$} \varphi^{\theta} = 1$ if and only if $\varphi$ has a sequential clinching implementation at $\theta$;

    \item $\scalebox{1.1}{$\#$} \varphi = 1$ if and only if $\varphi$ has a sequential clinching implementation at any problem $\theta$. Moreover, the mechanism can be implemented with a sequential clinching order that only depends on the set of available objects at each step (but otherwise does not depend on the problem). 
\end{enumerate}
\end{theorem}

Sequential clinching implementation is different from serial dictatorial implementation. The former is much more restrictive. In fact, serial dictatorships do not have a sequential clinching implementation for a 'generic' problem. We discuss these points further in Appendix \ref{appendix:seq-clinching-application}. Our notion of clinching is also different from that of \cite{pycia/troyan:23}. 

To highlight the limitations of sequential clinching implementation, we demonstrate that all these mechanisms violate basic efficiency notions when type reports include preferences over the objects (as in the previous sections' applications). In particular, we show that mechanisms with an auditability index of one do not satisfy the full range property. A mechanism $\varphi$ has \textbf{full range} if, for any allocation $\omega \in \Omega$, there is a problem $\theta$ such that $\varphi(\theta)(i) = \omega(i)$ for all $i \in \mathcal{I}$.\footnote{In the context of house allocation (Section \ref{section:house-allocation}), the condition is also known as citizen sovereignty \citep{papai:01}.} In other words, a mechanism has full range if every allocation may be selected for some problem. 

\begin{proposition}
\label{prop:full-range}
If $\scalebox{1.1}{$\#$} \varphi = 1$, then $\varphi$ does not have a full range. 
\end{proposition}

In our main applications (Section \ref{section:priority-based} and Section \ref{section:house-allocation}), the full range property is implied by a weak notion of efficiency, which says that a mechanism should always assign everyone to their most preferred objects whenever feasible.  

\section{Concluding Remarks}
\label{section:concluding_remarks}

We introduce a novel informational framework for evaluating how easily or hard participants can detect deviations from a mechanism. Our theory is tractable and informative, and our findings may contribute to discussions on choosing a mechanism for real-life problems. Our framework of auditability can be used to compare and characterize mechanisms in other important social choice applications. In Appendix \ref{appendix:other-applications}, we apply our framework to analyze auditability in other important applications, such as auctions, voting, and choice with affirmative action. 

We find that the standard first-price and all-pay auction mechanisms are maximally auditable for the auction problem. In contrast, the second-price auction mechanism is minimally auditable. For voting problems with a binary outcome, we characterize the dictatorial voting rule as the unique voting rule with an auditability index of one, and we characterize the majority voting rule as the unique most auditable, anonymous/symmetric voting rule. We also consider a choice with affirmative action problem (e.g., \cite{hafalir/yebmez/yildirim:13}, \cite{echenique/yenmez:15}, \cite{dur/kominers/pathak/sonmez:18}), and study two well-known implementations of reserves mechanisms: the \textit{reserved-seats-first} and the \textit{open-seats-first}. We establish that the reserved-seats-first mechanism is more auditable than the open-seats-first mechanism and that the difference is often quantitatively substantial.

We believe that testing the practicality of our auditability measure by lab experiments may be a fruitful research direction. Understanding the computation complexity of detecting deviations and measuring auditability is another open question that we leave for future research.

\bibliography{bibmatching}

\begin{thebibliography}{61}
\newcommand{\enquote}[1]{``#1''}
\expandafter\ifx\csname natexlab\endcsname\relax\def\natexlab#1{#1}\fi

\bibitem[\protect\citeauthoryear{Abdulkadiro\u{g}lu and Grigoryan}{Abdulkadiro\u{g}lu and Grigoryan}{2021}]{abdulkadiroglu/grigoryan:21}
\textsc{Abdulkadiro\u{g}lu, A. and A.~Grigoryan} (2021): \enquote{{Priority-Based Assignment with Reserves and Quotas},} {No. w28689. National Bureau of Economic Research}.

\bibitem[\protect\citeauthoryear{Abdulkadiro\u{g}lu and S\"{o}nmez}{Abdulkadiro\u{g}lu and S\"{o}nmez}{2003}]{abdulkadiroglu/sonmez:03}
\textsc{Abdulkadiro\u{g}lu, A. and T.~S\"{o}nmez} (2003): \enquote{{School Choice: A Mechanism Design Approach},} \emph{{A}merican {E}conomic {R}eview}, 93, 729--747.

\bibitem[\protect\citeauthoryear{Acquisti, Taylor, and Wagman}{Acquisti et~al.}{2016}]{acquisti2016economics}
\textsc{Acquisti, A., C.~Taylor, and L.~Wagman} (2016): \enquote{The economics of privacy,} \emph{Journal of economic Literature}, 54, 442--492.

\bibitem[\protect\citeauthoryear{Afacan}{Afacan}{2013}]{afacan2013alternative}
\textsc{Afacan, M.~O.} (2013): \enquote{Alternative characterizations of Boston mechanism,} \emph{Mathematical Social Sciences}, 66, 176--179.

\bibitem[\protect\citeauthoryear{Akbarpour and Li}{Akbarpour and Li}{2020}]{akbarpour/li:20}
\textsc{Akbarpour, M. and S.~Li} (2020): \enquote{{Credible Auctions: A Trilemma},} \emph{Econometrica}, 88, 425--467.

\bibitem[\protect\citeauthoryear{Alva and Manjunath}{Alva and Manjunath}{2019}]{alva2019stable}
\textsc{Alva, S. and V.~Manjunath} (2019): \enquote{Stable-dominating rules,} Tech. rep., Working paper, University of Ottawa.

\bibitem[\protect\citeauthoryear{Arnosti}{Arnosti}{2020}]{arnosti:20}
\textsc{Arnosti, N.} (2020): \enquote{{Deferred Acceptance is Unpredictable},} {Blog Post, https://nickarnosti.com/blog/da-predictability/}.

\bibitem[\protect\citeauthoryear{Ayoade and Pápai}{Ayoade and Pápai}{2023}]{ayoade/papai:23}
\textsc{Ayoade, N. and S.~Pápai} (2023): \enquote{{School Choice with Preference Rank Classes},} \emph{Games and Economic Behavior}, 137, 317--341.

\bibitem[\protect\citeauthoryear{Baliga, Corchon, and Sj{\"o}str{\"o}m}{Baliga et~al.}{1997}]{baliga1997theory}
\textsc{Baliga, S., L.~C. Corchon, and T.~Sj{\"o}str{\"o}m} (1997): \enquote{The theory of implementation when the planner is a player,} \emph{Journal of Economic Theory}, 77, 15--33.

\bibitem[\protect\citeauthoryear{Balinski and S\"{o}nmez}{Balinski and S\"{o}nmez}{1999}]{balinski/sonmez:99}
\textsc{Balinski, M. and T.~S\"{o}nmez} (1999): \enquote{{A Tale of Two Mechanisms: Student Placement},} \emph{Journal of Economic Theory}, 84, 73--94.

\bibitem[\protect\citeauthoryear{Benner and Boser}{Benner and Boser}{2018}]{benner/boser:18}
\textsc{Benner, M. and U.~Boser} (2018): \enquote{{Expanding Access to High Quality Schools (CAP Review)},} {Center for American Progress}.

\bibitem[\protect\citeauthoryear{Bester and Strausz}{Bester and Strausz}{2000}]{bester2000imperfect}
\textsc{Bester, H. and R.~Strausz} (2000): \enquote{Imperfect commitment and the revelation principle: the multi-agent case,} \emph{Economics Letters}, 69, 165--171.

\bibitem[\protect\citeauthoryear{Bonkoungou}{Bonkoungou}{2019}]{bonkoungou:19}
\textsc{Bonkoungou, S.} (2019): \enquote{{Deferred Acceptance under the French Tie-Breaking Rule},} {Working Paper}.

\bibitem[\protect\citeauthoryear{Brandt}{Brandt}{2001}]{brandt:01}
\textsc{Brandt, F.} (2001): \enquote{{Cryptographic Protocols for Secure Second-Price Auctions},} \emph{In International Workshop on Cooperative Information Agents: Springer Berlin Heidelberg}, 154--165.

\bibitem[\protect\citeauthoryear{Budish}{Budish}{2011}]{budish2011combinatorial}
\textsc{Budish, E.} (2011): \enquote{The combinatorial assignment problem: Approximate competitive equilibrium from equal incomes,} \emph{Journal of Political Economy}, 119, 1061--1103.

\bibitem[\protect\citeauthoryear{Canetti, Fiat, and Gonczarowski}{Canetti et~al.}{2023}]{canetti/fiat/gonczarowki:23}
\textsc{Canetti, R., A.~Fiat, and Y.~A. Gonczarowski} (2023): \enquote{{Zero-Knowledge Mechanisms},} {arXiv preprint arXiv:2302.05590}.

\bibitem[\protect\citeauthoryear{Chen and Kesten}{Chen and Kesten}{2017}]{chen/kesten:17}
\textsc{Chen, Y. and O.~Kesten} (2017): \enquote{{Chinese College Admissions and School Choice Reforms: A Theoretical Analysis},} \emph{Journal of Political Economy}, 125, 99--139.

\bibitem[\protect\citeauthoryear{Chitra, Ferreira, and Kulkarni}{Chitra et~al.}{2023}]{chitra/ferreir/kulkarni:23}
\textsc{Chitra, T., M.~V. Ferreira, and K.~Kulkarni} (2023): \enquote{{Credible, Optimal Auctions via Blockchains},} {arXiv preprint arXiv:2301.12532}.

\bibitem[\protect\citeauthoryear{Dubins and Freedman}{Dubins and Freedman}{1981}]{dubins/freedman:81}
\textsc{Dubins, L.~E. and D.~A. Freedman} (1981): \enquote{{Machiavelli and the Gale-Shapley algorithm},} \emph{{A}merican {M}athematical {M}onthly}, 88, 485--494.

\bibitem[\protect\citeauthoryear{Dur, Hammond, and Morrill}{Dur et~al.}{2019}]{dur/hammond/morrill:19}
\textsc{Dur, U., R.~G. Hammond, and T.~Morrill} (2019): \enquote{{The Secure Boston Mechanism: Theory and Experiments},} \emph{Experimental Economics}, 22, 918--953.

\bibitem[\protect\citeauthoryear{Dur, Kominers, Pathak, and Sönmez}{Dur et~al.}{2018}]{dur/kominers/pathak/sonmez:18}
\textsc{Dur, U., S.~D. Kominers, P.~A. Pathak, and T.~Sönmez} (2018): \enquote{{Reserve Design: Unintended Consequences and the Demise of Boston’s Walk Zones},} \emph{Journal of Political Economy}, 6, 2457--2479.

\bibitem[\protect\citeauthoryear{Dur, Pathak, Song, and S{\"o}nmez}{Dur et~al.}{2022{\natexlab{a}}}]{dur2022deduction}
\textsc{Dur, U., P.~A. Pathak, F.~Song, and T.~S{\"o}nmez} (2022{\natexlab{a}}): \enquote{Deduction dilemmas: The Taiwan assignment mechanism,} \emph{American Economic Journal: Microeconomics}, 14, 164--185.

\bibitem[\protect\citeauthoryear{Dur, Pathak, Song, and Sönmez}{Dur et~al.}{2022{\natexlab{b}}}]{dur/pathak/song/sonmez:22}
\textsc{Dur, U., P.~A. Pathak, F.~Song, and T.~Sönmez} (2022{\natexlab{b}}): \enquote{{Deduction Dilemmas: The Taiwan Assignment Mechanism},} \emph{American Economic Journal: Microeconomics}, 14, 164--185.

\bibitem[\protect\citeauthoryear{Echenique, Miralles, and Zhang}{Echenique et~al.}{2021}]{echenique2021constrained}
\textsc{Echenique, F., A.~Miralles, and J.~Zhang} (2021): \enquote{Constrained pseudo-market equilibrium,} \emph{American Economic Review}, 111, 3699--3732.

\bibitem[\protect\citeauthoryear{Echenique and Yenmez}{Echenique and Yenmez}{2015}]{echenique/yenmez:15}
\textsc{Echenique, F. and B.~M. Yenmez} (2015): \enquote{{How to control controlled school choice},} \emph{American Economic Review}, 105, 2679--2694.

\bibitem[\protect\citeauthoryear{Ehlers and Klaus}{Ehlers and Klaus}{2003}]{ehlers/klaus:03}
\textsc{Ehlers, L. and B.~Klaus} (2003): \enquote{{Consistent House Allocation},} \emph{Barcelona, Unpublished mimeo}.

\bibitem[\protect\citeauthoryear{Erdős and Rényi}{Erdős and Rényi}{1960}]{erdos/renyu:60}
\textsc{Erdős, P. and A.~Rényi} (1960): \enquote{{On the Evolution of Random Graphs},} \emph{Publications of the Mathematical Institute of the Hungarian Academy of Sciences}, 5, 17--60.

\bibitem[\protect\citeauthoryear{Ferreira and Weinberg}{Ferreira and Weinberg}{2020}]{ferreira/weinberg:20}
\textsc{Ferreira, M.~V. and M.~S. Weinberg} (2020): \enquote{{Credible, Truthful, and Two-Round (Optimal) Auctions via Cryptographic Commitments},} \emph{{In Proceedings of the 21st ACM Conference on Economics and Computation}}, 683--712.

\bibitem[\protect\citeauthoryear{Gale and Shapley}{Gale and Shapley}{1962}]{gale/shapley:62}
\textsc{Gale, D. and L.~S. Shapley} (1962): \enquote{{College Admissions and the Stability of Marriage},} \emph{{A}merican {M}athematical {M}onthly}, 69, 9--15.

\bibitem[\protect\citeauthoryear{Gangam, Mai, Raju, and Vazirani}{Gangam et~al.}{2023}]{gangam/mai/raju/vazirani:23}
\textsc{Gangam, R.~R., T.~Mai, N.~Raju, and V.~V. Vazirani} (2023): \enquote{{A Structural and Algorithmic Study of Stable Matching Lattices of Multiple Instances},} {arXiv preprint:2304.02590}.

\bibitem[\protect\citeauthoryear{Gonczarowski and Thomas}{Gonczarowski and Thomas}{2023}]{gonczarowski/thomas:23}
\textsc{Gonczarowski, Y.~A. and C.~Thomas} (2023): \enquote{{Structural Complexities of Matching Mechanisms},} {arXiv prepint: 2212.08709}.

\bibitem[\protect\citeauthoryear{Hafalir, Yenmez, and Yildirim}{Hafalir et~al.}{2013}]{hafalir/yebmez/yildirim:13}
\textsc{Hafalir, I.~E., M.~B. Yenmez, and M.~A. Yildirim} (2013): \enquote{{Effective Affirmative Action in School Choice},} \emph{Theoretical Economics}, 8, 325--363.

\bibitem[\protect\citeauthoryear{Hakimov and Raghavan}{Hakimov and Raghavan}{2023}]{hakimov2023improving}
\textsc{Hakimov, R. and M.~Raghavan} (2023): \enquote{Improving Transparency and Verifiability in School Admissions: Theory and Experiment,} Tech. rep., Working paper.

\bibitem[\protect\citeauthoryear{Haupt and Hitzig}{Haupt and Hitzig}{2023}]{haupt/hitzig:23}
\textsc{Haupt, A. and Z.~Hitzig} (2023): \enquote{{Contextually Private Mechanisms},} {arXiv preprint arXiv:2112.10812}.

\bibitem[\protect\citeauthoryear{He, Miralles, Pycia, and Yan}{He et~al.}{2018}]{he2018pseudo}
\textsc{He, Y., A.~Miralles, M.~Pycia, and J.~Yan} (2018): \enquote{A pseudo-market approach to allocation with priorities,} \emph{American Economic Journal: Microeconomics}, 10, 272--314.

\bibitem[\protect\citeauthoryear{Hylland and Zeckhauser}{Hylland and Zeckhauser}{1979}]{hylland/zeckhauser:79}
\textsc{Hylland, A. and R.~J. Zeckhauser} (1979): \enquote{{The Efficient Allocation of Individuals to Positions},} \emph{{J}ournal of {P}olitical {E}conomy}, 87(2), 293--314.

\bibitem[\protect\citeauthoryear{Imamura}{Imamura}{2020}]{imamura:20}
\textsc{Imamura, K.} (2020): \enquote{{Meritocracy versus diversity},} {Working Paper}.

\bibitem[\protect\citeauthoryear{Kesten}{Kesten}{2010}]{kesten:10}
\textsc{Kesten, O.} (2010): \enquote{{School Choice with Consent},} \emph{forthcoming, \textit{Quarterly Journal of Economics}}.

\bibitem[\protect\citeauthoryear{Li}{Li}{2017}]{li2017obviously}
\textsc{Li, S.} (2017): \enquote{Obviously strategy-proof mechanisms,} \emph{American Economic Review}, 107, 3257--3287.

\bibitem[\protect\citeauthoryear{McMichael}{McMichael}{2022}]{mcmichael:22}
\textsc{McMichael, B.~J.} (2022): \enquote{{Stealing Organs?}} \emph{{Indiana Law Journal}}, 97, 135--201.

\bibitem[\protect\citeauthoryear{Möller}{Möller}{2024}]{moller:24}
\textsc{Möller, M.} (2024): \enquote{{Transparent Matching Mechanisms},} {Working Paper}.

\bibitem[\protect\citeauthoryear{Ollar, Rostek, and Yoon}{Ollar et~al.}{2021}]{Ollar2021}
\textsc{Ollar, M., M.~Rostek, and J.~H. Yoon} (2021): \enquote{Privacy in Markets *,} Working paper.

\bibitem[\protect\citeauthoryear{Papai}{Papai}{2000}]{papai:00}
\textsc{Papai, S.} (2000): \enquote{{Strategyproof Assignment by Hierarchical Exchange},} \emph{{E}conometrica}, 68, 1403--1433.

\bibitem[\protect\citeauthoryear{Parkes, Rabin, Shieber, and Thorpe}{Parkes et~al.}{2006}]{parkes2006practical}
\textsc{Parkes, D.~C., M.~O. Rabin, S.~M. Shieber, and C.~A. Thorpe} (2006): \enquote{Practical secrecy-preserving, verifiably correct and trustworthy auctions,} in \emph{Proceedings of the 8th international conference on Electronic commerce: The new e-commerce: innovations for conquering current barriers, obstacles and limitations to conducting successful business on the internet}, 70--81.

\bibitem[\protect\citeauthoryear{Parkes, Rabin, and Thorpe}{Parkes et~al.}{2009}]{parkes2009cryptographic}
\textsc{Parkes, D.~C., M.~O. Rabin, and C.~Thorpe} (2009): \enquote{Cryptographic combinatorial clock-proxy auctions,} in \emph{Financial Cryptography and Data Security: 13th International Conference, FC 2009, Accra Beach, Barbados, February 23-26, 2009. Revised Selected Papers 13}, Springer, 305--324.

\bibitem[\protect\citeauthoryear{Pathak}{Pathak}{2017}]{pathak:17}
\textsc{Pathak, P.~A.} (2017): \enquote{{What Really Matters in Designing School Choice Mechanisms},} in \emph{Advances in Economics and Econometrics, 11th World Congress of the Econometric Society}, ed. by B.~Honore, M.~Piazessi, A.~Pakes, and L.~Samuelson, Cambridge University Press.

\bibitem[\protect\citeauthoryear{Pathak, Rees-Jones, and Sönmez}{Pathak et~al.}{2020}]{pathak/rees-jones/sonmez:20}
\textsc{Pathak, P.~A., A.~Rees-Jones, and T.~Sönmez} (2020): \enquote{{Immigration Lottery Design: Engineered and Coincidental Consequences of H-1B Reforms},} {National Bureau of Economic Research, Working Paper No. w26767}.

\bibitem[\protect\citeauthoryear{Pathak and S\"{o}nmez}{Pathak and S\"{o}nmez}{2013}]{pathak/sonmez:13}
\textsc{Pathak, P.~A. and T.~S\"{o}nmez} (2013): \enquote{{School Admissions Reform in Chicago and England: Comparing Mechanisms by their Vulnerability to Manipulation},} \emph{American Economic Review}, 103(1), 80--106.

\bibitem[\protect\citeauthoryear{Pathak, Sönmez, Ünver, and Yenmez}{Pathak et~al.}{2021}]{pathak/sonmez/unver/yenmez:21}
\textsc{Pathak, P.~A., T.~Sönmez, U.~M. Ünver, and B.~M. Yenmez} (2021): \enquote{{Fair Allocation of Vaccines, Ventilators and Antiviral Treatments: Leaving No Ethical Value Behind in Health Care Rationing},} {In Proceedings of the 22nd ACM Conference on Economics and Computation}.

\bibitem[\protect\citeauthoryear{Pittel}{Pittel}{1989}]{pittel:89}
\textsc{Pittel, B.} (1989): \enquote{{The Average Number of Stable Matchings},} \emph{SIAM Journal on Discrete Mathematics}, 2, 530--549.

\bibitem[\protect\citeauthoryear{Pittel}{Pittel}{1992}]{pittel:92}
---\hspace{-.1pt}---\hspace{-.1pt}--- (1992): \enquote{{On Likely Solutions of a Stable Marriage Problem},} \emph{The Annals of Applied Probability}, 2, 358--401.

\bibitem[\protect\citeauthoryear{Pycia and Troyan}{Pycia and Troyan}{2023}]{pycia/troyan:23}
\textsc{Pycia, M. and P.~Troyan} (2023): \enquote{{A Theory of Simplicity in Games and Mechanism Design},} \emph{Econometrica}, 91, 1495--1526.

\bibitem[\protect\citeauthoryear{Pycia and Ünver}{Pycia and Ünver}{2023}]{pycia/unver:23}
\textsc{Pycia, M. and U.~Ünver} (2023): \enquote{{Ordinal Simplicity and Auditability in Discrete Mechanism Design},} {Working Paper}.

\bibitem[\protect\citeauthoryear{Pycia and Ünver}{Pycia and Ünver}{2017}]{pycia/unver:17}
\textsc{Pycia, M. and U.~M. Ünver} (2017): \enquote{{Incentive compatible allocation and exchange of discrete resources},} \emph{Theoretical Economics}, 12, 286--329.

\bibitem[\protect\citeauthoryear{Pápai}{Pápai}{2001}]{papai:01}
\textsc{Pápai, S.} (2001): \enquote{{Strategyproof and Nonbossy Multiple Assignments},} \emph{Journal of Public Economic Theory}, 3, 257--271.

\bibitem[\protect\citeauthoryear{Roth}{Roth}{1982}]{roth:82}
\textsc{Roth, A.~E.} (1982): \enquote{{The Economics of Matching: Stability and Incentives},} \emph{{M}athematics of {O}perations {R}esearch}, 7, 617--628.

\bibitem[\protect\citeauthoryear{Satterthwaite and Sonnenschein}{Satterthwaite and Sonnenschein}{1981}]{satterthwaite1981strategy}
\textsc{Satterthwaite, M.~A. and H.~Sonnenschein} (1981): \enquote{Strategy-proof allocation mechanisms at differentiable points,} \emph{The Review of Economic Studies}, 48, 587--597.

\bibitem[\protect\citeauthoryear{Schuler}{Schuler}{2018}]{schuler:18}
\textsc{Schuler, N.} (2018): \enquote{{CPS OIG Uncovers Widespread Admissions Irregularities in K-8 Options for Knowledge Program},} {Office of Inspector General, Chicago Board of Education. Press Release, February 21}.

\bibitem[\protect\citeauthoryear{Shapley and Scarf}{Shapley and Scarf}{1974}]{shapley/scarf:74}
\textsc{Shapley, L. and H.~Scarf} (1974): \enquote{{On Cores and Indivisibility},} \emph{{J}ournal of {M}athematical {E}conomics}, 1, 23--28.

\bibitem[\protect\citeauthoryear{Svensson}{Svensson}{1999}]{svensson:99}
\textsc{Svensson, L.-G.} (1999): \enquote{{Strategy-Proof Allocation of Indivisible Goods},} \emph{{S}ocial {C}hoice and {W}elfare}, 16.

\bibitem[\protect\citeauthoryear{Woodward}{Woodward}{2020}]{woodward:20}
\textsc{Woodward, K.} (2020): \enquote{{Self-Auditable Auctions},} {Working Paper}.

\end{thebibliography}
\bibliographystyle{ecta}

\appendix

\newpage

\section{A General Framework for Auditability: Other Applications}
\label{appendix:other-applications}

\subsection{The General Model}
\label{appendix:general_model}

There is a set of individuals $\mathcal{I}$, a set of options $\Omega_i$ for each $i \in \mathcal{I}$, and a set of feasible outcomes $\Omega \subseteq \times_{i \in \mathcal{I}} \Omega_i$. Each individual $i$ has a type report $\theta_i \in \Theta_i$, and $\Theta \subseteq \times_{i \in \mathcal{I}} \Theta_i$ denotes the set of feasible type report profiles. We refer to an element $\theta \defeq (\theta_i)_{i \in \mathcal{I}} \in \Theta$ as a \textit{problem}. A mechanism is a mapping $\varphi: \Theta \rightarrow \Omega$ that gives a feasible outcome $\varphi(\theta)$ for every problem $\theta \in \Theta$.

The object allocation setup in the main text of this paper is a special case of the general model where $\Omega_i = \mathcal{O}$ for all $i \in \mathcal{I}$, and $\Omega \subseteq \mathcal{O}^{\mathcal{I}}$ is the space of bijections $\omega: \mathcal{I} \rightarrow \mathcal{O}$. 

A deviation from mechanism $\varphi$ at problem $\theta$ is a feasible outcome $\omega \neq \varphi(\theta)$.  A non-empty set of individuals $I \subseteq \mathcal{I}$ is a \textbf{detects the deviation} $\omega$  at problem $\theta$ if for any $\theta_{-I}' \in \Theta_{-I}$ with $(\theta_I, \theta_{-I}') \in \Theta$,
\begin{equation*}
\omega(i) \neq \varphi(\theta_I, \theta_{-I}')(i) \text{ for some } i \in I.
\end{equation*}

\begin{definition}
\label{def:auditability-problem-specific-general}
An auditability index of mechanism $\varphi$ at problem $\theta$ is 
\[
\scalebox{1.1}{$\#$} \varphi^{\theta} = \max_{\omega \neq \varphi(\theta)} \min \Big\{ |I| : I \subseteq \mathcal{I}, I \textit{ detects the deviation} \Big\}. 
\]
\end{definition}

\begin{definition}
\label{def:auditability:worst-case-general}
The worst-case auditability index of mechanism $\varphi$ is 
\[ \scalebox{1.1}{$\#$} \varphi = \max_{\theta \in \Theta} \varphi^{\theta}. \]
\end{definition}

\subsection{Single-Object Auctions}
 \label{section:auctions}
 
 Consider the problem of selling a single object to individuals in $\mathcal{I}$. The setup corresponds to the special case of our problem with $\Omega_i = \{0,1\} \times \mathbb{R}_{+}$ for all $i \in \mathcal{I}$. Given an $\omega \in \times_{i \in \mathcal{I}} \Omega_i$, we interpret $\omega(i) = (1,y)$ as $i$ receiving the object and paying $y$, and $\omega(i) = (0,y)$ as $i$ not receiving the object and paying $y$. The space of feasible outcomes $\Omega \subseteq \times_{i \in \mathcal{I}} \Omega_i$ is such that exactly one individual receives the object. A problem $\theta = (b_i)_{i \in \mathcal{I}} \in \mathbb{R}_{+}^{\mathcal{I}}$ is a vector of individuals' bids such that $b_i \neq b_{j}$ for any $i,j \in \mathcal{I}$.\footnote{Unequal bids are necessary for defining the common auction mechanisms. We may think that it is unlikely to have problems where some individuals submit the exact same bids. Alternatively, we may assume that there is a deterministic (observable) tie-breaking rule.} Let $\Theta$ be the space of all problems. 
 
We say an auction mechanism is a \textbf{fixed-pay auction}, if each individual's payment only depends on her bid and on whether she receives the object or not. Formally, a fixed-pay auction is a mechanism $\varphi: \Theta \rightarrow \Omega$ for which there are no problems $b,b' \in \Theta$ and an individual $i \in \mathcal{I}$ with $b_i = b_{i}'$, such that for $(x,y) \defeq \varphi(b)(i)$ and $(x',y') \defeq \varphi(b')(i)$, $x = x'$ and $y \neq y'$. The class of fixed-pay auctions includes two important mechanisms: the first-price auction and the all-pay auction. 

The \textbf{First-Price Auction (FPA)} is a fixed-pay auction mechanism such that for any problem $b \in \Theta$ and the corresponding outcome $\omega = FPA(b)$, 
\[
\omega(i) =
\begin{cases}
(1,b_i) & b_i = \max_{j \in \mathcal{I}} b_j, \\
(0,0) & \text{ otherwise.}
\end{cases}
\]

The \textbf{All-Pay Auction (APA)} is a fixed-pay auction mechanism such that for any problem $b \in \Theta$ and the corresponding outcome $\omega = APA(b)$, 
\[
\omega(i) =
\begin{cases}
(1,b_i) & b_i = \max_{j \in \mathcal{I}} b_j, \\
(0,b_i) & \text{ otherwise.}
\end{cases}
\]

Our first result is a characterization of the entire class of auctions mechanisms with a worst-case auditability index of one. We say a fixed-pay auction $\varphi$ is a \textbf{dual-dictatorship}, if there are individuals $i_1$ and $i_2$, and a subsets of bids $\bar{B} \subseteq \mathbb{R}_{+}$, such that for any problem $b \in \Theta$, and the corresponding outcomes $(x_1, y_1) = \varphi(b)(i_1)$ and $(x_2, y_2) = \varphi(b)(i_2)$, the following conditions hold:
\\ \text{} \; (1). $x_1 = 1$ if and only if $b \in \bar{B}$,
\\ \text{} \; (2). $x_2 = 1$ if and only if $b \notin \bar{B}$.

In other words, dual-dictatorships require that only two individuals $i_1$ and $i_2$ can receive the object, and moreover, $i_1$ receives the object whenever her bid is in some set $\bar{B}$, and otherwise, $i_2$ receives that object. 

\begin{theorem}
\label{thm:dual-dict}
$\scalebox{1.1}{$\#$} \varphi = 1$ if and only if the auction mechanism $\varphi$ is a dual-dictatorship.
\end{theorem}

\begin{proof}
    
We first prove the `if' direction. Suppose $\varphi$ is a dual-dictatorship. Consider an arbitrary problem $b \in \Theta$ and a deviation $\omega \neq \varphi(b)$. Let $i$ be some individual for whom $\omega(i) \neq \varphi(b)(i)$. Also, let $(x,y) = \omega(i)$ and $(x',y') = \varphi(b)(i)$. 
Consider cases:
\begin{enumerate}
    \item $x = x'$. We show that $i$ detects the deviation $\omega$. Suppose, for the sake of contradiction, that she does not detect the deviation, i.e., there is problem $\bar{b} \in \Theta$ with $\bar{b}_{i} = b_{i}$, such that $(x,y) = \varphi(\bar{b})(i)$. Since, $i$ has the same bid and allocation at problems $b$ and $\bar{b}$, and $\varphi$ is a fixed-pay auction, we should have that $y = y'$. This contradicts that $(x,y) = (x',y')$.

    \item $x \neq x'$. The condition means that $i$ receives the object under one of the outcomes $\omega$ or $\varphi(b)$, and she does not receive the object under the other outcome. Let $i_1$ be the first dictator in the definition of the dual-dictatorship. Then it should be that $i_1$ too receives the object under one of the outcomes $\omega$ or $\varphi(b)$, and she does not receive the object under the other outcome. However, by definition, whether $i_1$ receives an object or not is fully determined by her bid $b_{i_1}$. Hence, $i_1$ detects the deviation $\omega$.
\end{enumerate}

We proved that in both cases a single individual detects the deviation $\omega$. Since $b \in \Theta$ and deviation $\omega \neq \varphi(b)$ were arbitrary, we conclude that $\scalebox{1.1}{$\#$} \mathcal{\varphi} = 1$.

We now prove the `only if' direction. Suppose $\scalebox{1.1}{$\#$} \mathcal{\varphi} = 1$. 

\textbf{Claim 1.} $\varphi$ is a fixed-pay auction.

We prove the claim by contradiction. Suppose, there are problems $b,b' \in \Theta$ and an individual $i$ with $b_i = b_i'$, such that for $(x,y) \defeq \varphi(b)(i)$ and $(x',y') \defeq \varphi(b')(i)$, we have that $x = x'$ and $y \neq y'$. For the problem $b$, consider the deviation $\omega$ that differs from $\varphi(b)$ only by that $\varphi(b)(i) = (x',y')$. Then no single individuals detects this deviation, contradicting $\scalebox{1.1}{$\#$} \mathcal{\varphi} = 1$.

Let $\bar{I} \subseteq \mathcal{I}$ be the set of individuals that receive the object under some problem. If $\bar{I}$ is singleton, then $\varphi$ is a dual-dictatorship with $\bar{B} = \mathbb{R}_{+}$. For the rest of the proof, we assume that $\bar{I}$ is not a singleton. 

\textbf{Claim 2.} There is an $i_1 \in \bar{I}$ and a $\bar{B} \subseteq \mathbb{R}_{+}$, such that for all $b \in \Theta$ and $(x,y) = \varphi(b)(i_1)$,
\[ x = 1 \text{ if and only if } b_{i_1} \in \bar{B}. \]

We prove the claim by contradiction. Suppose, for all $i \in \bar{I}$ there is a bid $b_i \in \mathbb{R}_{+}$ and others' bids $b_{-i}'$ and $\tilde{b}_{-i}$, with $(x',y') = \varphi(b_i, b'_{-i})$ and $(\tilde{x}, \tilde{y}) = \varphi(b_i, \tilde{b}_{-i})$, such that $x = 1$ and $\tilde{x} = 0$. Fix the problem $b = (b_i)_{i \in \mathcal{I}}$, and let $i$ be the individual who receives the object at $\varphi(b)$. Consider a deviation $\omega \neq \varphi(b)$ such that some individual $i' \in \bar{I} \setminus \{i\}$ receives the object. Then no single individual detects the deviation $\omega$.

\textbf{Claim 3.} $|\bar{I}| = 2$.

We prove the claim by contradiction. Suppose $|\bar{I}| > 2$. 
For any $i \in \bar{I} \setminus \{i_1\}$, let the bid $b_i$ be such that for some bids of others $b_{-i}'$, $i$ receives the object at the outcome $\varphi(b_i, b_{-i}')$. Fix the bids of individuals in $\bar{I} \setminus \{i_1\}$ at $(b_i)_{i \in \bar{I} \setminus \{i_1\}}$ and set the bid of $i_1$ at some $b_{i_1} \notin \bar{B}$. Let $b$ denote the corresponding problem (where the bids of individuals in $\mathcal{I} \setminus \bar{I}$ are arbitrary). 
Then it should be that some individual in $i \in \bar{I} \setminus \{i_1\}$ receives the object. Consider a deviation $\omega \neq \varphi(b)$ such that some other individual $i' \in \bar{I} \setminus \{i_1, i\}$ receives the object (such a $i'$ exist since we assumed that $|\bar{I}| > 2$). Then no single individual detects the deviation $\omega$, which contradicts that $\scalebox{1.1}{$\#$} \mathcal{\varphi} = 1$.

By Claims 2 and 3, $i_1$ and $i_2 \defeq \bar{I} \setminus \{i_1\}$ are the dual-dictators of $\varphi$. This completes the proof of Theorem \ref{thm:dual-dict}. \end{proof}

%\begin{proof} See Appendix \ref{appendix:dual-dict}.  \end{proof}

From Theorem \ref{thm:dual-dict}, we conclude that the auditability indices of the FPA and APA shall be weakly larger than two. In our next result, we show their auditability indices are exactly two, which means that the mechanisms are maximally auditable among non-dual-dictatorial fixed-pay auctions. 

\begin{proposition}
\label{prop:FPA-APA}
$\scalebox{1.1}{$\#$} FPA = \scalebox{1.1}{$\#$} APA = 2$. 
\end{proposition}

The result is immediate given the following observation: under both auction formats, whenever an individual $i$ with the highest bid does not receive the object, and some individual $j$ with a lower bid receives it, then $i$ and $j$ detect the deviation. 

The final (non-fixed-pay) auction mechanism that we study is the \textbf{Second-Price Auction (SPA)}, defined as follows: for any problem $b \in \Theta$ and the corresponding outcome $\omega = SPA(b)$, 
\[
\omega(i) =
\begin{cases}
(1,\bar{b}) & b_i = \max_{j \in \mathcal{I}} b_j, \text{ and } \\
(0,0) & \text{ otherwise,}
\end{cases}
\]
where $\bar{b} = \max_{j \in \mathcal{I} \setminus \{i\}} b_j$ is the second highest bid. We show that the SPA is maximally not auditability for any problem.

\begin{proposition}
\label{prop:SPA}
For any problem $b \in \Theta$, $\scalebox{1.1}{$\#$} SPA^{b} = N$. 
\end{proposition}

Here is a quick proof of the result. 
Consider a deviation where an individual $i$ with the highest bid receives the object and pays a bid $\tilde{b}$ that is strictly larger than the second highest bid $\bar{b}$, but smaller than $b_i$. Then
no proper subset of individuals can detect the deviation, as they cannot know whether someone outside of the subset has bid $\tilde{b}$ or not.

\subsection{Voting with a Binary Outcome}
\label{sec:social_choice}

Consider the problem where individuals $\mathcal{I}$ collectively choose to implement one of two alternatives. This corresponds to the special case of our problem where $\Omega_i = \{0,1\}$ for all $i \in \mathcal{I}$, and the set of feasible outcomes $\Omega \subseteq \{0,1\}^{\mathcal{I}}$ is such that $\omega(i) = \omega(j)$ for all $\omega \in \Omega$ and $i,j \in \mathcal{I}$. Let $\theta_i \in \{0,1\}$ for all $i \in \mathcal{I}$. The set of feasible outcomes (problems) is $\Theta = \{0,1\}^{\mathcal{I}}$. 

We say a social choice mechanism is \textbf{dictatorial} if there is an individual $i \in \mathcal{I}$, such that $\varphi(\theta) = \varphi(\theta')$ for any problems $\theta$ and $\theta'$ with $\theta_i = \theta_i'$. 

\begin{theorem}
\label{thm:dictatorial-social-choice}
$\scalebox{1.1}{$\#$} \varphi = 1$ if and only if the social choice mechanism $\varphi$ is dictatorial.
\end{theorem}

\begin{proof}
The `if' part is trivial: if a social choice mechanism is dictatorial, then any deviation is detected by the dictator (i.e., the individual $\bar{i}$ in the definition of the dictatorial social choice mechanism). 

We now prove the `only if' part. The proof is by contraposition. Suppose the social choice mechanism $\varphi$ is not dictatorial. Then there is no individual $i \in \mathcal{I}$, such that $\varphi(\theta) = \varphi(\theta')$ for any problems $\theta$ and $\theta'$ with $\theta_i = \theta_i'$. Equivalently, for any individual $i \in \mathcal{I}$, there are two problems $\theta$ and $\theta'$ with $\theta_i = \theta_i' \defeq \bar{\theta}_i$, such that $\varphi(\theta) \neq \varphi(\theta')$.
Consider the problem $\bar{\theta} = (\bar{\theta}_i)_{i \in \mathcal{I}}$, and the deviation $\omega \neq \varphi(\bar{\theta})$. It is immediate from the construction of $\bar{\theta}$ that no individual detects the deviation $\omega$. Thus, $\scalebox{1.1}{$\#$} \varphi > 1$. 
\end{proof}

The dictatorial mechanism gives all the decision power to a single individual. It may be interesting to study mechanisms that treat individuals `symmetrically', which we do next. For the rest of this section, we assume that $N = |\mathcal{I}|$ is an odd number. We say a social choice mechanism $\varphi$ is \textbf{anonymous}, if $\varphi(\theta) = \varphi(\theta')$ for any $\theta \in \Theta = \{0,1\}^{\mathcal{I}}$, and $\theta'$ can be obtained from $\theta$ by permuting zeros and ones. In other words, an anonymous mechanism only considers the numbers of zeros and ones in the problem, and not the individuals' identities. 

An important example of an anonymous social choice mechanism is the \textit{majority voting} mechanism. A social choice mechanism is a \textbf{majority voting mechanism} for outcome $x \in \{0,1\}$, if for any $\theta \in \Theta = \{0,1\}^{\mathcal{I}}$, 
\[ \varphi(\theta)(i) = (x,x,\dots{},x) \text{ if and only if } \sum_{i \in \mathcal{I}} \theta_i \geq \frac{N+1}{2}.\]

It is easy to see that the auditability index of a majority voting mechanism is $(N+1)/2$. In the following theorem, we give a stronger result: we characterize majority voting as the unique most auditable anonymous social choice mechanism. 

\begin{theorem}
\label{thm:majority-voting}
If $\varphi$ is an anonymous social choice mechanism, then $\scalebox{1.1}{$\#$} \varphi \geq \frac{N+1}{2}$. Moreover, $\scalebox{1.1}{$\#$} \varphi = \frac{N+1}{2}$ if and only if $\varphi$ is the majority voting mechanism. 
\end{theorem}

\begin{proof}
The proof that a majority voting mechanism has a worst-case auditability index of $(N+1)/2$ is immediate, and therefore omitted. To prove the theorem, it is sufficient to show that any other anonymous mechanism has a worst-case auditability index of strictly larger than $(N+1)/2$.

Consider an anonymous social choice mechanism $\varphi$, which is not the majority voting mechanism. Then there are problems $\theta$ and $\theta'$ with either
\[ n \defeq \sum_{i \in \mathcal{I}} \theta_i < \sum_{i \in \mathcal{I}} \theta_i' \defeq n' \leq \frac{N-1}{2},\] 
or
\[ n \defeq \sum_{i \in \mathcal{I}} (1-\theta_i) < \sum_{i \in \mathcal{I}} (1-\theta_i') \defeq n' \leq \frac{N-1}{2},\] 
such that $\varphi(\theta) = (x,x,\dots{},x) \neq (y,y,\dots{},y) = \varphi(\theta')$. Without loss of generality, (by renaming zeros and ones if necessary) suppose that $n \defeq \sum_{i \in \mathcal{I}} \theta_i < \sum_{i \in \mathcal{I}} \theta_i' \defeq n' \leq \frac{N-1}{2}$. 
Moreover, again without loss of generality (by choosing a larger $n$ and a smaller $n'$), suppose that $n$ and $n'$ are consecutive numbers, i.e., $n' = n + 1$.

Consider the deviation $(y,y,\dots{},y) \neq \varphi(\theta)$. We argue that no subset of individuals $I \subseteq \mathcal{I}$ with $|I| < N - n$ can detect this deviation. Since $I \subseteq \mathcal{I}$, it should be that
\[ \sum_{i \in I} \theta_i \leq  \sum_{i \in \mathcal{I}} \theta_i = n.\]
Therefore, there are type reports $\tilde{\theta}_{-I}$ of individuals in $\mathcal{I} \setminus I$, such that $\sum_{i \in I} \theta_i + \sum_{i \in \mathcal{I} \setminus I} \tilde{\theta}_i = n'$.
Namely, this condition is satisfied for $\tilde{\theta}_{-I}$ that differs from $\theta_{-I}$ by only that one additional individual in $\mathcal{I} \setminus I$ has a type report equal to one, instead of zero. Since, $\sum_{i \in I} \theta_i + \sum_{i \in \mathcal{I} \setminus I} \tilde{\theta}_i = n' = \sum_{i \in \mathcal{I}} \theta_i'$, and $\varphi$ is anonymous, we get that \[\varphi(\theta_I, \tilde{\theta}_{-I}) = \varphi(\theta') = (y,y,\dots{},y).\] Thus, $I$ does not detect the deviation $(y,y,\dots{},y)$. 

Since $I, |I| < N - n$ was arbitrary, by definition of the auditability index we get that
\[\scalebox{1.1}{$\#$} \varphi^{\theta} \geq N - n > N - \frac{N-1}{2} = \frac{N+1}{2}.\]
This completes the proof of Theorem \ref{thm:majority-voting}. \end{proof}

The characterization results in Theorems \ref{thm:dictatorial-social-choice} and \ref{thm:majority-voting} crucially rely on using the notion of worst-case auditability. Specifically, some anonymous (and hence non-dictatorial) social choice mechanisms may have an auditability index of one for some (or most) problems. Consider for instance the `veto mechanism' that implements the outcome $1$ unless some individual's type report is $0$. Here, one can easily see that for any problem for which at least one individual has a type report equal to $0$, the veto mechanism has an auditability index of one. However, also note that for the problem where each individual's type report is $1$, the veto mechanism has an auditability index of $N$.

\subsection{Choice with Affirmative Action}
\label{section:reserves}

We study the problem of selecting a subset of individuals for up to $Q < |\mathcal{I}|$ positions, subject to meeting certain distributional targets. The individuals are partitioned into two (disjoint) subsets $I_L$ and $I_H$, where $I_L$ denotes the individuals from low-income status, and $I_H$ denotes the individuals of high-income status. The type report $\theta_i = \rho_i \in \mathbb{R}_{+}$ of individual $i \in \mathcal{I}$ is her priority score. The space of options for each $i$ is $\Omega_i = \{0,1\}$, where $1$ denotes being selected, and $0$ denotes not being selected. The set of feasible outcomes $\Omega \subseteq \{0,1\}^{\mathcal{I}}$ are those $\omega$ for which $\sum_{i \in \mathcal{I}} \omega(i) = Q$ and $\sum_{i \in I_L} \omega(i) \geq R$. The former condition can be interpreted as non-wastefulness of the outcome, and the latter condition can be interpreted as meeting the distributional targets.\footnote{Restricting the set of feasible outcomes to those that are non-wasteful and meet the distributional targets is crucial for our results. More specifically, our results do not extend to the environment where the designer could choose a deviation violating these conditions. Our modelling choice is natural in an environment where non-wastefulness and meeting distributional objectives are hard requirements.}
The model covers a myriad of important applications, including all those mentioned in the introduction. 

In these problems, an outcome is typically chosen by \textbf{reserves} mechanisms, where low-income applicants are preferentially treated for $R$ of the (reserved) positions. There are two well-studied reserves mechanisms, which we discuss below. 
\begin{itemize}
    \item \textbf{Reserved-seats-first (RSF) mechanism:} For a given $\theta \in \Theta$, $REG(\theta)$ is determined as follows:
    
    In the first step, choose up to $R$ highest priority individuals from $I_L$, and let $\bar{I}$ denote the set of chosen individuals. In the second step, choose up to $Q - |\bar{I}|$ highest priority individuals from $\mathcal{I} \setminus \bar{I}$. Let $I$ denote the set of chosen individuals after these two steps. Then for all $i \in \mathcal{I}$,
    \[ REG(\theta)(i) = \mathbbm{1}[i \in I].\]
    
    \item \textbf{Open-seats-first (OSF) mechanism:} For a given $\theta \in \Theta$, $OSF(\theta)$ is determined as follows:
    
    In the first step, choose up to $Q - R$ highest priority individuals from $\mathcal{I}$, and let $\bar{I}$ denote the set of chosen individuals. In the second step, choose up to $R$ highest priority individuals from $I_L \setminus \bar{I}$, and let $\tilde{I}$ denote the set of chosen individuals in this step. In the third step, choose up to $R - |\tilde{I}|$ highest priority individuals from $\mathcal{I} \setminus (\bar{I} \cup \tilde{I})$. 
    Let $I$ denote the set of chosen individuals after these three steps. Then for all $i \in \mathcal{I}$,
    \[ OSF(\theta)(i) = \mathbbm{1}[i \in I].\]
\end{itemize}

There are well-known results about the properties of the RSF and OSF rule. The RSF mechanism minimizes priority violations (i.e., situations where a lower priority individual is chosen, and a higher priority one is not) subject to meeting the reserves \citep{echenique/yenmez:15,abdulkadiroglu/grigoryan:21}. On the other hand, the OSF mechanism provides an additional boost to low-income applicants: the mechanism first chooses some of the highest priority low-income applicants from the general pool (step one), and only after that, it chooses an additional $R$ of the remaining (lower priority) low-income applicants (step two). In this way, OSF admits more low-income applicants than RSF \citep{dur/kominers/pathak/sonmez:18}. 
Choosing more low-income applicants under OSF comes at the cost of creating priority violations, and the same distributional outcomes can be achieved by the RSF with a higher reserves ratio. \cite{abdulkadiroglu/grigoryan:21} mention that the RSF mechanism solves this tradeoff in a more \textit{transparent} way; any priority violation can be explained by diversity objectives under RSF, but not under OSF. Our auditability theory provides a very different angle for comparing the two reserves mechanisms.

\begin{proposition}
\label{prop:reserves}
$\scalebox{1.1}{$\#$} RSF = R + 2$ and $\scalebox{1.1}{$\#$} OSF \geq  \max\{Q - R + 1, R + 2\}$.
\end{proposition}

\begin{proof}

The proof has four parts. In Part 1 we show that $\scalebox{1.1}{$\#$} RSF \leq R + 2$. In Part 2 we construct a problem $\theta$ such that $\scalebox{1.1}{$\#$} RSF^{\theta} \geq R + 2$. Parts 1 and 2 jointly establish that $\scalebox{1.1}{$\#$} RSF = R + 2$. 
In Part 3 we construct a problem $\theta$ such that $\scalebox{1.1}{$\#$} OSF^{\theta} \geq R + 2$. Finally, in Part 4 we construct a problem $\theta$ such that $\scalebox{1.1}{$\#$} OSF^{\theta} \geq Q - R + 1$. Parts 3 and 4 jointly establish that $\scalebox{1.1}{$\#$} OSF \geq \max \{Q - R + 1, R + 2\}$.

\textbf{Part 1.} $\scalebox{1.1}{$\#$} RSF \leq R + 2$.

First, we introduce several definitions and state a lemma about the RSF mechanism. For a given outcome $\omega \in \Omega$, and individuals $i,j \in \mathcal{I}$, we say that $i$'s \textbf{priority is violated by} $j$, if $\omega(i) = 0, \omega(j) = 1$, and $\rho_i > \rho_j$. We say an outcome $\omega$ is \textbf{within-type priority compatible}, if whenever $i$'s priority is violated by $j$, it should be that
\[ i \in I_H \text{ and } j \in I_L. \]
We say that an outcome $\omega$ is \textbf{saturated priority compatible}, if whenever $i$'s priority is violated by $j$, $i \notin I_L$ and $j \in I_L$, then it should be that 
\[ \big| \big\{ i' \in I_L : \omega(i') = 1 \big\} \big| = R. \]

We say a choice mechanism $\varphi$ is within-type priority compatible (or saturated priority compatible), if the outcome $\varphi(\theta)$ is within-type priority compatible (or saturated priority compatible) for all $\theta \in \Theta$.

The following result is a special case of Theorem 3 in \cite{imamura:20}. 
\begin{lemma}
\label{lemma:imamura}
In the choice with affirmative action setting,  RSF is the unique within-type priority compatible and saturated priority compatible mechanism.
\end{lemma}

Now we prove  $\scalebox{1.1}{$\#$} RSF \leq R + 2$. 
Consider an arbitrary problem $\theta$ and an arbitrary deviation $\omega \neq RSF(\theta)$. By Lemma \ref{lemma:imamura}, the outcome $\omega$ is either not within-type priority compatible, or not saturated priority compatible. We consider each case separately.

First, suppose $\omega$ is not within-type priority compatible. Then there are individuals $i$ and $j$, such that $i$'s priority is violated by $j$, and one of the following holds: $(i).$ $i,j \in I_L$, $(ii).$ $i,j \in I_H$, or $(iii).$ $i \in I_L$ and $j \in I_H$. None of these cases can happen under the RSF mechanism, hence, $\{i,j\}$ detects the deviation $\omega$. 

Now, suppose $\omega$ is not saturated priority compatible. Then there are individuals $i$ and $j$, such that $i$'s priority is violated by $j$, $i \in I_H, j \in I_L$, and
\[ \big| \big\{ i' \in I_L : \omega(i') = 1 \big\} \big| > R. \]
Let $\tilde{I} \subseteq \big\{ i' \in I_L(\theta) : \omega(i') = 1 \big\}$ be the lowest priority $R+1$ individuals in the subset. It is immediate from the description of RSF that $i$ should have been chosen over some individual in $\tilde{I}$. Hence, $\tilde{I} \cup \{i\}$ (whose size is $R + 2$) detects the deviation $\omega$. 

Since the problem $\theta$ and deviation $\omega$ was arbitrary, we conclude that $\scalebox{1.1}{$\#$} RSF \leq R + 2$.

\textbf{Part 2.} There is problem $\theta$, such that $\scalebox{1.1}{$\#$} RSF^{\theta} \geq R + 2$.

Consider a problem $\theta$, where all individuals in $I_L$
have lower priorities than all individuals in $I_H$. 

For this problem, RSF chooses $R$ individuals with the highest priorities from $I_L$ and $Q-R$ individuals with the highest priorities from $I_H$. Consider the deviation $\omega \neq RSF(\theta)$, that chooses all $R+1$ highest priority individuals in $I_L$ and only $Q-R+1$ highest priority individuals from $I_H$. Then no subset of individuals of size strictly less than $R+2$ detects the deviation $\omega$. Hence, $\scalebox{1.1}{$\#$} OSF^{\theta} \geq R + 2$.

\textbf{Part 3.} There is a problem $\theta$, such that $\scalebox{1.1}{$\#$} OSF^{\theta} \geq R + 2$.

The proof of this part is identical to that of Part 2. Namely, we use the same problem and deviation to prove the result.

\textbf{Part 4.} There is a problem $\theta$, such that $\scalebox{1.1}{$\#$} OSF^{\theta} \geq Q - R + 1$. 

Consider a problem $\theta$ where $\min \{ Q - R, |I_L| - R \}$ individuals in $I_L$ have highest priorities among all individuals $\mathcal{I}$, and the remaining individuals in $I_L$ have the lowest priorities among all individuals $\mathcal{I}$. For this problem, the OSF mechanism would chose $\min \{ Q - R, |I_L| - R \} + R = \min \{ Q, |I_L| \}$ highest priority individuals in $I_L$. 

Let $i$ be the highest priority non-chosen individual in $I \setminus I_L$ under $OSF(\theta)$, and let $\tilde{i}$ be the lowest priority chosen individual in $I_L$. Consider the deviation $\omega$ that differs from $OSF(\theta)$ by that $\omega(i) = 1$ and $\omega(\tilde{i}) = 0$. Let $I \subseteq \mathcal{I}$ be an arbitrary subset of individuals with $|I| < Q - R + 1$. Consider cases:

$(i)$. Suppose $I \subseteq I_L$. Since the outcome of individuals in $I \setminus \{\tilde{i}\}$ is the same under $OSF(\theta)$ and $\omega$, this subset does not detect the deviation. 

$(ii)$. Suppose $I \cap I_H \neq \emptyset$. Since $I_H \neq \emptyset$ and $|I| < Q - R + 1$, we have that $|I \cap I_L| < Q - R$. Consider the problem $\theta'$, such that $\theta_I'= \theta_I$, and individuals in $\big(I \cap I_L \setminus \{\tilde{i}\} \big) \cup \{i\}$ have higher priorities than everyone in $\mathcal{I} \setminus I$. Then the set of chosen individuals from $I$ is the same under both $\omega$ and $OSF(\theta')$. Hence, $I$ does not detect the deviation $\omega$. This completes the proof of Proposition \ref{prop:reserves}.
\end{proof}

If the proportion of reserves $R/Q$ is small, there may be a large gap between the auditabilities of RSF and OSF. For example, if this proportion is 20\%, the RSF will have about four times smaller auditability index than the OSF. Some examples of small reserves ratios in real-life problems include allocation of H1-B visas with about 23\% \citep{pathak/rees-jones/sonmez:20}, and pandemic rationing with about 10-40\% \citep{pathak/sonmez/unver/yenmez:21}.\footnote{A more detailed account of the reserves policy pandemic rationing can be found here: https://www.covid19reservesystem.org/policy-impact.}

\section{Additional Results and Discussions for Allocation Problems}
\label{appendix:ADDITIONAL_results}

\subsection{Direct Descriptions of of IA and AR}
\label{app:algorithms}

\textit{\textbf{Immediate Acceptance (IA) for input $\theta$:} Start with Step 1.}

\textit{ \textbf{Step $t \geq 1$}. Each available individual claims her most preferred available object. Each object that is claimed by some individuals, is matched to the the claimant with the highest priority score for the object. The matched objects become unavailable. If there are no available objects, the algorithm terminates. Otherwise, we proceed to Step $t + 1$.}

Given $\theta$, denote the outcome of the IA mechanism by $IA(\theta)$. 

The \textit{application-rejection} mechanism with \textit{permanency-execution period $e$}, denoted by $AR_e$, selects allocations as follows. We follow the description in \cite{chen/kesten:17}. Take any problem $\theta$: 

\textbf{Round $t=0$.}
\textit{Each individual claims her most preferred object. Each object that is claimed by some individuals, is matched to the claimant with the highest priority and rejects the rest.}

\textit{ (in general)
Rejected individuals that have not yet claimed their $e$-th most preferred object, claim their most preferred objects that has not rejected them. In case an individual is rejected from her top $e$ most preferred objects, she becomes passive in this round. Each object that is claimed by some individuals, is tentatively matched to the claimant with the highest priority and rejects the rest. 
The round terminates when each individual is matched or passive. All tentatively matched individuals become unavailable, and we proceed to the next round.}

\textbf{Round $t\geq 1$.}
\textit{Each available individual claims her $te$+1-st most preferred object. Each object with claimants, is matched to the one with the highest priority and rejects the rest. }

\textit{(in general) Rejected individuals that have not yet claimed their $te+ e$-th most preferred objects, claim their most preferred objects that has not rejected them. In case an individual is rejected from her top $te + e$ most preferred objects, she remains passive in this round. Each object that is claimed by some individuals, is tentatively matched to the claimant with the highest priority and rejects the rest.  The round terminates either when each individual is either tentatively matched or passive. All tentatively matched individuals become unavailable, and we proceed to the next round.}

The algorithm terminates when each individual has been matched to an object. In this case, all assignments become final and the resulting outcome is $AR_e(\theta)$.

\subsection{Additional Information Disclosure under AR}
\label{prop:ARalternative}

Here, we give an alternative proof of Proposition \ref{prop:AR}.

\begin{proof}
Consider an arbitrary $AR_e$ with $e > 1$. To prove that $\scalebox{1.1}{$\#$} AR_e = N$, we need to construct a problem $\theta$ and a deviation $\omega \neq AR_e(\theta)$, such that no proper subset of individuals detects this deviation.

Suppose the individuals and objects are indexed, i.e., $\mathcal{I} = \{i_1,i_2,\dots,i_N\}$ and $\mathcal{O} = \{o_1,o_2,\dots,o_N\}$. Consider the following problem $\theta = (P,r)$:
\begin{itemize}
    \item $o_1 P_{i_1} o_2 P_{i_1} o_n$ and $o_2 P_{i_2} o_1 P_{i_2} o_n$ for all $n \in \{3,\dots,N\}$,
   \item $o_1 P_{i_n} o_2 P_{i_n} o_n P_{i_n} o_m$ for all $n \in \{3,\dots,N\}$ and for all $m \neq n$,
    \item $r_{i_2 o_1} > r_{i_1 o_1} > r_{i_n o_1} $ and $r_{i_1 o_2} > r_{i_2 o_2} > r_{i_n o_2} $ for all $n \in \{3,\dots,N\}$,
\item otherwise, the preferences and priority scores are arbitrary.
\end{itemize}

Then we have, $AR_e(\theta)(i_n) = o_n$ for all $n \in \{1,2,\dots,N\}$. Now consider the deviation $\omega \neq AR_e(\theta)$ that differs from $AR_e(\theta)$ by that $\omega(i_1) = o_2$ and $\omega(i_2) = o_1$. 

Let $I \subsetneq \mathcal{I}$ be an arbitrary proper subset of individuals. We will show that $I$ does not detect the deviation. Consider the following cases:

(i) $i_1 \notin I$ or $i_2 \notin I$. Without loss of generality, suppose $i_1 \notin I$. 
Consider the problem $\tilde{\theta} = (\tilde{P},\tilde{r})$ that differs from $\theta$ by only that $r_{i_1 o_1}>\tilde{r}_{i_1 o_1}$ such that $r_{i_2 o_1} >  r_{i o_1} > \tilde{r}_{i_1 o_1} $ for some $i \in \mathcal{I}\setminus{i_1}$. Then $AR_e(\theta_{I}, \tilde{\theta}_{-I}) = AR_e(\tilde{\theta}) = \omega$. Thus, $I$ does not detect the deviation $\omega$.

(ii) $i_1, i_2 \in I$.  Consider an arbitrary $i \in \mathcal{I} \setminus \big(I \cup \{i_1,i_2\} \big)$. Such an $i$ exists because $I \subsetneq \mathcal{I}$ and $i_1,i_2 \in I$. Consider the problem $\tilde{\theta} = (\tilde{P},\tilde{r})$ that differs from $\theta$ by only that $\tilde{r}_{i o_1}> r_{i o_1}$ such that $r_{i_2 o_1} > r_{i o_1} > \tilde{r}_{i_1 o_1}$. Then $AR_e(\theta_{I}, \tilde{\theta}_{-I}) = AR_e(\tilde{\theta}) = \omega$. Thus, $I$ does not detect the deviation $\omega$. This completes the proof.
\end{proof}

A quick glance at this alternative proof of Proposition \ref{prop:AR} reveals that for each $I \subsetneq \mathcal{I}$,  $\theta$ differs from $\tilde{\theta}$ only with regard to a small variation in the score ranking of an individual outside of $I$. Thus,  jointly and publicly disclosing all reported preference rankings, the entire allocation $\omega$ along with the cutoff for each object at $\omega$ has no influence on the result.

\subsection{Stable-Dominating Mechanisms}
\label{appendix:stable-dominating}

\begin{definition}
\label{def:stable-dominating}
We say a mechanism $\varphi$ weakly Pareto dominates a mechanism $\psi$, if for any problem $\theta = (P,r)$, and any $i \in \mathcal{I}$, either $\varphi^{c}(\theta)(i) P_i \psi(\theta)(i)$ or $\varphi^{c}(\theta)(i) = \psi(\theta)(i)$. We say a mechanism $\varphi$ is \textbf{stable-dominating}, if there is stable mechanism $\psi$ such that $\varphi$ stable dominates $\psi$.
\end{definition}

\begin{proposition}
If $\varphi$ is a stable-dominating mechanism, then  $\scalebox{1.1}{$\#$} \varphi = N$. 
\end{proposition}

\begin{proof}

Let $\varphi$ be an arbitrary stable-dominating mechanism. We want to show that $\scalebox{1.1}{$\#$} \varphi = N$.

Suppose individuals and objects are indexed, i.e., $\mathcal{I} = \{i_1,\dots{},i_N\}$ and $\mathcal{O} = \{o_1,\dots{},o_N\}$,
and consider a problem $\theta$ where for all
$n \in \{1,\dots{},N\}$, $i_n$ ranks $o_{n+1}$ as her most preferred object, and she ranks $o_n$ as her second most preferred object. Here, we define $o_{N+1} \defeq o_1$. These preferences are illustrated in the table below. 
     
\begin{center}
\begin{tabular}{ccccc}
$i_1$ & $i_2$ & $\cdots{}$ & $i_{N-1}$ & $i_N$ \tabularnewline
\hline 
$o_2$ & $o_3$ & $\dots{}$ & $o_N$ & $o_1$ \tabularnewline
$o_1$ & $o_2$ & $\dots{}$ & $o_{N-1}$ & $o_N$ \tabularnewline
$\dots{}$  & $\dots{}$  & $\dots{}$  & $\dots{}$  & $\dots{}$  \tabularnewline
\end{tabular}
\par\end{center}

Moreover, suppose that for all $n \in \{1,\dots{},N\}$, $i_{n}$ has the highest priority score at $o_n$, and the second highest priority score at $o_{n+1}$. 

In this problem, there are two stable allocations, one that assigns each individual to their most preferred object, and one that assigns everyone their second most preferred object. 
Hence, these are exactly the only two allocations that a stable-dominating mechanism can select at problem $\theta$. We will study each case one-by-one. 

\textbf{Case 1.} Suppose $\varphi$ assigns every individual to their most preferred object, that is, $\varphi(\theta)(i_n) = o_{n+1}$ for all $n \in \{1,2,\dots,N\}$. Consider the deviation $\omega \neq \varphi(\theta)$, such that $\omega(i_n) = o_n$ for all $n \in \{1,2,\dots,N\}$. Let $I \subsetneq \mathcal{I}$ be an arbitrary proper subset of individuals (i.e. $|I| < N$). We will show that $I$ does not detect the deviation $\omega$. Take an arbitrary $i_n \notin I$, and let $\theta_{i_n}'$ differ from $\theta_{i_n}$ by that $i_n$ ranks $o_n$ as her first choice. Then since $\omega$ is the unique stable allocation at problem $(\theta_{-i_n},\theta_{i_n}')$, $\omega$ is Pareto efficient, and $\varphi$ is stable-dominating, it should be that $\varphi(\theta_{-i_n},\theta_{i_n}') = \omega$. Hence, $I$ does not detect the deviation $\omega$.

\textbf{Case 2.} Suppose $\varphi$ assigns every individual to their second preferred object, that is, $\varphi(\theta)(i_n) = o_{n}$ for all $n \in \{1,2,\dots,N\}$. Consider the deviation $\omega \neq \varphi(\theta)$, such that $\omega(i_n) = o_{n+1}$ for all $n \in \{1,2,\dots,N\}$. Let $I \subsetneq \mathcal{I}$ be an arbitrary proper subset of individuals. We will show that $I$ does not detect the deviation $\omega$. Take an arbitrary $i_n \notin I$, and let $\theta_{i_n}' = (P_{i_n}',r_{i_n}')$ differ from $\theta_{i_n}$ by that $r_{i_n o_n} < r_{i_{n-1} o_{n}}$ (where $i_0 \defeq i_N$). Then since $\omega$ is the unique stable allocation at problem $(\theta_{-i_n},\theta_{i_n}')$, $\omega$ is Pareto efficient at $(\theta_{-i_n},\theta_{i_n}')$, and $\varphi$ is stable-dominating, it should be that $\varphi(\theta_{-i_n},\theta_{i_n}') = \omega$. Hence, $I$ does not detect the deviation $\omega$.
\end{proof}

\subsection{DA-Representation and Other Classes}
\label{appendix:da-representable}

In our definition of DA-representable mechanisms in section \ref{section:priority-based}, we require that the mapping $\tau$ satisfies three conditions: Independence of Irrelevant Alternatives, Monotonicity, and Equal Treatment. 

In this subsection, we study related classes of DA-representable 
mechanisms, namely, the Taiwan mechanisms \citep{dur/pathak/song/sonmez:22} and the 
Preference Rank Partitioned (PRP) mechanisms \citep{ayoade/papai:23}.
We identify the logical relations of these classes to our DA-representable mechanisms, and the possibility of extending our Theorem \ref{thm:DA-representable-full-characterization-two} to those alternatives.

\subsubsection{Preference Rank Partitioned (PRP) Mechanisms}

In this subsection, we study the class of PRP mechanisms introduced by \cite{ayoade/papai:23} and show that it is not in our class of DA-representable mechanisms. We do this by means of an example. Roughly speaking, a PRP mechanism
operates as a DA, but objects may prioritize individuals based on priority tiers and preference tiers (as explained below), in addition to the priority scores. 

 Let $\tau: \Theta \rightarrow (\mathbb{R}^N)^N$ be a mapping that corresponds to the modified prioritization.\footnote{The PRP class allows a more general partition of priority and preference tiers, and moreover, preference tier partitioning may not treat individuals symmetrically. An interested reader can find the details of the PRP class in \cite{ayoade/papai:23}.}
Given any problem $\theta=(P,r)$, let $\tau(\theta)$ reflect the following restrictions using the terminology of \cite{ayoade/papai:23} of preference and priority tiers: 

\begin{itemize}
    \item all individuals in the highest priority tier are prioritized over all individuals in the lowest priority tier, 
    \item within a priority tier, individuals in the highest preference tier are prioritized over all individuals in the lowest preference tier,
    \item within a given preference and priority tier, individuals with higher priority scores are prioritized over individuals with lower priority scores.  
\end{itemize}
The outcome of the PRP mechanism $\varphi$ that operates on $\tau$ is such that for any problem $\theta = (P,r)$ is equal to $DA(P,\tau(\theta))$. 

For the formal definition of the class, we kindly refer to \cite{ayoade/papai:23}.
\begin{proposition}
\label{prop:PRP}
There exists a PRP mechanism that is not DA-representable.
\end{proposition}

\begin{proof}
We first construct a specific mechanism that is in the PRP class and in a second step show that it is not DA-representable.

Consider a PRP mechanism corresponding to the following $\tau$.
For some object $o \in \mathcal{O}$, the modified priority scores with respect to $o$ at every problem $\theta$ satisfy the following two conditions: 
\begin{itemize}
    \item individuals who have the highest priority score at $o$ are in the highest priority tier, and everyone else is in the lowest priority tier, 
    \item individuals who rank $o$ as on the first position in their preference ranking  are in the highest preference tier, and everyone else is is the lowest preference tier.
\end{itemize}
The modified priority scores with respect to all objects in $\mathcal{O} \setminus \{o\}$ are unaffected (i.e., are equal to the original ones). 

We will show that $\tau$ does not satisfy the Independence of Irrelevant Alternatives condition as introduced in Section \ref{section:priority-based}, and hence, the PRP mechanism mentioned above is not DA-representable. 

Consider a setting with three individuals $\{i,j,k\}$ and three objects $\{o,o_1,o_2\}$, and a problem $\theta = (P,r)$, such that (i) $P_i$ ranks $o_1$ on the first position and $o$ on the second position, (ii) $P_j$ ranks $o$ on the first position, and $o_1$ on the second position, (iii) $r_{io} > r_{jo} > r_{ko}$, (iv) $P_k$ ranks $o_2$ on the first position, and (v) $r_{ko_2} > r_{io_2},r_{jo_2}$. The rest of the problem is set arbitrarily. Then for $\tau(\theta)=\hat{r}$, we have $\hat{r}_{io} > \hat{r}_{jo}$, because $i$ is in the highest priority tier, and $j$ is in the lowest priority tier. 

Now consider another problem $\theta' = (P',r')$ that differs from $\theta$ by only that $r_{ko}' > r_{io}' > r_{jo}'$. Then with the modified scores $\hat{r}' = \tau(\theta')$, it holds $\hat{r}_{io}' < \hat{r}_{jo}'$, since both $i$ and $j$ are in the lowest priority tier, $j$ is in the highest preference tier, and $i$ is in the lowest preference tier. Since the preferences and priority score rankings of $i$ and $j$ are unchanged across the two problems $\theta$ and $\theta'$, and $\hat{r}_{io} > \hat{r}_{jo}$ and $\hat{r}_{io}' < \hat{r}_{jo}'$, we conclude that $\tau$ does not satisfy the Independence of Irrelevant Alternatives condition. \end{proof}

\subsubsection{Taiwan Mechanisms}

\cite{dur/pathak/song/sonmez:22} introduce a general class of mechanisms that are defined through some deduction rule. In a nutshell, the deduction rule $\lambda: \{1,2,\dots,N\} \rightarrow \mathbb{R}_{+}$ specifies a `penalty' for ranking objects at worse positions, and otherwise, it operates as a DA with the deduction priority scores. Namely, for a problem $\theta = (P,r)$, the Taiwan mechanism for a deduction rule $\lambda$ corresponds to applying the DA to the problem $(P,\hat{r})$, where the modified priority scores $\hat{r}$ satisfy the following condition: for any $i \in \mathcal{I}$ and $o \in \mathcal{O}$, if $i$ ranks $o$ at the $n$-th position, then
\[
\hat{r}_{io} = r_{io} - \lambda(n).
\]

Let $\tau: \Theta \rightarrow (\mathbb{R}^N)^N$ be the mapping that gives the modified priority score $\hat{r}$ (specified above) for any problem $r$. We will show that the Taiwan mechanism is not DA-representable, by showing that $\tau$ does not satisfy the Independence of Irrelevant Alternatives condition. To see this, let $\lambda(n) = N - n$ and consider the following problems $\theta = (P,r)$ and $\theta' = (P',r')$: for some $i,j \in \mathcal{I}$ and $o \in \mathcal{O}$, (i) $o$ is ranked as the second preferred object by $i$ according to both $P_i$ and $P'_{i}$, (ii) $o$ is ranked as the most preferred object by $j$ according to both $P_j$ and $P'_{j}$, (iii) $r_{io} - r_{jo} > 1$, (iv) $r_{io}' - r_{jo}' \in (0,1)$.

For these problems $\theta$ and $\theta'$, with  $\hat{r} = \tau(\theta)$ and $\hat{r}' = \tau(\theta')$, we get
$\hat{r}_{io} > \hat{r}_{jo}$ and $\hat{r}_{io}' < \hat{r}_{jo}'$, which violates the Independence of Irrelevant Alternatives condition. 

Despite the fact that the Taiwan mechanisms are not included in our class of DA-representable mechanisms, a close investigation of our Theorem \ref{thm:DA-representable-full-characterization-two} would reveal that the characterization also holds for all Taiwan mechanisms. Below we give a stronger result. We show that our characterization in Theorem \ref{thm:DA-representable-full-characterization-two} holds for an alternative class of DA-representable mechanisms, that also includes the Taiwan mechanisms.

Consider the following conditions on $\tau: \Theta \rightarrow (\mathbb{R}^N)^N$.

\begin{itemize}
    \item \textbf{Own-Type Dependence.} Let $\theta$ and $\theta'$ be arbitrary problems, and let $\hat{r} = \tau(\theta)$ and $\hat{r}' = \tau(\theta')$ be the corresponding modified priority scores.
    Then for any $i \in \mathcal{I}$,
    \[
    \theta_i = \theta_i' \implies \hat{r}_i = \hat{r}_i'.
    \]

\item \textbf{First-Choice Monotonicity.} Let $\theta = (P,r)$ and $\theta' = (P',r)$ be arbitrary problems, and let $\hat{r} = \tau(\theta)$ and $\hat{r}' = \tau(\theta')$ be the corresponding modified priority scores. Then for any $i \in \mathcal{I}$ and $o \in \mathcal{O}$, if $P_i$ (potentially) differs from $P_i'$ by that $o$ is ranked as the most preferred object under $P_i$, we have that
\[
\hat{r}_{io} \geq \hat{r}_{io}'.
\]

\item \textbf{Continuity.} Let $\theta = (P,r)$ be an arbitrary problem, and let $\hat{r} = \tau(\theta)$ be the corresponding modified priority scores. Then for any $i,j \in \mathcal{I}, o \in \mathcal{O}$, and any $\epsilon > 0$, there is a problem $\theta'$, such that for the modified priority scores $\hat{r}' = \tau(\theta')$,
\[
\hat{r}_{jo}' \in (\hat{r}_{io}-\epsilon, \hat{r}_{io}).
\]

\end{itemize}

\begin{definition}
\label{def:da-representation*}
We say a mechanism $\varphi$ is \textbf{DA-representable$^*$}, if there is a mapping $\tau$ satisfying Own-Type Dependence, First-Choice Monotonicity, and Continuity, such that for any problem $\theta = (P,r)$ with $\tau(\theta) \in \mathcal{R}$,\footnote{Unlike in the main text, we did not put any restriction on the range of $\tau$. Hence, we restrict the domain to make sure that we have strict priorities and the DA is well-defined.}
\[ \varphi(\theta) = DA(P, \tau(\theta)).\]    
\end{definition}

The following result is immediate from the description of Taiwan mechanisms. 

\begin{proposition}
\label{prop:taiwan}
Taiwan mechanisms are DA-representable$^*$.
\end{proposition}
We now state the main result of this section.

\begin{theorem}
\label{thm:DA*-representable}
Let $\varphi$ be an arbitrary DA-representable$^*$ mechanism. Then,

\begin{enumerate}
    \item For a given problem $\theta = (P,r)$, $\scalebox{1.1}{$\#$} \varphi^{\theta} = 2$ if and only if either $\varphi(\theta)$ is the unique stable allocation at problem $(P,\tau(\theta))$, or there are stable allocation other than $\varphi(\theta)$, but any such allocation is sufficiently undesirable for some pair of individuals. 

    \item $\scalebox{1.1}{$\#$} \varphi = 2$ if and only if point 1 holds for all problems. 
\end{enumerate}

\end{theorem}

\begin{proof}
The proof of Theorem \ref{thm:DA*-representable} is almost identical to the proof of Theorem \ref{thm:DA-representable-full-characterization-two}, except that we use the alternative conditions on $\tau
$ in our arguments.
\end{proof}

\subsection{Asymptotic Results}

\subsubsection{Problem-Specific Auditability of the DA}
\label{appendix:DA-problem-specific}

In this section we explore the problem-specific auditability index of the DA. For a given $M \in \{1,2,\dots,N\}$ and problem $\theta$, we give a sufficient condition for $\scalebox{1.1}{$\#$} \varphi^{\theta} \geq M$.

For a problem $\theta$ and a subset of individuals $I \subseteq \mathcal{I}$ and subset of objects $O \subseteq \mathcal{O}$, let $DA^{I,O}(\theta)$ denote the outcome of the DA in the restricted market that only consists of individuals $I$ and objects $O$.

\begin{proposition}
\label{prop:DA-aud>M}
    For a given problem $\theta = (P,r)$, $\scalebox{1.1}{$\#$} DA^{\theta} \geq M$ if there is a stable allocation $\omega \neq DA(\theta)$ such that for any $I \subseteq \mathcal{I}$ with $|I| < M$, one of the following conditions holds: (i) $DA^{I,O}(\theta)(i) = \omega(i)$, for all $i \in \mathcal{I}$, or (ii)
    there is an object in $o \in \mathcal{O} \setminus \{\omega(i)\}_{i \in I}$ with $\omega(i) P_i o$ for all $i \in \mathcal{I}$.
\end{proposition}

\begin{proof}
For a problem $\theta = (P,r)$, suppose that there is a stable allocation $\omega \neq \varphi(\theta)$ such that for all $I \subseteq \mathcal{I}$ with $|I| < M$, we have (i) $DA^{I,O}(\theta)(i) = \omega(i)$ for all $i \in \mathcal{I}$, or (ii) there is an object in $o \in \mathcal{O} \setminus \{\omega(i)\}_{i \in I}$ with $\omega(i) P_i o$ for all $i \in \mathcal{I}$. To prove $\scalebox{1.1}{$\#$} DA^{\theta} \geq M$, it is sufficient to show that no subset $I \subseteq \mathcal{I}, |I| < M$ detects the deviation $\omega$. Let $I \subseteq \mathcal{I}$ be an arbitrary subset with $|I| < M$.  We consider the two cases separately.

\textbf{Case 1.} Suppose $DA^{I,O}(\theta)(i) = \omega(i)$, for all $i \in \mathcal{I}$. Consider the type reports $\theta_{-I}'$ of individuals in $\mathcal{I} \setminus I$, such that for every $i \in \mathcal{I} \setminus I$, $\omega(i)$ is ranked as the most preferred object, and $i$ has the highest priority score for that object. Otherwise, $\theta_{-I}'$ can be arbitrary.  

Then $DA(\theta_{I}, \theta_{-I}') = \omega$, and $I$ does not detect the deviation $\omega$. 

\textbf{Case 2.} Suppose there is an object $\bar{o} \in \mathcal{O} \setminus \{\omega(i)\}_{i \in I}$ with $\omega(i) P_i \bar{o}$ for all $i \in I$. Let $\bar{i} \notin I$ be the individual with $\omega(\bar{i}) = \bar{o}$. 

Consider the following type reports $\theta_{-I}'$ of individuals in $\mathcal{I} \setminus I$,

\begin{itemize}
    \item Every individual $i \in \mathcal{I} \setminus I$, $ i \neq \bar{i}$, ranks $\omega(i)$ as her most preferred object, and has the highest priority score for that object. 

    \item $\bar{i}$ ranks objects in $\{\omega(i)\}_{i \in I}$ as her $|I|$ most preferred objects (in any order), and her priorities $r_{\bar{i}o}'$ for any object $o \in \{\omega(i)\}_{i \in I}$ is slightly lower than $r_{io}$, where $\omega(i) = o$. 

    \item $\bar{i}$ ranks $\bar{o}$ as her $|I| + 1$-th most preferred object, and she has the highest priority for that object. 

    \item $\theta_{-I}'$ is otherwise arbitrary.

\end{itemize}

Then $DA(\theta_I, \theta_{-I}) = \omega$, and $I$ does not detect the deviation.
\end{proof}

Proposition \ref{prop:DA-aud>M} can be used to show that DA will `generically' have a large auditability index. We will study the auditability property of the DA in the uniform random economy model of \cite{pittel:89}, and evaluate its index asymptotically as the market size $N$ goes to infinity. For a randomly drawn problem $\theta$, we will establish that $\scalebox{1.1}{$\#$} DA^{\theta}$ is in the order of $\sqrt{N} \cdot \ln N$, with a probability close to one. In contrast, from Proposition \ref{cor:IA} we know that $\scalebox{1.1}{$\#$} IA^{\theta} = 2$ for every problem $\theta$.

\begin{proposition}
\label{prop:DA-generic-index}
For any $\epsilon > 0$, the proportion of problems for which DA has an auditability index of at least $\sqrt{N} \cdot \ln N (1-\epsilon)$ converges to one as $N \rightarrow \infty$. 
\end{proposition}

\begin{proof}
For a uniform randomly chosen problem $\theta$, let $\omega$ denote the \textit{egalitarian stable assignment} defined in \cite{pittel:92}. Informally stated, $\omega$ minimizes the sum of ranks of matched objects for the individuals and the priority ranks of the individuals for the matched objects. This assignment will be different from the DA outcome with a probability close to one. 

To establish that $\#DA^{\theta} \geq \sqrt{N} \cdot \ln N (1-\epsilon)$ (with a probability close to one), by Proposition \ref{prop:DA-aud>M}, it is sufficient to show that for any $I \subseteq \mathcal{I}$ with $|I| < \#DA^{\theta} \geq \sqrt{N} \cdot \ln N (1-\epsilon)$, there is an object $o \in \mathcal{O} \setminus \{\omega(i)\}_{i \in I}$ such that $\omega(i) P_i o$ for all $i \in I$.

Consider an arbitrary subset of individuals $I \subseteq \mathcal{I}$ with $|I| < \#DA^{\theta} \sqrt{N} \cdot \ln N (1-\epsilon)$. 
By Theorem 6.3 of \cite{pittel:92}, the worst rank of matched objects for the individuals at $\omega$ is in the order of $\frac{\sqrt{N}}{\ln N}$. Hence, with probability close to one, for any $\epsilon_1 < \frac{1}{1-\epsilon} - 1$,
, there are at most $(1+\epsilon_1) \cdot \frac{\sqrt{N}}{\ln N}$ objects that are weakly preferred to $\omega(i)$ for any given $i \in I$. If we add up the size of all these sets of  objects for all individuals in $I$, we will get an upper bound on the size of the set $\big\{ o \in \mathcal{O} : o \text{ is weakly preferred to } \omega(i) \text{ for some } i \in I \big\}$. That upper bound would be 
\[
\sqrt{N} \cdot \ln N (1-\epsilon) \cdot (1+\epsilon_1) \cdot \frac{\sqrt{N}}{\ln N} = N (1-\epsilon) \cdot (1+\epsilon_1) < N.
\]
Hence, there is an object $o$ such that $\omega(i) P_i o$ for all $i \in I$. This completes the proof of Proposition \ref{prop:DA-generic-index}.
\end{proof}

\subsubsection{Problem-Specific Auditability of the Serial Dictatorship}
\label{appendix:seq-clinching-application}

In this section, we elaborate on the auditability properties of serial dictatorships which are formally introduced in section \ref{section:house-allocation}. We have already established that these mechanisms have a worst-case auditability index of two (Theorem \ref{theorem:TTCcharacterization}) since each serial dictatorship is a vice ownership mechanism in the sense of Definition \ref{def:viceownership}. The question is whether there are some problems for which serial dictatorships have an auditability index of one, instead of two. We show that this `rarely' happens. We will use the sequential clinching characterization result for auditability index of one (Theorem \ref{characterization:pr-sp index_one}) to establish this result. More specifically, we will show that for a `generic' problem, a serial dictatorship mechanism does not have a sequential clinching implementation. 

Consider an arbitrary serial dictatorship $\varphi$, and let us index individuals based on the fix dictatorial ordering $\mathcal{I} = \{i_1, i_2, \dots, i_N\}$.

\begin{proposition}
\label{prop:seq-generic-convergence}
The proportion of problems for which $\varphi$ has an audiability index of one converges to zero as $N \rightarrow \infty$.
\end{proposition}

Proposition \ref{prop:seq-generic-convergence} follows from a lemma that we state and prove next. 

In what follows we fix a problem $\theta$. For each $n \in \{1,2,\dots,N\}$, let $\bar{O}_n$ denote the $n$ most preferred objects of $i_n$. 

\begin{lemma}
\label{lemma:problems-with-aud-index-one-serial}
A serial dictatorship $\varphi$ has a sequential clinching implementation at problem $\theta$ if and only if $\bar{O}_n \subsetneq \bar{O}_{n+1}$, for all $n \in \{1,2,\dots,N-1\}$.
\end{lemma}

\begin{proof}
Let $O_{i_n}(\theta_{i_n})$ be the set of possible objects for $i_n$ at type report $\theta_{i_n}$.
From the description of the serial dictatorship, we can conclude that each $i_n$ is guaranteed to be matched to one of her $n$ most preferred objects. Moreover, she may be matched to any of the $n$ most preferred objects, depending on the type reports of earlier dictators. Hence, for each $n \in \{1,2,\dots,N\}$, we have that 
\[
O_{i_n}(\theta_{i_n}) = \bar{O}_n.
\]

We now prove the Lemma, and we start with the proof of the `if' part. Suppose $\bar{O}_n \subsetneq \bar{O}_{n+1}$ for all $n \in \{1,2,\dots,N-1\}$. We will show that the serial dictatorial implementation at problem $\theta$ is the desired sequential clinching implementation of $\varphi$. 

The proof is by induction. 
It is immediate that $i_1=i_p$ (the first dictator) is clinching her most preferred object $o_1$. Now suppose the claim holds for up to some step $t \geq 1$. Consider the step $t+1$ of the serial dictatorial implementation. Since $i_{t}$ clinches some object among available ones at step $t$ (by the induction hypothesis), it should be that the clinched object is the unique available object in $O_{i_t}(\theta_{i_t})$ (by definition of clinching). Hence, $i_t$ clinches the last remaining available object at $O_{i_t}(\theta_{i_t})$ at step $t$, which in turn implies that at step $t+1$, no object in $O_{i_t}(\theta_{i_t})$ is available. Since, $\bar{O}_t \subsetneq \bar{O}_{t+1} = O_{i_{t+1}}(\theta_{i_{t+1}})$, we conclude that $O_{i_{t+1}}(\theta_{i_{t+1}}) \setminus \bar{O}_t$
is a singleton. In other words, there is a unique available object in $O_{i_{t+1}}(\theta_{i_{t+1}})$, which should be clinched by $i_{t+1}$ at step $t+1$. This completes the proof of the `if' part. 

We now prove the `only if' part. The proof is by contraposition. Suppose, there is an $n \in \{1,2,\dots,N\}$, such that $\bar{O}_n \setminus \bar{O}_{n+1} \neq \emptyset$. Let us pick the smallest such $n$. 

 With the similar induction proof as in the `if' part, but only applied up until to step $t = n$, we can see that at step $t+1$ all objects $\bar{O}_t$ are unavailable, and all objects in $\mathcal{O} \setminus \bar{O}_t$ are available. Because $i_{t+1}$ is the dictator at step $t+1$, no individual other than $i_{t+1}$ can clinch an object among $\mathcal{O} \setminus \bar{O}_t$ at this step. Thus, $i_{t+1}$ is the only potential candidate for being a clincher. Yet, we will show that $i_{t+1}$ too cannot clinch an object among $\mathcal{O} \setminus \bar{O}_t$. By the choice of $t$, we know that $\bar{O}_t \setminus \bar{O}_{t+1} \neq \emptyset$. Since $\bar{O}_t \subsetneq \bar{O}_{t+1} = O_{i_{t+1}}(\theta_{i_{t+1}})$, there should be at least two available objects in $O_{i_{t+1}}(\theta_{i_{t+1}}) \setminus \bar{O}_{t+1}$. Hence, $i_{t+1}$ does not clinch any object among $\mathcal{O} \setminus \bar{O}_t$. This completes the proof of the `only if' part of the Lemma. \end{proof}

Now, Proposition \ref{prop:seq-generic-convergence} follows from the fact that the proportion of problems that satisfy the conditions of Lemma \ref{lemma:problems-with-aud-index-one-serial} goes to zero as $N \rightarrow \infty$.

\section{For Online Publication: Omitted Proofs}
\label{app:ommitedproofs}

\subsection{Proof of Theorem \ref{characterization:pr-sp index_one}}
\label{appendix:proof-of-theorem-char-pr-one}

\textit{Part 1.} Consider an arbitrary $\theta \in \Theta$. Our goal is to show that $\scalebox{1.1}{$\#$} \varphi^{\theta} = 1$ if and only if $\varphi$ has a sequential clinching implementation at $\theta$.

First, we prove the `if' part. Suppose $\varphi$ is a sequential clinching mechanism. Consider an arbitrary problem $\theta$, and an arbitrary deviation $\omega \neq \varphi(\theta)$. We show that some individual $i$ detects the deviation $\omega$. 

Let $i$ be the first individual in the sequential clinching implementation such that $\varphi(\theta)(i) \neq \omega(i) \defeq o$ (that is, who gets the `wrong' object under $\omega$). Let $O$ be the set of available objects at the step that $i$ clinches $\varphi(\theta)(i)$. By the choice of the step, $o \in O$. (since all individuals in previous steps have received the `correct' objects). By definition of sequential clinching, $\varphi(\theta)(i)$ is the only possible object among $O$ for $i$ at $\theta_i$. Hence, $i$ detects the deviation $\omega$. 

We now prove the `only if' part. Suppose $\scalebox{1.1}{$\#$} \varphi^{\theta} = 1$.
For an arbitrary $\theta$, we will find a sequence of individuals $(i_1,i_2,\dots{},i_N)$ that would correspond to the sequential clinching implementation of $\varphi$ at problem $\theta$. 

Let $\omega \defeq \varphi(\theta)$. For an individual $i$, let $O_{i}(\theta_{i})$ denote the set of possible objects for $i$ at $\theta_{i}$. First, we show that there is an individual $i_1$ such that $|O_{i_1}(\theta_{i_1})| = 1$. Suppose, for the sake of contradiction, that there is no such individual. Or equivalently, suppose that  $| O_i(\theta_i) | \geq 2$ for all $i \in \mathcal{I}$. Consider a directed graph whose vertices are individuals $\mathcal{I}$, and where we have a directed edge from vertex $i$ to vertex $j$ if and only if there is an object $o \in O_i(\theta_i)$ such that $\omega(j) = o$. Since each vertex has an outdegree of at least one, there is a directed cycle. Select one such cycle, and let the individuals trade their objects at $\omega$ according to this cycle. The resulting outcome, call it $\omega'$, gives a possible object to each individual.  Hence, the deviation $\omega' \neq \omega$ cannot be detected by any single individual. This contradicts that $\scalebox{1.1}{$\#$} \varphi = 1$.

Thus, $i_1$ is the desired individual in step $1$ of the sequential clinching implementation. Let $o_1$ be $\varphi(\theta)(i_1)$. Consider the restriction of the problem to individuals $\mathcal{I} \setminus \{i_1\}$ and objects $\mathcal{O} \setminus \{o_1\}$. Again, we can use a proof by contradiction to show that there is an  individual an $i_2$ such that $|O_{i_2}(\theta_{i_2}) \setminus \{o_1\}| = 1$. 

We construct the rest of the individuals $(i_3,\dots{}, i_N)$ in the same way, completing the proof of Part 1.

\textit{Part 2.} Next, we prove the following: $\scalebox{1.1}{$\#$} \varphi = 1$ if and only if $\varphi$ has a sequential clinching implementation at any problem $\theta$, and moreover, the mechanism can be implemented with a sequential clinching order that only depends on the set of available objects at each step (but otherwise does not depend on the problem). 

The first part of the previous statement (sentence) is a direct consequence of Part 1. Our goal is to prove the second part of the statement, that is, to find a sequential clinching implementation of $\varphi$, where the selection of individuals only depends on the set of available objects at each step. First, we state and prove an intermediate result.

\textbf{Claim.} Let $O$ and $I$ be any set of available objects and individuals that can be achieved by some sequential clinching implementation of $\varphi$ at some problem $\theta$. Then there is an individual $i \in I$ such that $\big| O_i(\theta_i') \cap O \big| = 1$ for all $\theta_i'$.

\begin{proof}
 The proof of the Claim resembles the proof of the `only if' part of Theorem \ref{characterization:pr-sp index_one}. Suppose, for the sake of contradiction, that $O$ and $I$
can be achieved by some sequential clinching implementation of $\varphi$ at some problem $\theta$, and for any $i \in I$, there is a type report $\theta_i'$ such that $\big| O_i(\theta_i') \cap O \big| \geq 2$. Consider the problem $(\theta_I', \theta_{-I})$, and let $\omega \defeq \varphi(\theta_I', \theta_{-I})$. By definition of the sequential clinching 
implementation, individuals in $\mathcal{I} \setminus I$ are matched to the same objects in $\mathcal{O} \setminus O$ at both problems $\theta$ and $(\theta_I', \theta_{-I})$. 
Consider a directed graph whose vertices are individuals $I$, and where we have a directed edge from vertex $i$ to vertex $j$ if and only if there is an object $o \in O_i(\theta_i) \cap O$ such that $\omega(j) = o$. Since each vertex has an outdegree of at least one, there is a directed cycle. Select one such cycle, and let the individuals trade their objects at $\omega$ according to this cycle. The resulting outcome, call it $\omega'$, gives a possible object to each individual. Hence, the deviation $\omega' \neq \omega$ cannot be detected by any single individual. This contradicts that $\scalebox{1.1}{$\#$} \varphi = 1$.   
\end{proof}

Now, consider the following sequential clinching implementation of $\varphi$ for some arbitrary $\theta$. Let the first clincher $i_1$ be an individual for whom $| O_{i_1}(\theta_{i_1})| = 1$ for all type reports $\theta_{i_1}'$. In general, let the step $t$ clincher $i_t$ be the available individual for whom $\big|O_{i_t}(\theta_{i_t}') \cap O_t \big| = 1$ for all $\theta_{i_t}'$, where $O_t$ is the set of available objects at step $t$. At any step, if there is more than one such available individual, select one with some fixed tie-breaking rule. The sequential clinching order in this implementation only depends on the set of objects. This completes the proof of Theorem \ref{characterization:pr-sp index_one}.

\subsection{Proof of Proposition \ref{prop:full-range}}
\label{appendix:full-range}

Suppose $\scalebox{1.1}{$\#$} \varphi = 1$. By the second part of Theorem \ref{characterization:pr-sp index_one}, there is a sequential clinching implementation of $\varphi$ where the identity of each step's clincher only depends on the set of available objects. Let $(i_1, i_2, \dots, i_N)$ be the sequence of individuals in this sequential clinching implementation. 

Suppose, for the sake of contradiction, that $\varphi$ has full range.

\textbf{Claim.} For any problem $\theta$, 
$|O_{i_n}(\theta_{i_n})| \geq n$ for all $n \in \{1,2,\dots,N\}$. 

\begin{proof}
Consider an arbitrary problem $\theta$ and an arbitrary $n \in \{1,2,\dots,N\}$. We will show that $|O_{i_n}(\theta_{i_n})| \geq n$. 

Suppose, for the sake of contradiction, that $|O_{i_n}(\theta_{i_n})| \leq n - 1$. Since $\varphi$ has full range, there is a problem $\theta'$ such that $\varphi(\theta')$ assigns all objects in $O_{i_n}(\theta_{i_n})$ to the first $|O_{i_n}(\theta_{i_n})|$ individuals in $\{i_1, i_2, \dots, i_{n-1}\}$. Consider the problem $(\theta_{i_n}, \theta_{-i_n}')$. Then 
by definition of the sequential clinching implementation, it should be that $\varphi(\theta_{i_n}, \theta_{-i_n}')$
also assigns all objects $O_{i_n}(\theta_{i_n})$ to the first $|O_{i_n}(\theta_{i_n})|$ individuals in $\{i_1, i_2, \dots, i_{n-1}\}$. Therefore, $\varphi(\theta_{i_n}, \theta_{-i_n'})(i_n) \notin O_{i_n}(\theta_{i_n})$, which is a contradiction. \end{proof}

We now prove Proposition \ref{prop:full-range}. 
If $O_{i_2}(\theta_{i_2}) \neq \mathcal{O}$, fix some $o \in \mathcal{O} \setminus O_{i_2}(\theta_{i_2})$. If $O_{i_2}(\theta_{i_2}) = \mathcal{O}$, $o$ can be arbitrary. Since $\varphi$ has full range, there is a problem $\bar{\theta}$ such that $\varphi(\bar{\theta})(i_1) = o$. Let $\omega \defeq \varphi(\bar{\theta}_{i_1}, \theta_{-i_1})$. Since $|O_{i_2}(\theta_{i_2})| \geq 2$ (by the Claim), and by the choice of $o$, we have $|O_{i_2}(\theta_{i_2}) \setminus \{o\}| \geq 2$. Also, since for each $n \in \{3,\dots{},N\}$, $|O_{i_n}(\theta_{i_n})| \geq n \geq 3$ (by the Claim), we have that $|O_{i_n}(\theta_{i_n}) \setminus \{o\}| \geq 2$.

Consider a directed graph whose vertices are individuals $\mathcal{I} \setminus \{i_1\}$, and where we have a directed edge from vertex $i$ to vertex $j$ if and only if there is an object $o_1 \in O_i(\theta_i) \setminus \{o\}$ such that $\omega(j) = o_1$. Since each vertex has an outdegree of at least one, there is a directed cycle. Select one such cycle, and let the individuals trade their objects at $\omega$ according to this cycle. The resulting outcome, call it $\omega'$, gives a possible object to each individual. Hence, the deviation $\omega' \neq \omega$ cannot be detected by any single individual. This contradicts that $\scalebox{1.1}{$\#$} \varphi = 1$.

\subsection{Proof of Theorem \ref{thm:DA-representable-full-characterization-two}}
\label{appendix:DA-representable-two}

Let $\varphi$ be an arbitrary DA-representable mechanism through a mapping $\tau: \Theta \rightarrow (\mathbb{R}^N)^N$. We first prove a claim. 

\textbf{Claim.} For any problem $\theta$, $\scalebox{1.1}{$\#$} \varphi^{\theta} > 1$.

\begin{proof}
By Theorem \ref{characterization:pr-sp index_one}, it is sufficient to show that $\varphi$ does not have a sequential clinching implementation at $\theta$.
More specifically, no individual $i$ can clinch an object $o$ at $\theta$. To see this, consider a type report $\theta_j'$ of some individual $j \neq i$, where $j$ ranks $o$ as her most preferred object, and she has a strictly higher priority score at $o$ than $i$. Define $\hat{r}' = \tau(\theta_{-j}, \theta_j')$. Since $\tau$ satisfies monotonicity and equal treatment, it should be that $\hat{r}_{jo}' > \hat{r}_{io}'$. Hence, the DA implementation at problem $(P, \hat{r}')$ can never assign $o$ to $i$ instead of $j$. 
\end{proof}

We now prove Theorem \ref{thm:DA-representable-full-characterization-two}. 

The second part of Theorem \ref{thm:DA-representable-full-characterization-two} follows directly from the first part of the Theorem, and the Claim above. In what follows we prove the first part of Theorem \ref{thm:DA-representable-full-characterization-two}. 

We first prove the `if' direction. Let $\omega \neq \varphi(\theta)$ be an arbitrary deviation, and suppose that either (i) $\omega$ is not stable at $(P,\tau(\theta))$, or (ii) it is stable at $(P,\tau(\theta))$ and it is sufficiently undesirable for some pair of individuals.
By the Claim above, $\scalebox{1.1}{$\#$} \varphi^{\theta} > 1$. Hence, to show $\scalebox{1.1}{$\#$} \varphi^{\theta} = 2$, it is sufficient to show that some two individuals can detect the deviation $\omega$. We will prove this for the cases (i) and (ii) separately.

Suppose that it is not stable at problem $(P,\hat{r})$. Then by definition of stability, there are two individuals $i,j \in \mathcal{I}$ and an object $o \in \mathcal{O}$ such that (1) $o \ P_i \ \omega(i)$, (2)  $\omega(j) = o$, and (3) $\hat{r}_{io} > \hat{r}_{jo}$. However, since $\varphi$ is DA-representable, there are no type reports $\theta_{-\{i,j\}}$ of individuals in $\mathcal{I} \setminus \{i,j\}$ for which $\varphi$ produces such an outcome. Hence, $i$ and $j$ detect the deviation. 

Now suppose that $\omega \neq \varphi(\theta)$ is stable at $(P,\hat{r})$, and it is sufficiently undesirable for some pair of individuals. That is, there are individuals $i$ and $j$ that prefer each others' objects at $\omega$ to their own, and there is no object $o \notin \{\omega(i), \omega(j)\}$ that is less preferred by both individuals than their respective objects at $\omega$.

Suppose, for the sake of contradiction, that there are type reports $\theta_{-\{i,j\}}'$ of individuals in $\mathcal{I} \setminus \{i,j\}$, such that $\varphi(\theta_{\{i,j\}}, \theta_{-\{i,j\}}') = \omega$. 

Let $\hat{r}' = \tau(\theta_{\{i,j\}}, \theta_{-\{i,j\}}')$. Since $\varphi$ is DA-representable, $\omega$ should be the outcome of the DA implementation at the modified problem $(P_i, P_j, P_{-\{i,j\}}', \hat{r}')$.

Consider the following mutually exclusive and collectively exhaustive scenarios that can happen during the DA implementation in the modified problem: (i) $j$ is matched to $\omega(j)$ before $i$ is matched to $\omega(i)$, (ii) $i$ is matched to $\omega(i)$ before $j$ is matched to $\omega(j)$, (iii) $i$ and $j$ are matched to their respective objects at $\omega$ in the same step. We study each of these cases separately. 

\textbf{Case 1.} Suppose $j$ is matched to $\omega(j)$ before $i$ is matched to $\omega(i)$. Let 
$t$ be the step of the DA implementation at the modified problem, when $i$ claims and is matched to $\omega(i)$, and $j$ has already been tentatively matched to $\omega(j)$. Since $j$ prefers $\omega(i)$ over $\omega(j)$, it means that at the beginning of step $t$ some individual in $\mathcal{I} \setminus \{i,j\}$ is tentatively matched to $\omega(i)$. 

Therefore, at the beginning of step $t$ we have that no object is tentatively matched to $i$, and both objects $\{\omega(i), \omega(j)\}$ are tentatively matched to some individuals. Since $|\mathcal{I}| = |\mathcal{O}|$, there should be some object $o \notin \{\omega(i), \omega(j)\}$ that is not tentatively matched to anyone. Since no object is simultaneously less preferred by both $i$ and $j$ than their own objects at $\omega$, then either $i$ prefers $o$ to $\omega(i)$, or $j$ prefers $o$ more than $\omega(j)$. 
Either way, we arrive at a contradiction: $i$ cannot prefer $o$ more than $\omega(i)$ because she is claiming $\omega(i)$ at step $t$ when $o$ is available, and $j$ cannot prefer $o$ more than $\omega(j)$, because $\omega(j)$ is tentatively matched to $j$, despite that $o$ is not tentatively matched to anyone (and hence, has not been claimed by anyone in any previous step). 

\textbf{Case 2.} Suppose $i$ is matched to $\omega(i)$ before $j$ is matched to $\omega(j)$. This case is symmetric to Case 1, and hence, the proof is similar (identical). 

\textbf{Case 3.} $i$ and $j$ are matched to their respective objects at $\omega$ in the same step. The proof is similar to Case 1. Let $t$ be the step of the DA implementation, when $i$ and $j$ claim and are matched to their respective objects at $\omega$. Since $i$ prefers $\omega(j)$ over $\omega(i)$ and $j$ prefers $\omega(i)$ over $\omega(j)$, it should be that both objects are tentatively matched to some individuals in $\mathcal{I} \setminus \{i,j\}$. Therefore, at the beginning of step $t$, we have that no object is tentatively matched to $i$ and $j$, and both objects $\{\omega(i), \omega(j)\}$ are tentatively matched to some individuals other than $i$ and $j$. Since $|\mathcal{I}| = |\mathcal{O}|$, there should be some object $o \notin \{\omega(i), \omega(j)\}$ that is not tentatively matched to anyone (in fact, there are at least two such objects). Since no object is simultaneously less preferred by both $i$ and $j$ than their own objects at $\omega$, then either $i$ prefers $o$ to $\omega(i)$, or $j$ prefers $o$ more than $\omega(j)$. 
Either way, we arrive at a contradiction: both $i$ and $j$ are claiming their respective objects at $\omega$ in step $t$, despite that $o$ is not tentatively matched to anyone (and hence, has not been claimed by anyone in any previous step).

We now prove the `only if' direction.
The proof is by contraposition. Suppose that there is an allocation $\omega \neq \varphi(\theta)$ which is stable at $(P, \hat{r})$, and which is not sufficiently undesirable for any pair of individuals. We will show that no two individuals detect the deviation $\omega$, which would imply that $\scalebox{1.1}{$\#$} \varphi^{\theta} > 2$.

Consider arbitrary two individuals $i$ and $j$. By the supposition above, $\omega$is not sufficiently undesirable for $i$ and $j$. Namely, either one of the individuals in $\{i,j\}$ does not prefer the other individual's object at $\omega$ to her own, or both individuals in $\{i,j\}$ prefer the other individual's object at $\omega$ to their own, and there is an object $o \notin \{\omega(i), \omega(j)\}$ such that $\omega(i) P_i o$ and $\omega(j) P_j o$. We will prove $\scalebox{1.1}{$\#$} \varphi^{\theta} > 2$ for each of these cases separately. 

\textbf{Case 1.} Suppose that one of the individuals in $\{i,j\}$ does not prefer the other individual's object at $\omega$ to her own. Without loss of generality, suppose $\omega(i) P_i \omega(j)$.

    Consider the type reports $\theta_{-\{i,j\}}' = (P_{-\{i,j\}}', r_{-\{i,j\}}')$ of individuals in $\mathcal{I} \setminus \{i,j\}$ where every individual in this set ranks her matched object at $\omega$ as her first choice. Otherwise, the preferences and priorities agree with those in problem $\theta$.
    To establish that $i$ and $j$ do not detect the deviation $\omega$, it is sufficient to show that $\varphi(\theta_{\{i,j\}}, \theta_{-\{i,j\}}') = \omega$. 

    Let $\hat{r}' = \tau(\theta_{\{i,j\}}, \theta_{-\{i,j\}}')$ denote the modified priorities at problem $(\theta_{\{i,j\}}, \theta_{-\{i,j\}}')$.

    Since $\omega$ is a stable allocation at problem $(P,\hat{r})$, it should be that any object that $i$ prefers to $\omega(i)$ is matched to an individual with a higher $\hat{r}$ priority score than $i$ at $\omega$. Similarly, any object that $j$ prefers to $\omega(j)$ is matched to an individual with a higher $\hat{r}$ priority score than $j$ at $\omega$. 
    Since $\tau$ is monotone, it should be that these priority scores' comparisons are preserved for $\hat{r}'$. Hence, during the implementation of DA at problem $(P_{\{i,j\}}, P_{-\{i,j\}}', \hat{r}')$, individuals $i$ and $j$ will be rejected by all objects that they prefer more than $\omega(i)$ and $\omega(j)$, respectively. 
    Hence, the DA implementation at problem $(P_{\{i,j\}}, P_{-\{i,j\}}', \hat{r}')$ produces $\omega$. Or equivalently, $\varphi(\theta_{\{i,j\}}, \theta_{-\{i,j\}}') = \omega$.

\textbf{Case 2.} Suppose that both $i$ and $j$ prefer the other individual's objects at $\omega$ to their own, and there is an object $o \notin \{\omega(i), \omega(j)\}$ such that $\omega(i) P_i o$, and $\omega(j) P_j o$.  

   Since $\omega(j) P_i \omega(i)$, $\omega(i) P_j \omega(j)$, 
and $\omega$ is stable at problem $(P, \hat{r})$, it should be that $i$ has a higher $\hat{r}$ priority at $\omega(i)$ than $j$, and $j$ has a higher $\hat{r}$ priority at $\omega(j)$ than $i$. 

Let $k$ be the individual with $\omega(k) = o$, where $o$ is as defined above. Consider type reports $\theta_{-\{i,j\}}' = (P_{-\{i,j\}}', r_{-\{i,j\}})$ of individuals in $\mathcal{I} \setminus \{i,j\}$, such that all individuals in this set other than $k$ rank their matched objects at $\omega$ as their first choice, and otherwise, the preference rankings and priority scores agree with $\theta_{-\{i,j\}}$. We construct the type report $\theta_k' = (P_k', r_k')$ of $k$ as follows. 
\begin{itemize}
    \item $k$ shares the same ranking of objects as $i$ until (and including) $\omega(i)$, but she ranks $o$ as her next choice right after $\omega(i)$, 
 \item $r_k'$ satisfies $r_{k\tilde{o}}' < r_{i\tilde{o}}$ for all $o_1$ with $o_1 P_i \omega(i)$ (or equivalently, for all $o_1 P_k' \omega(i)$),
 
 \item If $r_{i\omega(i)} > r_{j\omega(i)}$, then $r_{k\omega(i)}'$ is an arbitrary number in $(r_{j\omega(i)}, r_{i\omega(i)})$. Otherwise, $r_{k\omega(i)}'$ is an arbitrary number strictly smaller than $r_{i\omega(i)}$. 

\end{itemize}

Let $\hat{r}' = \tau(\theta_{\{i,j\}}, \theta_{-\{i,j\}}')$ denote the modified priorities at problem $\theta_{\{i,j\}}, \theta_{-\{i,j\}}'$. 

\textbf{Claim.} $\hat{r}_{k\omega(i)}' < \hat{r}_{i\omega(i)}'$ and $\hat{r}_{k\omega(i)}' > \hat{r}_{j\omega(i)}'$.

\begin{proof}
The first inequality in the claim directly follows from the construction of $\theta'_k$ and the equal treatment property of $\tau$ applied to $i$ and $k$ at problem $(\theta_{\{i,j\}}, \theta_{-\{i,j\}}')$. 

We now prove $\hat{r}_{k\omega(i)}' > \hat{r}_{j\omega(i)}'$. The proof is by contradiction. Suppose $\hat{r}_{k\omega(i)}' < \hat{r}_{j\omega(i)}'$. Consider an alternative problem $\theta'' = (P'', r'')$ that differs from $(\theta_{\{i,j\}}, \theta_{-\{i,j\}}')$ by only that $k$'s priority score $r_{k\omega(i)}''$ at object $\omega(i)$ is as follows. If $r_{i\omega(i)} > r_{j\omega(i)}$, then $r_{k\omega(i)}''$ is an arbitrary number strictly greater than $r_{i\omega(i)}$. Otherwise, $r_{k\omega(i)}''$ is an arbitrary number in $(r_{i\omega(i)}, r_{j\omega(i)})$. Let $\hat{r}'' = \tau(\theta'')$. 

When moving from problem $(\theta_{\{i,j\}},\theta_{-\{i,j\}}')$ to problem $\theta''$, both the preference rankings of $j$ and $k$, and the priority score comparison between $j$ and $k$ are unchanged. Hence, by the independence of irrelevant alternatives property of $\tau$, and by our supposition, we have that $\hat{r}_{k\omega(i)}'' < \hat{r}_{j\omega(i)}''$. Note that the type reports of $i$ and $j$ are unchanged across problems $\theta$ and $\theta''$. Therefore, by the monotonicity property of $\tau$, and by that $\hat{r}_{j\omega(i)} < \hat{r}_{i\omega(i)}$ (which we established in the beginning of the proof of Case 2 of the `only if' part), we get $\hat{r}_{j\omega(i)}'' < \hat{r}_{i\omega(i)}''$. Finally, by the equal treatment property of $\tau$ applied to $i$ and $k$ at problem $\theta''$, we get $\hat{r}_{i\omega(i)}''  < \hat{r}_{k\omega(i)}''$.  Combing three inequalities above, we have,
\[
\hat{r}_{k\omega(i)}'' < \hat{r}_{j\omega(i)}'' < \hat{r}_{i\omega(i)}'' < \hat{r}_{k\omega(i)}'',
\]
a contradiction. \end{proof}

We now complete the proof of Case 2 of the `only if' part. 

Since $\omega$ is  stable at $(P, \hat{r})$, it should be that any object that $i$ prefers to $\omega(i)$ is matched to an individual with a higher $\hat{r}$ priority than $i$ at $\omega$. By the monotonicity property of $\tau$ is a monotone mapping, this is also true for $\hat{r}'$ priority scores. Similarly, any object that $j$ that prefers to $\omega(j)$ is matched to an individual with a higher $\hat{r}$ priority than $j$ at $\omega$, and this is also true for $\hat{r}'$ priority scores.

Since $k$ has a lower $\hat{r}'$ priority scores than $i$ at all objects weakly more preferred than $\omega(i)$ (according to $P_k'$), during the implementation of DA at the modified problem $(P_{\{i,j\}}, P_{-\{i,j\}}, \hat{r}')$, she will be rejected by all objects strictly more preferred than $\omega(i)$, and she will be tentatively matched at $\omega(i)$ at some step. Since $\hat{r}_{k\omega(i)}' > \hat{r}_{j\omega(i)}'$ (by the second part of the Claim), $j$ will be eventually rejected by $\omega(i)$.

Continuing the DA implementation, $i$ and $j$ will be eventually rejected by all objects that 
they prefer strictly more than $\omega(i)$ and $\omega(j)$, respectively. Since $\hat{r}_{k\omega(i)}' > \hat{r}_{i\omega(i)}'$ (by the first part of the Claim), 
$i$ will be tentatively matched to $\omega(i)$, and $k$ will be rejected by that object, and she will be matched to $o$. In conclusion, the DA implementation at the modified problem $(P_{\{i,j\}}, P_{-\{i,j\}}, \hat{r}')$ would yield the outcome $\omega$. Hence, $\varphi(\theta_{\{i,j\}}, \theta_{-\{i,j\}}') = \omega$, and $i$ and $j$ do not detect the deviation. 

Since $i$ and $j$ were arbitrary, we conclude that $\scalebox{1.1}{$\#$} \varphi^{\theta} > 2$. This complete the proof of Theorem \ref{thm:DA-representable-full-characterization-two}.

\subsection{Proof of Proposition \ref{prop:TTC}}
\label{appendix:prop:TTC}
Take any $\varphi^c$ that satisfies the supposition Proposition of \ref{prop:TTC}. We show that there is a deviation $\omega$ such that no $\hat{I} \subsetneq \mathcal{I}$ with $|\hat{I}|<n$ can detect it. First, since $\varphi^c$ is non-bossy and strategy-proof, we can find $\sigma$ and $\theta_{\sigma_I}$ such that for each $k \in \sigma_I$, $\sigma(k)$ is ranked highest under $\theta_k$, so that under any profile $(\theta_{\sigma_I},\tilde{\theta}_{\bar{I}(\sigma)}) \in \Theta$, where $\tilde{\theta}_{\bar{I}(\sigma)}$ is chosen arbitrarily the TTC algorithm arrives at suballocation $\sigma_{t^*-1}(\theta_{\sigma_I},\tilde{\theta}_{\bar{I}(\sigma)})=\sigma$ in some step $t^*$ of the TTC algorithm, while at the same time $|c_{\sigma}(\bar{I}(\sigma))| \geq n$.\footnote{A mechanism $\varphi$ is \textit{non-bossy}, if for all $\theta \in \Theta$, there is no $i \in I$, and $\theta'_i \in \Theta_i$, such that $\varphi^{c}(\theta)(i) =\varphi^{c}(\theta'_i,\theta_{-i})(i)$, while $\varphi^{c}(\theta) \neq \varphi^{c}(\theta'_i,\theta_{-i})$.  As shown by \cite{papai:00}, $\varphi^c$ satisfies non-bossiness \citep{satterthwaite1981strategy}.} 
Next, pick arbitrary $n$ individuals $ C \defeq \{ i_1,i_2,\dots i_n \} \subseteq c_{\sigma}(\bar{I}(\sigma))$ and for each $n' \in \{1,\dots, n\}$, let $i_{n'}$ be the owner of object $o_{n'} \in \bar{O}_{\sigma}$.  Now take $\theta_C$ for individuals in $C$, as it is presented in the table below. Let $o_{n+1} \defeq o_1$.
     
\begin{center}
\begin{tabular}{ccccc}
$\theta_{i_1}$ & $\theta_{i_2}$ & $\cdots{}$ & $\theta_{i_{n-1}}$ & $\theta_{i_n}$ \tabularnewline
\hline 
$o_2$ & $o_3$ & $\dots{}$ & $o_n$ & $o_1$ \tabularnewline
$o_1$ & $o_2$ & $\dots{}$ & $o_{n-1}$ & $o_n$ \tabularnewline
$\dots{}$  & $\dots{}$  & $\dots{}$  & $\dots{}$  & $\dots{}$  \tabularnewline
\end{tabular}
\par\end{center}

Denote $K= \mathcal{I}\setminus (\sigma_I \cup C)$ and let $\theta_K$ be specified arbitrarily. Thus, under problem $\theta$, $\varphi^c$ assigns each individual in $\sigma_I \cup C$ her most preferred object.  
Consider also $\theta'_C$ for individuals in $C$: 
\begin{center}
\begin{tabular}{ccccc}
$\theta'_{i_1}$ & $\theta'_{i_2}$ & $\cdots{}$ & $\theta'_{i_{n-1}}$ & $\theta'_{i_n}$ \tabularnewline
\hline 
$o_1$ & $o_2$ & $\dots{}$ & $o_{n-1}$ & $o_n$ \tabularnewline
$o_2$ & $o_3$ & $\dots{}$ & $o_{n}$ & $o_1$ \tabularnewline
$\dots{}$  & $\dots{}$  & $\dots{}$  & $\dots{}$  & $\dots{}$  \tabularnewline
\end{tabular}
\end{center}

Consider deviation $\omega=\varphi^c(\theta_{\sigma_I},\theta'_C,\theta_{K})\neq \varphi^c(\theta_{\sigma_I},\theta_C,\theta_{K}) $.  Finally, note that for each $n' \in \{1,\dots, n\}$, $\omega=\varphi^c(\theta_{\sigma_I},\theta'_C,\theta_{K}) = \varphi^c (\theta_{\sigma_I},\theta_{C \setminus \{i_{n'}\}},\theta'_{i_{n'}},\theta_K)$. Thus, any $\hat{I} \subsetneq \mathcal{I}$ with $|\hat{I}|< n$ does not detect $\omega$. This completes the proof.

\subsection{Proof of Theorem \ref{theorem:TTCcharacterization}}
\label{appendix:TTC-full-characterization}

For the rest of this section, fix an arbitrary hierarchical exchange mechanism $\varphi^c$. For expositional clarity, we make the following assumptions throughout this section: First, we remain agnostic about how $c$ is specified for suballocations that are not on-path. This is without loss of generality since for any consistent ownership structure $c'\neq c$ that also implements $\varphi^c$, ownership structure $c'$ implies the same on-path suballocations as under $c$, the same ownership mapping as $c$, and the specification of a consistent ownership structure $c$ for suballocations that are not on-path have no influence on outcomes. In the following, when we refer to a generic step $t$, then we mean a step $t$ of the TTC algorithm running on consistent ownership structure $c$.

Let $I_p$ denotes the set of principal owners (i.e. the owners at the empty suballocation $\sigma_{\emptyset}$) and let $I_v$ be the set of vice owners.   

We use the following equivalent definition of vice ownership mechanisms. 

\begin{lemma}\label{lemma:viceownership}
Hierarchical exchange mechanism $\varphi^c$ is a vice ownership mechanism as defined in Definition \ref{def:viceownership} if and only if the following conditions are satisfied:
\begin{enumerate}
\item[(1)] For each $\sigma$, there are at most two owners at $\sigma$.
\item[(2)] For any pair $\sigma, \hat{\sigma}$, if $\bar{O}(\sigma)=\bar{O}(\hat{\sigma})$ and $\bar{I}(\sigma)=\bar{I}(\hat{\sigma})$, then $c_{\sigma}=c_{\hat{\sigma}}$. 
\item[(3)]
 \begin{enumerate}
      \item[a.] 
      There are at most two principal owners and at most two vice owners.
        \item[b.]  If there are two principal owners, then for any pair $\sigma, \hat{\sigma}$, with $|\sigma_I|=|\hat{\sigma}_I|=1$, if a vice owner  $i$ owns an object $o$ at $\sigma$ and $i$ and $o$ are also unmatched at $\hat{\sigma}$, then $i$ also owns $o$ at $\hat{\sigma}$ if $o$ is not owned by a principal owner.
        \item[c.] If there are two principal owners and there exists $\sigma$, with $|\sigma_I|=1$, such that a single principal owner $k$ is matched, while a vice owner $j$ owns some $o'$ owned by $k$ at $\sigma_{\emptyset}$, then, at any suballocation $\hat{\sigma}$, such that $\sigma \subsetneq \hat{\sigma}$, where both principal owners are matched and $j$ is unmatched, we have that $j$  must own any remaining object $o$ that $k$ owned at $\sigma_{\emptyset}$.
        \item[d.] For each $\sigma$, if $|\sigma_I|\geq 3$ , then all principal owners are matched under $\sigma$. 

 \end{enumerate} 
\item[(4)]
\begin{enumerate}
    \item[a.] If there exists $\sigma'$, with $|\sigma'_I| \leq N-4$ such that there are two owners $i,j \notin I_p$ at $\sigma'$, then $j$ is the unique owner at $\hat{\sigma}$ provided that $j$ is unmatched and $\hat{\sigma}_I=\sigma_I \cup \{i\}$ for some $\sigma$, with $|\sigma_I|=|\sigma'_I|$.  
    \item[b.] For each $\sigma$, if  $l \notin I_p\cup I_v$ owns an object at $\sigma$, then all principal owners and vice owners are matched under $\sigma$.
    \item[c.] For any pair $\sigma, \hat{\sigma}$ with  $|\hat{\sigma}_I|\leq N-3$ and $|\sigma_I| \leq |\hat{\sigma}_I|$, if an individual $i \in \bar{I}(\sigma)\cap\bar{I}(\hat{\sigma})$, $i \notin I_p \cup I_v$ owns an object $o \in \bar{O}(\sigma)\cap\bar{O}(\hat{\sigma})$ at $\sigma$, then $i$ also owns $o$ at $\hat{\sigma}$, if $o$ is not owned by an individual $j \in I_p \cup I_v$ at $\hat{\sigma}$.  
   \item[d.] If there exists $\sigma'$, with $|\sigma'_I| = N-3$ such that there are two owners $i,j \notin I_p$ at $\sigma'$, then $j$ is the unique owner at $\hat{\sigma}$ provided that $j$ is unmatched and $\hat{\sigma}_I=\sigma_I \cup \{i\}$ for some $\sigma$, with $|\sigma_I|=|\sigma'_I|$.
    \end{enumerate}
\end{enumerate}
\end{lemma}

\begin{proof} It is clear that a vice ownership in the sense of Definition \ref{def:viceownership} implies all conditions of Lemma \ref{lemma:viceownership}. 

For the remaining direction, first note that Lemma \ref{lemma:viceownership} (2) and Definition \ref{def:viceownership} (2) are identical. The same holds for condition (3) b. and Definition \ref{def:viceownership} (4) i., as well as for Lemma \ref{lemma:viceownership} (3) c. and Definition \ref{def:viceownership} (4) ii.

Also, the conditions (2), (3), (4) of Lemma \ref{lemma:viceownership} together imply Definition \ref{def:viceownership} (5). If additionally Lemma \ref{lemma:viceownership} (1) holds, then Definition \ref{def:viceownership} (1) holds. Finally, given that all conditions of Lemma \ref{lemma:viceownership} hold, then Lemma \ref{lemma:viceownership} (4) c. implies that if $i$ and $j$ are regular owners in the same level for some $n$ with $n<N-2$, then there must exist $\sigma$ where $i$ and $j$ are both owners. Thus, since there are at most two level-$n$ owners, Lemma \ref{lemma:viceownership} (4) a. - (4) d. and $n< N-2$ imply that for any $\sigma'$ where $j$ owns an object and for any $\sigma''$ with $\sigma''_I= \sigma'_I \cup \{j\}$, we must have that $i$ is the owner of all remaining objects at $\sigma''$ (even if she is not an owner at $\sigma'$). Together with the remaining conditions of Lemma \ref{lemma:viceownership}, this implies Definition \ref{def:viceownership} (3). This completes the proof.
\end{proof}

We start with the `only if' direction of Theorem \ref{theorem:TTCcharacterization}. Suppose that $\varphi^c$ is not a vice ownership mechanism. We go through a series of cases, of which each assumes that a specific condition in Lemma \ref{lemma:viceownership} is violated. In each case, we show that there is a deviation that no proper subset with at most two individuals can detect it. We follow the ordering of the conditions as presented in Lemma \ref{lemma:viceownership}. At each case, we assume that the conditions from Lemma \ref{lemma:viceownership} of previous cases are satisfied. Throughout, we often make specific selections of suballocations that appear in the conditions of the Lemma. The way we select these suballocations will be without loss of generality, since the consistency of the ownership structure guarantees that ownership rights under a suballocation only depend on the mapping of the suballocation itself and not the underlying problem to reach it in the TTC algorithm.

\paragraph{Case 1: \textit{$\varphi^c$ violates Lemma \ref{lemma:viceownership} (1)}.} By Proposition \ref{prop:TTC}, we know that $\# \varphi>2$. 

\paragraph{Case 2: \textit{$\varphi^c$ violates Lemma \ref{lemma:viceownership} (2)}.} We proceed in two steps.

\textit{Step 1:}
There must exist a pair $\sigma^1, \sigma^2$ with $\bar{O}(\sigma^1)=\bar{O}(\sigma^2)$ and $\bar{I}(\sigma^1)=\bar{I}(\sigma^2)$ such that $c_{\sigma^1}\neq c_{\sigma^2}$. Thus, $c_{\sigma^1}(o_1) = i$, $c_{\sigma^2}(o_1) = j$ for some $i,j  \in \bar{I}(\sigma^1) \cap \bar{I}(\sigma^2)$ with $o_1 \in \bar{O}(\sigma^1) \cap \bar{O}(\sigma^2)$. By consistency of $c$ and since $\sigma^1 \neq \sigma^2$, it is clear that $|\sigma_I^1|= |\sigma_I^2|\geq 2$.  Now, consider two problems $\hat{\theta},\tilde{\theta}$ such that $\sigma^1=\sigma_{t}(\hat{\theta})$ and $\sigma^2=\sigma_{t'}(\tilde{\theta})$ for some steps $t$ and $t'$. 
Let $o_2\neq o_1$ with $o_2\in \bar{O}(\sigma^1) \cap \bar{O}(\sigma^2)$.

\textit{Step 2:} Let $o_2 \in  \bar{O}(\sigma^1) \cap \bar{O}(\sigma^2)$ with $o_1\neq o_2$. Consider a problem $\hat{\theta}^{*}$ such that for all $i^{*} \notin \bar{I}(\sigma^1)$, object $\varphi^{c}(\hat{\theta})(i^{*})$ is ranked first on $\hat{\theta}^{*}_{i^{*}}$, while $\hat{\theta}^{*}_i$ and $\hat{\theta}^{*}_j$ rank $o_1$ first and $o_2$ second. Furthermore, for all $k^{*} \in \mathcal{I} \setminus \{i,j\}$ and each $\bar{o} \notin  \{o_1,o_2\}$, let $\bar{o} \ \hat{\theta}^{*}_{k^{*}} \ o_1$ and $\bar{o} \ \hat{\theta}^{*}_{k^{*}} \ o_2$. Thus, by strategy-proofness and non-bossiness of $\varphi^c$ and consistency of $c$, we must reach $\sigma_{t^*}(\hat{\theta}^{*})=\sigma^1$ for some step $t^*$. Also, $\varphi^{c}(\hat{\theta}^{*})(i)=o_1$, $\varphi^{c}(\hat{\theta}^{*})(j)=o_2$. 

Consider $\theta$ such that for all $k^{*} \in \mathcal{I} \setminus \{i,j\}$, object $\varphi^{c}(\hat{\theta}^{*})(k^{*})$ ranks first on $\theta_{k^{*}}$, while the rest of $\theta$ is identical to $\hat{\theta}^*$. Again, by strategy-proofness and non-bossiness of $\varphi^c$ and consistency of $c$, $\varphi^{c}(\theta)=\varphi^{c}(\hat{\theta}^{*})$.

Next, consider deviation $\omega \neq \varphi^{c}(\theta)$ at $\theta$, where for all $k^{*} \in \mathcal{I} \setminus \{i,j\}$, $\omega(k^{*})=\varphi^{c}(\theta)(k^{*})$, $\omega(i)=o_2$ and $\omega(j)=o_1$. 

Take any $\hat{I}\subsetneq \mathcal{I}$ with $|\hat{I}|= 2$. We show that $\hat{I}$ does not detect $\omega$ considering the following exhaustive case distinction:

\begin{itemize}

\item Let $\hat{I}\cap \{i,j\} = \emptyset$. For all $k^{*} \in \mathcal{I} \setminus \{i,j\}$, we have $\omega(k^{*})=\varphi^{c}(\theta)(k^{*})$. Thus, $\hat{I}$ does not detect $\omega$. 

\item Let $\hat{I}= \{i,j\}$. For each $\tilde{\theta}^{*}$ such that for all $j^{*} \notin \bar{I}(\sigma^2)$, $\varphi^{c}(\tilde{\theta})(j^{*})$ ranks first on $\tilde{\theta}^{*}_{j^{*}}$, we must reach $\sigma^2= \sigma_{t''}(\tilde{\theta}^{*})$ for some step $t''$. Now, consider problem $\theta'$ where for all $j^{*} \notin \bar{I}(\sigma^2)$, $\varphi^{c}(\tilde{\theta})(j^{*})$ ranks first on $\theta'_{j^{*}}$. Let $\theta'_i=\theta_i$,  $\theta'_j=\theta_{j}$ and for all $\bar{j} \in \bar{I}(\sigma^2) \setminus \{i,j\}$, let objects $o_1,o_2$ rank last on $\theta'_{\bar{j}}$. Thus, $\omega(i)=\varphi^{c}(\theta')(i)$ and $\omega(j)=\varphi^{c}(\theta')(j)$.

\end{itemize}

We proceed with the remaining subcases. We only consider the subcases for which  $i \in \hat{I}$. The remaining subcases (i.e., cases where $j \in \hat{I}$) are omitted since we can apply similar arguments. 

We start with some helpful observations: First, neither $i$ nor $j$ can own $o_1$ at $\sigma_{\emptyset}$. Second, for each problem, at least one principal owner is assigned in the first step. Third, neither $i$ nor $j$ can be assigned in step $\hat{t}=1$ for each problem $\hat{\theta}$ such that $\sigma_{\hat{t}}(\hat{\theta})=\sigma^1$ or $\sigma_{\hat{t}}(\hat{\theta})=\sigma^2$, since $|\sigma_I^1|=|\sigma_I^2|\geq 1$. Fourth, not both of $i$ and $j$ can be in $I_p$, since Lemma \ref{lemma:viceownership} (1) implies, that there are at most two individuals in $I_p$ and the third observation ensures that $i$ and $j$ are not assigned at step 1. The remaining subcases are:

\begin{itemize}
\item Let $\hat{I}= \{i,i_p\}$ with $i_p \in I_p$ and suppose $i_p$ owns $o_1$ at $\sigma_{\emptyset}$. 

\begin{itemize}
    \item Suppose  $i_p$ owns $\omega(i_p)$ at $\sigma_{\emptyset}$. Then there exists a problem $\bar{\theta}$, with $\bar{\theta}_i=\theta_i$ and $\bar{\theta}_{i_p}=\theta_{i_p}$, where $o_1$ is owned by some $k \in \mathcal{I} \setminus \{i,j,i_p\}$ at a suballocation $\sigma^3=\sigma_t(\bar{\theta})$ for some step $t$, while $i \in \bar{I}(\sigma^3)$ and $o_2 \in \bar{O}(\sigma^3)$. We further refine $\bar{\theta}$ as follows: $\bar{\theta}_k$ ranks $o_1$ first and for all $l^{*} \notin \{i,j,k,i_p\}$, $o_1$, $o_2$, $\omega(i_p)$ are ranked last on $\bar{\theta}_{l^{*}}$. Then, $\omega(i_p)=\varphi^{c}(\bar{\theta})(i_p)$ and $\omega(i)=\varphi^{c}(\bar{\theta})(i)=o_2$.

\item Suppose $i_p$ does not own $\omega(i_p)$ at $\sigma_{\emptyset}$. Thus, $\omega(i_p)$ is owned by some $k \in \mathcal{I} \setminus \{i,j,i_p\}$. Consider $\bar{\theta}$ exactly as in the last paragraph. Then, $\omega(i_p)=\varphi^{c}(\bar{\theta})(i_p)$ and $\omega(i)=\varphi^{c}(\bar{\theta})(i)=o_2$.
\end{itemize}
\item Let $\hat{I}= \{i,i_p\}$, with $i_p \in I_p$ and $i_p$ does not own $o_1$ at $\sigma_{\emptyset}$ or. Our four observations from above imply that there is $k \notin \{i,j,i_p\}$ with $k \in I_p$ such that $k$ owns $o_1$ at $\sigma_{\emptyset}$. In this scenario, consider $\bar{\theta}$ as in the previous paragraph to conclude that $\omega(i_p)=\varphi^{c}(\bar{\theta})(i_p)$ and $\omega(i)=\varphi^{c}(\bar{\theta})(i)$. 
\item Let $\hat{I}= \{i,\hat{j}\}$ with $\hat{j} \notin I_p$. Then consider $\bar{\theta}$ such that for all $l^{*} \notin \hat{I}$, $o_1$ is ranked first on $\bar{\theta}_{l^{*}}$, while $\omega(\hat{j})$ and $o_2$ rank at the last two positions. Let $\bar{\theta}_i=\theta_i$ and $\bar{\theta}_{\hat{j}}=\theta_{\hat{j}}$. Then $\omega(i) =\varphi^{c}(\bar{\theta})(i)$ and $\omega(\hat{j}) =\varphi^{c}(\bar{\theta})(\hat{j})$.

\end{itemize}
We conclude that in all instances above, $\hat{I}$ does not detect $\omega$. Thus, $\# \varphi > 2$.
This completes the arguments for Case 2. 

\paragraph{Case 3:} 

We begin with some additional notation that will be useful for the rest of the proof. Given any $i \in \mathcal{I}$, $o \in \mathcal{O}$, $\theta \in \Theta$, let $n(\theta,o)(i)$ be the smallest size of a suballocation, such that $i$ becomes the owner of $o$ under the TTC algorithm with input $\theta$.\footnote{Note that $n(\theta,o)(i)\leq N$ and hence $n(\theta,o)(i)$ is well-defined.} Moreover, given any $\theta \in \Theta$, and any $n \in \{1,\dots, N\}$, let $I^*_n(\theta)$ be the set of individuals that are owners of an object for a suballocation of size $n-1$ under the TTC algorithm with input $\theta$ for some step $t$. If no such suballocation exists, then this set is empty. Let $I^*_n \defeq \cup_{\theta \in \Theta} \ I^*_n(\theta)$. Note that $I_p=I^*_1$.

\textbf{\textit{Case 3.1: $\varphi^c$ violates Lemma  \ref{lemma:viceownership}} (3) a.} It is clear that by condition (1), there are at most two principal owners. In this case, we show that if there is a single principal owner, then there are at most two vice owners.\footnote{The remaining subcase where we have two principal owners will be treated at a later stage.} Concretely, given that $I_p$ is a singleton, we have $|(I^*_2 \setminus I_p)|> 2$.
We start with some useful remarks: First, if $I_p$ is a singleton, then $I^*_2=I^*_2 \setminus I_p$ and hence $I^*_2(\theta)= I^*_2(\theta) \setminus I_p$ for each $\theta$. Second, $|I^*_2(\theta)|\leq 2$ for each $\theta$ by Lemma \ref{lemma:viceownership} (1). 

\textit{Case 3.1.1.} Let $I^*_2(\theta)$ be singleton for each $\theta$.

\begin{itemize}
\item Suppose (i) there exists $j \in I^*_2$ with $j \notin I^*_3$, or (ii) $I^*_2= I^*_3$. To start, in case that (ii) applies, then since Lemma \ref{lemma:viceownership} (3) a. is violated, we must have $|I^*_2|= |I^*_3|> 2$. Therefore, if (i) or (ii) apply, then we can select $j \in I^*_2$ with $n(\hat{\theta},o_1)(j)\geq 4$ for some $o_1 \in \mathcal{O}$ and some problem $\hat{\theta}$. Let $i \in I^*_3$ be such that $n(\hat{\theta},o)(i)=3$.  Since $j \in I^*_2$ and since we know that $I^*_2(\hat{\theta})$ is a singleton, there must exist $\tilde{\theta}$ such that $n(\tilde{\theta},o_1)(j)=2 > n(\tilde{\theta},o_1)(i)$. 

\item Suppose there exist $i' \in I^*_3$ with $i' \notin I^*_2$.
Thus, we can select $i \in I^*_3$ and $\hat{\theta}$ such that $n(\hat{\theta},o_1)(i)=3$ for some $o_1 \in \mathcal{O}$ and $i \notin I^*_2$. Because $I^*_2$ is not a singleton, we can select $j \in I^*_2$ with $n(\hat{\theta},o_1)(j)\geq 4$.   Moreover, there exists $\tilde{\theta}$ such that $n(\tilde{\theta},o_1)(j)=2 > n(\tilde{\theta},o_1)(i)$.  
\end{itemize}

For each of the two subcases above, we can select $I=\{i,j\}$, $\sigma^1=\sigma_{2}(\hat{\theta})$ and $\sigma^2=\sigma_{1}(\tilde{\theta})$. Since $N\geq 6$ and a single object is matched in $\sigma^2$ and only two objects are matched in $\sigma^1$, we have $|\bar{O}(\sigma^1) \cap \bar{O}(\sigma^2)| \geq 2$.  Now apply Step 2 of Case 1, taking into account that not all subcases of Case 1 are relevant since, in the current case, there is a single principal owner and Case 1 additionally allows for two principal owners.

\textit{Case 3.1.2.} Suppose that $|I^*_2(\hat{\theta})|=2$ for some $\hat{\theta}$. Denote $\{j,k\}$ as the individuals in $I^*_2(\hat{\theta})$. Since $|I^*_2|>2$, there must exist $\tilde{\theta}$ with $i \in I^*_2(\tilde{\theta})$ and $i  \notin I^*_2(\hat{\theta})$. Denote $\sigma^1=\sigma_{1}(\hat{\theta})$, $\sigma^2=\sigma_{1}(\tilde{\theta})$. Then there is $o_1$ such that $c_{\sigma^1}(o_1) \in \{j,k\}$ and $c_{\sigma^2}(o_1)=i$. W.l.o.g, let $c_{\sigma^1}(o_1)= j$ (the arguments for $c_{\sigma^1}(o_1)= k$ are symmetric). Also, since $N\geq 6$, there exists $o\neq o_1$, such that $o \in  \bar{O}(\sigma^1) \cap \bar{O}(\sigma^2)$.

\underline{\textit{Case 3.1.2.1:}}
Suppose there is $o_2 \neq o_1$, $o_2 \in  \bar{O}(\sigma^1) \cap \bar{O}(\sigma^2)$ such that $c_{\sigma^2}(o_2) =j$ and $c_{\sigma^1}(o_2)=k$. 
Next, consider $\theta''$ for which $\sigma^1= \sigma_{1}(\theta'')$ and where $\theta''_j$ ranks $o_2$ first, $\theta''_k$ ranks $o_1$ first and some $o_3 \in \bar{O}(\sigma^1) \cap \bar{O}(\sigma^2)$ with $o_3\notin \{o_1,o_2\}$ is ranked second on $\theta''_k$. Such $o_3$ exists since $N\geq 6$. Let $\theta''_i$ rank $o_1$ first and $o_3$ second and assume that for all $k^{*} \notin \{i,j,k\}$ and for each $\bar{o} \notin  \{o_1,o_2, o_3\}$, $\bar{o} \ \theta''_{k^{*}} \ o_1$, $\bar{o} \ \theta''_{k^{*}} \ o_2$ and $\bar{o} \ \theta''_{k^{*}} \ o_3$. Also, let $o_3$ rank last on $\theta''_{j}$. Hence, $\varphi^c(\theta'')(i)=o_3$, $\varphi^c(\theta'')(j)=o_2$ and $\varphi^c(\theta'')(k)=o_1$.

Consider $\theta$ such that for all $k^{*} \notin \{i,j,k,l\}$, $\varphi^c(\theta'')(k^{*})$ ranks first on $\theta_{k^{*}}$, while the remaining details of $\theta''$ are the same as under $\theta''$. Thus, $\varphi^c(\theta)=\varphi^c(\theta'')$.

Deviation $\omega \neq \varphi^c(\theta)$ at $\theta$ is specified as follows: 
For all $k^{*} \in \mathcal{I} \setminus \{i,j,k\}$, let $\omega(k^{*})=\varphi^c(\theta)(k^{*})$, while $\omega(i)=o_1$ and $\omega(j)=o_2$ and $\omega(k)=o_3$. 

Take any $\hat{I}\subsetneq \mathcal{I}$ with $|\hat{I}|= 2$. 
If $\hat{I}\cap \{i,j,k\} = \emptyset$, then $\hat{I}$ does not detect $\omega$. For the remaining cases, construct a problem $\theta'$ as follows:

\begin{itemize}

\item If $\hat{I}= \{i,j\}$, then, for all $j^{*} \notin \{i,j\}$, $\varphi^c(\tilde{\theta})(j^{*})$ ranks first on $\theta'_{j^{*}}$. Thus, $\omega(i)=\varphi^c(\theta')(i)$ and $\omega(j)=\varphi^c(\theta')(j)$.
\item If $\hat{I}= \{i,k\}$, then $i_p \in I_p$ ranks $\sigma_2(i_p)$ first on $\theta_{i_p}'$, while for all $j^{*}\neq i_p$, $\theta'_{j^*}=\theta_{j^*}$  Thus, $\omega(i)=\varphi^c(\theta')(i)$ and $\omega(k)=\varphi^{c}(\theta')(k)$. 
\item If $\hat{I}= \{j,k\}$, then $i_p$ ranks $\sigma_2(i_p)$ first on $\theta_{i_p}'$ while for all $j^{*} \neq i_p$,  $\theta'_{j^*}=\theta_{j^*}$. Thus, $\omega(j)=\varphi^c(\theta')(j)$ and $\omega(k)=\varphi^c(\theta')(k)$.
\item If $\hat{I}= \{i,l^*\}$ with $l^* \notin \{i,j,k\}$.  Then $i_p \in I_p$ ranks $\sigma_2(i_p)$ first on $\theta_{i_p}'$, while for all $j^{*}\neq i_p$, $\theta'_{j^*}=\theta_{j^*}$ Thus, $\omega(l^*)=\varphi^c(\theta')(l^*)$ and $\omega(i)=\varphi^c(\theta')(i)$. 
\item If $\hat{I}= \{j,l^*\}$ with $l^* \notin \{i,j,k\}$, then, for all $j^{*} \notin \hat{I}$, $o_1$ is ranked last on $\theta'_{j^{*}}$. Let $\theta'_j=\theta_j$ and $\theta'_{l^*}=\theta_{l^*}$.  Thus, $\omega(l^*)=\varphi^c(\theta')(l^*)$ and $\omega(j)=\varphi^c(\theta')(j)$. 
\item If $\hat{I}= \{k,l^*\}$ with $l^* \notin \{i,j,k\}$, then, $o_1$ be ranked first on $\theta'_{j}$. For all $j^{*} \notin \hat{I}$, $o_3$ ranks last on $\theta'_{j^*}$. Also let $\theta'_k=\theta_k$ and $\theta'_{l^*}=\theta_{l^*}$. Thus, $\omega(l^*)=\varphi^c(\theta')(l^*)$ and $\omega(k)=\varphi^c(\theta')(k)$.
\end{itemize}

Hence, in all instances above, $\hat{I}$ does not detect $\omega$. Thus, $\# \varphi^c > 2$.
This completes the arguments for this subcase.

\underline{\textit{Case 3.1.2.2:}}
Suppose there exists $o_2 \in  \bar{O}(\sigma^1) \cap \bar{O}(\sigma^2)$ such that $c_{\sigma^2}(o_2) =k$ and $c_{\sigma^1}(o_2)=k$. 
Consider a problem $\theta''$ such that $\sigma^1=\sigma_1(\theta'')$. Let $\theta''_j$ rank $o_2$ first, while $\theta''_k$ ranks $o_1$ first and some $o_3 \in \bar{O}(\sigma^1) \cap \bar{O}(\sigma^2)$ with $o_3 \notin \{o_1,o_2\}$ in the second position. Such $o_3$ exists since $N\geq 6$. Moreover, assume that for all $k^{*} \notin \{i,j,k\}$ and for each $\bar{o} \notin  \{o_1,o_2, o_3\}$, $\bar{o} \ \theta''_{k^{*}} \ o_1$, $\bar{o} \ \theta''_{k^{*}} \ o_2$ and $\bar{o} \ \theta''_{k^{*}} \ o_3$ . Furthermore, $\theta''_i$ ranks $o_2$ first and $o_3$ second. Hence $\varphi^c(\theta'')(i)=o_3$, $\varphi^c(\theta'')(j)=o_2$ and $\varphi^c(\theta'')(k)=o_1$.

Consider $\theta$ where for all $k^{*} \notin \{i,j,k,l\}$, object $\varphi^c(\theta'')(k^{*})$ ranks first on $\theta_{k^{*}}$, while the remaining details of $\theta$ are the same as under $\theta''$. Thus, $\varphi^c(\theta)=\varphi^c(\theta'')$.

The deviation $\omega \neq \varphi^c(\theta)$ at problem $\theta$ is such that for all $k^{*} \in \mathcal{I} \setminus \{i,j,k\}$, $\omega(k^{*})=\varphi^c(\theta)(k^{*})$, while $\omega(i)=o_1$ and $\omega(j)=o_2$ and $\omega(k)=o_3$.

Take any $\hat{I}\subsetneq \mathcal{I}$ with $|\hat{I}|= 2$. First, if $\hat{I}\cap \{i,j,k\} = \emptyset$, then $\hat{I}$ does not detect $\omega$. For the remaining cases, construct a problem $\theta'$ as follows:

\begin{itemize}
\item If $\hat{I}= \{i,j\}$, then for all $j^{*} \notin \{i,j\}$, $\omega(j^{*})$ ranks first on $\theta'_{j^{*}}$. Thus, $\omega(i)=\varphi^c(\theta')(i)$ and $\omega(j)=\varphi^c(\theta')(j)$.  

\item If $\hat{I}= \{i,k\}$, then individual $i_p$ ranks $\sigma_2(i_p)$ first on $\theta_{i_p}'$, while for all $j^{*} \notin \{i,k\}$. Thus, $\omega(i)=\varphi^c(\theta')(i)$ and $\omega(k)=\varphi^c(\theta')(k)$.

\item If $\hat{I}= \{j,k\}$, then individual $i_p$ ranks $\sigma_2(i_p)$ first on $\theta_{i_p}'$, while for all $j^{*} \notin \{j,k, i_p\}$, $\omega(j^{*})=\varphi^c(\theta)(j^{*})$ ranks first. Thus, $\omega(j)=\varphi^c(\theta')(j)$ and $\omega(k)=\varphi^c(\theta')(k)$.

    \item If $\hat{I}= \{i, l^*\}$ with $l^* \notin \{i,j,k\}$, then for all $j^{*} \notin \hat{I}$, $o_1$ and $\omega(l^*)$ are ranked at the last two positions on $\theta'_{j^{*}}$. Let $\theta'_i=\theta_i$ and $\theta'_{l^*}=\theta_{l^*}$. Thus, $\omega(l^*)=\varphi^c(\theta')(l^*)$ and $\omega(i)=\varphi^c(\theta')(i)$. 
\item If $\hat{I}= \{j,l^*\}$ with $l^* \notin \{i,j,k\}$, then, for all $j^{*} \notin \hat{I}$, $o_1$ and $\omega(l^*)$ are ranked at the last two positions on $\theta'_{j^{*}}$. Let $\theta'_j=\theta_j$ and $\theta'_{l^*}=\theta_{l^*}$.  Thus, $\omega(j)=\varphi^c(\theta')(j)$ and $\omega(l^*)=\varphi^c(\theta')(l^*)$.

\item If $\hat{I}= \{k,l^*\}$ with $l^* \notin \{i,j,k\}$, then $o_1$ is ranked first on $\theta'_{j}$ and for all $j^{*} \notin \hat{I}\cup \{j\}$, $\theta'_{j^*}=\theta_{j^*}$. Let $\theta'_k=\theta_k$ and $\theta'_{l^*}=\theta_{l^*}$. Thus, $\omega(k)=\varphi^c(\theta')(k)$ and $\omega(l^*)=\varphi^c(\theta')(l^*)$.

\end{itemize}
Hence, in all instances above, $\hat{I}$ does not detect $\omega$. Thus, $\# \varphi^c > 2$.
This completes the arguments for this subcase.

\underline{\textit{Case 3.1.2.3:}}
Suppose that for all $o \in  \bar{O}(\sigma^1) \cap \bar{O}(\sigma^2)$ such that $o \neq o_1$, it holds  $c_{\sigma^2}(o) =k$ and $c_{\sigma^1}(o)=l$ with $l \notin \{i,j,k\}$. 
In this case, we can slightly extend the arguments of Case 3.1.2.1 to reach the desired conclusion. More specifically, take a problem $\theta$ such that $j$ and $k$ form a cycle at step 2, such that $j$ receives $o_1$ and $k$ receives $o_2$. At $\theta$, $i$ ranks $o_1$ first and $o_3$ second, while $l$ ranks $o_2$ and $o_4$ second. All other individuals rank the four objects $o_1,o_2,o_3,o_4$ at the last four positions. The deviation assigns $i$ to $o_1$ and $l$ to $o_2$, while $j$ gets $o_3$ and $k$ gets $o_4$. Then one can use similar arguments as for Case 3.1.2.1 and Case 3.1.2.2.

\textbf{\textit{Case 3.2: $\varphi^c$ violates Lemma  \ref{lemma:viceownership}} (3) b..}
Thus, there must exist a pair $\sigma^1, \sigma^2$ with $|\sigma_I^1|=|\sigma_I^2|=1$ such that $i \notin I_p$ owns $o_1  \in \bar{O}(\sigma^1) \cap \bar{O}(\sigma^2)$ at $\sigma^1$ and $i$ and $o_1$ are unmatched at $\sigma^2$, but there is $j \neq i$, with $j \notin I_p$ who owns $o_1$ at $\sigma^2$. Thus, $c_{\sigma^1}(o_1) = i$, $c_{\sigma^2}(o_1) = j$. Let $o_2\neq o_1$ with $o_2\in \bar{O}(\sigma^1) \cap \bar{O}(\sigma^2)$ and consider $\hat{\theta},\tilde{\theta}$ with $\sigma^1=\sigma_{t}(\hat{\theta})$ and $\sigma^2=\sigma_{t'}(\tilde{\theta})$ for some steps $t$ and $t'$. 
Apply the arguments of Case 1 Step 2.

\textbf{\textit{Case 3.3: $\varphi^c$ violates Lemma  \ref{lemma:viceownership}} (3) c.}
Thus, there are four individuals $i,j,k,l$, with $i,k \in I_p$ and $j,l \in I_v$ such that there exists problem $\theta''$, where 
\begin{itemize}
    \item $i \in I_p$ owns some object $o_2$ at $\sigma^1= \sigma_1(\theta'')$ with $|\sigma_I^1|=1$,
    \item $i$ does not own $o_2$ at $\sigma_{\emptyset}$, and
    \item $i$ does not own some $o_3\neq o_2$ at $\sigma^1$ (thus $o_3$ is not owned by $i$ at $\sigma_{\emptyset}$). 
\end{itemize}  
Together, this implies that there is $k \in I_p$ with $k \neq i$, who owns at least three objects, including the two objects $o_2$ and $o_3$. This $k$ must be matched to some $o \notin \{o_2,o_3\}$ that $k$ owns at $\sigma^{\emptyset}$.  Furthermore, since Lemma \ref{lemma:viceownership} (3) c. is violated, there is $j \in I_v$ who owns some object at $\sigma^1$ that $k$ owns at $\sigma^{\emptyset}$. W.l.o.g., let this object be $o_3$. Finally, there is also an object $\hat{o}$ owned by $k$ at $\sigma^1$, that is not owned by $j$ at $\sigma^2=\sigma^1\cup \{i,\bar{o}\}$, where $\bar{o}$ is an object owned by $i$ at $\sigma^1$. Denote $\bar{o}:=o_4$ and w.l.o.g let $\hat{o}:=o_2$.

We further specify $\theta''$ as follows: $\theta''_k$ ranks $o_1$ first, while $\theta''_i$ ranks $o_3$ first and $o_4$ second. $\theta''_j$ ranks $o_2$ first, $o_3$ second and $o_1$ last. $\theta''_l$ ranks $o_2$ first, $o_4$ second and $o_1$ last. For all $k^{*} \notin \{i,j,k,l\}$ and for each $o' \notin  \{o_1,o_2, o_3,o_4\}$, and $o'' \in \{o_1,o_2, o_3,o_4\}$, $o' \ \theta''_{k^{*}} \ o''$. Note that this specification of $\theta''$ does not contradict our arguments in the last paragraph by strategy-proofness, non-bossiness of $\varphi^c$ and consistency of $c$. Hence $\varphi^c(\theta'')(i)=o_3$, $\varphi^c(\theta'')(j)=o_2$, $\varphi^c(\theta'')(k)=o_1$, $\varphi^c(\theta'')(l)=o_4$.

Consider $\theta$ where for all $k^{*} \notin \{i,j,k,l\}$, 
$\varphi^c(\theta'')(k^{*})$ ranks first on $\theta_{k^{*}}$, while the remaining details of $\theta$ are the same as under $\theta''$. Thus, $\varphi^c(\theta)=\varphi^c(\theta'')$.

Let $\omega \neq \varphi^c(\theta)$ at $\theta$ be such that for all $k^{*} \in \mathcal{I} \setminus \{i,j,l\}$, $\omega(k^{*})=\varphi^c(\theta)(k^{*})$, while
$\omega(i)=o_4$, $\omega(j)=o_3$ and $\omega(l)=o_2$. 

Now, take any $\hat{I}\subsetneq \mathcal{I}$ with $|\hat{I}|= 2$. If $\hat{I}\cap \{i,j,l\} = \emptyset$, then the deviation will not be detected.  
For the remaining cases, construct a problem $\theta'$ as follows:

\begin{itemize}

\item If $\hat{I}= \{i,j\}$, then $k$ ranks $o_2$ first on $\theta_{k}'$, while for all $j^{*}\neq k$, $\theta_{j^*}'=\theta_{j^*}$. Thus, $\omega(i)=\varphi^c(\theta')(i)$ and $\omega(j)=\varphi^c(\theta')(j)$.
\item If $\hat{I}= \{i,l\}$, then $j$ ranks $o_3$ first on $\theta_{j}'$, while for all $j^{*} \neq j$,  $\theta_{j^*}'=\theta_{j^*}$. Thus, $\omega(i)=\varphi^c(\theta')(i)$ and $\omega(l)=\varphi^c(\theta')(l)$. 
\item If $\hat{I}= \{j,l\}$, then $i$ ranks $o_4$ first on $\theta_{i}'$, while for all $j^{*}\neq i$, $\theta_{j^*}'=\theta_{j^*}$. Thus, $\omega(j)=\varphi^c(\theta')(j)$ and $\omega(l)=\varphi^c(\theta')(l)$.
\item If $\hat{I}= \{i,l^*\}$ with $l^* \notin \{j,l\}$, then $j$ ranks $o_3$ first on $\theta_{j}'$, while for all $j^{*}\neq j$,  $\theta_{j^*}'=\theta_{j^*}$. Thus, $\omega(i)=\varphi^c(\theta')(i)$ and $\omega(l^*)=\varphi^c(\theta')(l^*)$.
\item If $\hat{I}= \{j,l^*\}$ with $l^* \notin \{i,l\}$, then $i$ ranks $o_4$ first on $\theta_{i}'$, while for all $j^{*} \neq i$,  $\theta_{j^*}'=\theta_{j^*}$. Thus, $\omega(j)=\varphi^c(\theta')(j)$ and  $\omega(l^*)=\varphi^c(\theta')(l^*)$.
\item If $\hat{I}= \{l,l^*\}$ with $l^* \notin \{i,j\}$, $i$ ranks $o_4$ first on $\theta_{i}'$ while for all $j^{*}\neq i$, $\theta_{j^*}'=\theta_{j^*}$. Thus, $\omega(l)=\varphi^c(\theta')(l)$ and $\omega(l^*)=\varphi^c(\theta')(l^*)$.
\end{itemize}

Hence, in all these instances above, $\hat{I}$ does not detect $\omega$. Thus, $\# \varphi^c > 2$. This completes the arguments for Case 3.3.

\textbf{\textit{Case 3.4: $\varphi^c$ violates Lemma  \ref{lemma:viceownership}} (3) d.}

Thus, there exists $\sigma^1$ with $|\sigma_I^1|=2$, such that there is an individual $l \notin I_p$ who owns an object $o_1$ at $\sigma^1$, but there is an individual  $i \neq l$, with $i \in I_p$ that is unmatched under $\sigma^1$. 

This implies that there is $j \in I_p$ with $j \neq i$, who owns at least three objects at $\sigma_{\emptyset}$. Furthermore, $j$ is matched to an object $o_1$ that she owns at $\sigma_{\emptyset}$.  Second, there must exist $k \in I_v$ who receives ownership for at least two objects that $j$ owns at $\sigma_{\emptyset}$ and $k$ must be matched to one of these objects at $\sigma^1$. Otherwise, $l$ cannot own an object that is owned by $j$ at $\sigma_{\emptyset}$ at $\sigma^1$ ( This follows from consistency of $c$ and the fact that that $\sigma^1$ has size two and $I_2^*(\theta)<2$ for each $\theta$. Let $\sigma^1(k)=:o_2$ and $o_3 \in \bar{O}(\sigma_1)$ be owned by $j$ at $\sigma_{\emptyset}$ such that $l$ owns $o_3$ at $\sigma^1$. Finally, let $o_4 \in \bar{O}(\sigma_1)$,  which $i$ owns at $\sigma_{\emptyset}$.

Next, consider $\theta''$ such that we reach $\sigma^1=\sigma_1(\theta'')$. Let $\theta''_j$ rank $o_1$ first, while $\theta''_k$ ranks $o_2$ first. Furthermore, $\theta''_l$ ranks $o_3$ first and some $o_5$ in the second position. Such $o_5$ exists since $N\geq 6$. Also, $\theta''_i$ ranks $o_3$ first and $o_5$ second. For all $k^{*} \notin \{i,j,k,l\}$, for each $\bar{o} \notin  \{o_1,o_2, o_3,o_4,o_5\}$ and $o' \in \{o_1,o_2, o_3,o_4,o_5\}$, $\bar{o} \ \theta''_{k^{*}} \ o'$. Hence $\varphi^c(\theta'')(i)=o_5$, $\varphi^c(\theta'')(j)=o_1$, $\varphi^c(\theta'')(k)=o_2$, $\varphi^c(\theta'')(l)=o_3$.

Consider $\theta$ such that for all $k^{*} \notin \{i,j,k,l\}$, object $\varphi^c(\theta'')(k^{*})$ ranks first on $\theta_{k^{*}}$, while the remaining details of $\theta$ are the same as under $\theta''$. Thus, $\varphi^c(\theta)=\varphi^c(\theta'')$.

Consider a deviation $\omega \neq \varphi^c(\theta)$ at $\theta$ such that for all $k^{*} \in \mathcal{I} \setminus \{i,l\}$, $\omega(k^{*})=\varphi^c(\theta)(k^{*})$, while $\omega(i)=o_3$ and $\omega(l)=o_5$. Let $\hat{I}\subsetneq \mathcal{I}$ with $|\hat{I}|= 2$ be arbitrary. If $\hat{I}\cap \{i,l\} = \emptyset$, then for all $k^{*} \in \mathcal{I} \setminus \{i,l\}$, $\omega(k^{*})=\varphi^c(\theta)(k^{*})$. For the remaining cases, construct a problem $\theta'$ as follows:
\begin{itemize}
    \item If $\hat{I}= \{i,l\}$, then $o_4$ ranks first on $\theta'_{j}$. For $i'$ with $\varphi^c(\theta)(i')=o_4$,  $o_1$ ranks first on $\theta'_{i'}$. For all $j^{*} \notin \{i,j,l,i'\}$, $\varphi^c(\tilde{\theta})(j^{*})$ ranks first on $\theta'_{j^{*}}$. Let $\theta'_i=\theta_i$ and $\theta'_{l}=\theta_{l}$.  Thus, $\omega(i)=\varphi^c(\theta')(i)=o_3$ and $\omega(l)=\varphi^c(\theta')(l)=o_5$.
    \item  If $\hat{I}= \{i,k^*\}$ with $k^* \neq l$, then for all $j^{*} \notin \hat{I}$, $o_1$ is ranked last on $\theta'_{j^{*}}$. Let $\theta'_i=\theta_i$ and $\theta'_{k^*}=\theta_{k^*}$.  Thus, $\omega(i)=\varphi^c(\theta')(i)$ and $\omega(k^*)=\varphi^c(\theta')(k^*)$. 
    \item   If $\hat{I}= \{j,l\}$, $o_3$ ranks first on $\theta_{k}'$. Also, $o_2$ ranks first on $\theta'_{i}$. For all $j^*\notin \{i,k\}$,  $\theta'_{j^*}=\theta_{j^*}$ Thus, $\omega(j)=\varphi^c(\theta')(j)$ and $\omega(l)=\varphi^{c}(\theta')(l)$.
\item If $\hat{I}= \{k^*,l\}$ with $k^* \notin \{i,j\}$, then $o_3$ first on $\theta_{j}'$. Also, $o_1$ ranks first on $\theta'_{i}$. For all $j^*\notin \{i,j\}$,  $\theta'_{j^*}=\theta_{j^*}$. Thus, $\omega(k^*)=\varphi^c(\theta')(k^*)$ and $\omega(l)=\varphi^{c}(\theta')(l)$.
\end{itemize}

This completes the arguments for Case 3.4. Also note that Case 3.1 - to Case 3.4 imply that there are at most two vice owners. Thus, Lemma \ref{lemma:viceownership} (3) a. needs to be satisfied also in the case that there are two principal owners.

\paragraph{Case 4:}

\textbf{\textit{Case 4.1: $\varphi^c$ violates Lemma  \ref{lemma:viceownership}} (4) a.}

Let $\sigma^1$ be the smallest suballocation such that $|\sigma_I^1|\leq N-4$  and for which there are two owners $i,j \notin I_p$ and where $j$ does not own some $o' \in \bar{O}(\sigma^1)$ at $\sigma^2=\sigma^1 \cup \{(i,o)\}$ for some $o$.

We first show that at least two objects are owned by $i$ at $\sigma^1$ and that one of these objects is $o$. By contradiction, assume that $i$ only owns $o$ at $\sigma^1$. This implies that $j$ owns all remaining objects at $\sigma^2$ by consistency of the ownership structure and Lemma \ref{lemma:viceownership} (1). Moreover, if $i$ does not own $o$, then $j$ cannot be unmatched under $\sigma^2$, which leads to a contradiction with respect to our selection of $\sigma^1$. Consider $o_1,o_2,o_3,o_4 \in \bar{O}(\sigma^1)$, where $o \defeq o_3$, $o':=o_1$, while $o_1,o_2,o_4 \in \bar{O}(\sigma^2)$. Furthermore, $c_{\sigma^1}(o_1)=i$, $c_{\sigma^1}(o_3)=i$,  $c_{\sigma^1}(o_4)=j$ and $c_{\sigma^2}(o_1)=k$. Note that this implies that neither $k$ nor $i$ can own $o_1$ at $\sigma^{\emptyset}$.

Next, consider $\theta''$ such that $\sigma^1=\sigma_t(\theta'')$ and $\sigma^2=\sigma_{t'}(\theta'')$ for some steps $t$ and $t'$.
Specifically, $\theta''_i$ ranks $o_3$ in the first position, $\theta''_j$ ranks $o_3$ in the first position and $o_1$ in the second position. $\theta''_k$ ranks $o_1$ in the first, $o_2$ in the second position and $o_4$ ranks first on $\theta''_l$. For all $k^* \notin \{i,j,k,l\}$, let $\theta''_{k^*}$ be such that for each $\bar{o} \notin  \{o_1,o_2, o_3, o_4\}$, $\bar{o} \ \theta''_{k^{*}} \ o_1$, $\bar{o} \ \theta''_{k^{*}} \ o_2$, $\bar{o} \ \theta''_{k^{*}} \ o_3$ and $\bar{o} \ \theta''_{k^{*}} \ o_4$. Thus, $\varphi^c(\theta'')(i)=o_3$, $\varphi^c(\theta'')(j)=o_2$, $\varphi^c(\theta'')(k)=o_1$ and $\varphi^c(\theta'')(l)=o_4$. 

Consider $\theta$ such that for all $k^{*} \notin \{i,j,k,l\}$, $\varphi^c(\theta'')(k^{*})$ ranks first on $\theta_{k^{*}}$ while the rest of $\theta$ is identical to $\theta''$. Thus, $\varphi^c(\theta)=\varphi^c(\theta'')$.

Consider deviation $\omega \neq \varphi^c(\theta)$ such that for all $k^{*} \in \mathcal{I} \setminus \{j,k\}$, $\omega(k^{*})=\varphi^c(\theta)(k^{*})$, while $\omega(j)=o_1$ and $\omega(k)=o_2$. Take any $\hat{I}\subsetneq \mathcal{I}$ with $|\hat{I}|= 2$. If $\hat{I}\cap \{j,k\} = \emptyset$, then for all $k^{*} \in \mathcal{I} \setminus \{j,k\}$, $\omega(k^{*})=\varphi^c(\theta)(k^{*})$. For the remaining cases, construct a problem $\theta'$ as follows:

\begin{itemize}
\item  If $\hat{I}= \{j,k\}$, then let $\theta'_i$ be such that $o \ \theta'_i \ o_4$, if and only if, $o \in \mathcal{O}\setminus \{o_4\}$ is matched under $\sigma^1$. Thus, $o_4 \ \theta_i' \ o_1$. For each $j^* \neq i$, let $\theta'_{j^*}=\theta_{j^*}$. Then $\omega(j)=\varphi^c(\theta)(j)$ and $\omega(k)=\varphi^c(\theta)(k)$.

\item If $\hat{I}= \{j,l^*\}$ with $l^* \neq k$, then $\theta_{k}'$ ranks $o_2$ on the first position and for each $j^* \neq k$, let $\theta'_{j^*}=\theta_{j^*}$. Thus, $\omega(j)=\varphi^c(\theta')(j)$ and $\omega(l^*)=\varphi^c(\theta')(l^*)$.

\item If $\hat{I}= \{k,l^*\}$ with $l^* \in I_p \setminus \{i,j\}$, then $\theta_{i}'$ ranks $o_1$ first and $\theta'_j$ ranks $o_4$ first. For each $j^{*} \notin \{i, j\}$, $\theta'_{j^*}=\theta_{j^*}$. Thus, $\omega(k)=\varphi^c(\theta')(k)$ and $\omega(l^*)=\varphi^c(\theta')(l^*)$.

\item Suppose that $\hat{I}= \{k,l^*\}$ with (i) $l^* \notin I_p \setminus \{j\}$, or (ii) $l^*=i$ and $i \in I_p$. Then for $i_p$ that owns $o_1$ at $\sigma_{\emptyset}$, let $\theta_{i_p}'$ rank $o_1$ first, while $\varphi^c(\theta)(i_p)$ ranks first on $\theta'_j$.\footnote{Recall that neither $k$ nor $i$ can own $o_1$ at $\sigma^{\emptyset}$. The same holds for $j$ and $l^*$.} For each $j^{*} \notin \{i_p,j\}$, $\theta'_{j^*}=\theta_{j^*}$. Thus, $\omega(k)=\varphi^c(\theta')(k)$ and $\omega(l^*)=\varphi^c(\theta')(l^*)$.

\end{itemize}

Hence, in all these instances above, any $\hat{I}$ does not detect $\omega$. Thus, $\# \varphi^c > 2$.
This completes Case 4.1.

\textbf{\textit{Case 4.2: $\varphi^c$ violates Lemma  \ref{lemma:viceownership}} (4) b.}
Condition 4) b. follows from the conditions of Lemma \ref{lemma:viceownership} that are treated in previous subcases. Specifically, if (4) b. would be violated, then there is $j \in I_p\cup I_v$ such that $j$ is unmatched at some suballocation $\sigma$, while there exists some $i \notin I_p\cup I_v$ that is an owner of some object $o$ under $\sigma$. First, if $j$ is a principal owner, then condition (3) d. is violated. 

Second, consider the remaining subcases where $j$ is a vice owner: Suppose that there are two principal owners and two vice owners $j,k$ (recall that there are at most two principal owners and at most two vice owners). Then there exists $\sigma$ such that both, $j$ and $k$ own an object at $\sigma^1$: To see this, note that if there are two principals and two vice owners, then if there exists a problem where each vice owner gets the ownership for the all remaining objects of a particular principal owner, then each principal owner owns at least two objects at $\sigma^{\emptyset}$. Then by (3) b., there must exist $\sigma$ such that both $j$ and $k$ own an object. Together with (4) a. this implies that (4) b. cannot be violated. Next, if there is single vice owner, then we can immediately apply condition (3) b. to reach the desired conclusion. Finally, suppose there is single principal owner. Then, if there is a single vice owner, condition (4) b. cannot be violated. If there are two vice owners and there is a suballocation, where both vice owners are owners, then condition (4) a. implies that (4) b. cannot be violated using the same arguments as for the case with two principal owners above. Finally, if at any suballocation at most one vice owner owns an object, then one can use similar arguments as in Case 3.1.1 to reach the desired conclusion.

This completes the arguments for Case 4.3.

\textbf{\textit{Case 4.3: $\varphi^c$ violates Lemma  \ref{lemma:viceownership}} (4) c.}

Let $\sigma^1,\sigma^2$ be the smallest suballocations such that (i) there are at least three individuals and three objects unmatched, (ii) there is $i \notin I_p \cap I_v$ with $i \in \bar{I}(\sigma^1)\cup \bar{I}(\sigma^2)$ that owns $o \in \bar{O}(\sigma^1)\cup \bar{O}(\sigma^2)$, and there is $j\neq i$, $j \notin I_p \cap I_v$ that owns $o$ under $\sigma^2$. Let $\hat{\theta}$ and $\tilde{\theta}$ be such that $\sigma_{t}(\hat{\theta})= \sigma^1$ and $\sigma_{t'}(\tilde{\theta}) = \sigma^2$ for some $t, t' \in \{1,\dots,N-1\}$ .

 Let $|\sigma_I^1|=n$ for some $n \leq N-3$. Then, by our selection of $\sigma^2$, $j$ cannot own $o$ under any submatching $\sigma^3$ such that $\sigma^3=\sigma_{\hat{t}}(\tilde{\theta})$ for some $\hat{t}<t'$, since otherwise $\sigma^2$ is not selected according to the smallest size. Together these arguments imply that $|\sigma_I^2|=n$.

Furthermore, it holds $N-n-1 = |\bar{O}(\sigma^1) \cap \bar{O}(\sigma^2)|$, since if this equation does not hold, then one can find $\hat{\sigma}^1$, $\hat{\sigma}^2$ that satisfy all previous conditions, and that both have the same size as $\sigma^1$ and $\sigma^2$.

Next, since $n \leq N-3$, there exists $o_2\neq o_1$ with  $o_1,o_2 \in \bar{O}(\sigma^1) \cap \bar{O}(\sigma^2)$. Apply the arguments of Case 1 Step 2. 

This completes the arguments for Case 4.3.
 
\textbf{\textit{Case 4.4: $\varphi^c$ violates Lemma  \ref{lemma:viceownership}} (4) d.}

First, there exists a problem $\theta''$ with  $\sigma^1=\sigma_{t}(\theta'')$ such that $|\sigma_I^1|=N-3$ for some step $t$, such that there is an individual $i$ that owns some object $o_1$ at $\sigma^1$ and another individual $j$ that owns two objects $o_2,o_3$ at $\sigma^1$. By the remaining conditions of Lemma \ref{lemma:viceownership}, $i$ and $j$ do not own any object at a smaller suballocation than $\sigma^1$ (in particular, condition (4) a. and (4) c.). Second, there exists $l \notin \{i,j\}$ that owns $o_2$ at $\sigma^2 = \sigma^1 \cup \{(j,o_3)\}$.\footnote{The symmetric case would be that $i$ owns $o_2$ at $\sigma^1$ and $l$ owns $o_2$ at $\sigma^2 = \sigma^1 \cup \{(i,o_1)\}$. The arguments are identical.}

We further specify $\theta''$ as follows: $\theta''_i$ ranks $o_2$ first, $o_1$ second, and $o_3$ last. Also, $\theta''_j$ ranks $o_1$ first and $o_3$ second and let $\theta''_l$ rank $o_1$ first, $o_2$ second and $o_3$. For all $k^{*} \notin \{i,j,l\}$, each $\bar{o} \notin  \{o_1,o_2, o_3\}$ and $o' \in \{o_1,o_2, o_3\}$, $\bar{o} \ \theta''_{k^{*}} \ o'$. Hence $\varphi^c(\theta'')(i)=o_2$, $\varphi^c(\theta'')(j)=o_1$, $\varphi^c(\theta'')(l)=o_3$.

Consider $\theta$ where for all $k^{*} \notin \{i,j,l\}$, $\varphi^c(\theta'')(k^{*})$ ranks first on $\theta_{k^{*}}$ while the rest of $\theta$ is identical to $\theta''$. Thus, $\varphi^c(\theta)=\varphi^c(\theta'')$. 

Take the deviation $\omega \neq \varphi^c(\theta)$ at problem $\theta$such that
for all $k^{*} \in \mathcal{I} \setminus \{i,j,l\}$, $\omega(k^{*})=\varphi^c(\theta)(k^{*})$ and $\omega(i)=o_1$, $\omega(j)=o_3$ and $\omega(l)=o_2$.

Next, take any $\hat{I}\subsetneq \mathcal{I}$ with $|\hat{I}|= 2$. If $\hat{I}\cap \{i,j,l\} = \emptyset$, then for all $k^{*} \in \mathcal{I} \setminus \{i,j,l\}$, we have $\omega(k^{*})=\varphi^c(\theta)(k^{*})$. For the remaining cases, construct a problem $\theta'$ as follows:

\begin{itemize}
\item If $\hat{I}= \{i,j\}$, then consider some $k \notin \{i,j,l\}$ that ranks $o_2$ first on $\theta'_{k}$. For all $j^{*} \neq k$, $\theta_{j^*}'=\theta_{j^*}$. Thus, $\omega(i)=\varphi^c(\theta')(i)$ and $\omega(j)=\varphi^c(\theta')(j)$.

\item If $\hat{I}= \{i,l\}$, then let $j$ rank $o_3$ first on $\theta_{j}'$. For all $j^{*} \neq j$, $\theta_{j^*}'=\theta_{j^*}$. Thus, $\omega(i)=\varphi^c(\theta')(i)$ and $\omega(l)=\varphi^c(\theta')(l)$.

\item If $\hat{I}= \{j,l^*\}$ such that $l^*\neq i$, then $i$ ranks $o_1$ first on $\theta_{i}'$, while for all $j^{*}\neq i$, $\theta_{j^*}'=\theta_{j^*}$. Thus, $\omega(j)=\varphi^c(\theta')(j)$ and $\omega(l^*)=\varphi^c(\theta')(l^*)$.

\item If $\hat{I}= \{i,l^*\}$ with $l^* \notin \{j,l\}$, then $j$ ranks $o_2$ first on $\theta_{j}'$, while for all $j^{*} \neq j$, $\theta_{j^*}'=\theta_{j^*}$. Thus, $\omega(i)=\varphi^c(\theta')(i)$ and $\omega(l^*)=\varphi^c(\theta')(l^*)$. 

\item If $\hat{I}= \{l,l^*\}$ with $l^* \notin \{i,j\}$, then $i$ ranks $o_1$ first on $\theta_{i}'$, while for all $j^{*}\neq i$, $\theta_{j^*}'=\theta_{j^*}$. Thus, $\omega(l)=\varphi^c(\theta')(l)$ and $\omega(l^*)=\varphi^c(\theta')(l^*)$.
\end{itemize}

Hence, in all these instances above, $\hat{I}$ does not detect $\omega$. Thus, $\# \varphi^c > 2$.

This completes the arguments for Case 4.  Hence, this completes the proof for the `only if' direction.

We next turn to the proof of the `if' direction. Consider an arbitrary vice ownership mechanism $\varphi^c$.  To show that $\# \varphi^c = 2$, consider an arbitrary problem $\theta$ and an arbitrary deviation $\omega\neq \varphi^c(\theta)$. We have to show that there is a group of two individuals that can detect the deviation. Consider the smallest suballocation $\sigma^1\defeq \sigma_{t}(\theta)$ such that $\sigma^2\defeq\sigma_{t+1}(\theta)$ implies that $\varphi^c(\theta)\neq \omega$. Let $\sigma^1$ be of size $n-1$.

Let $I(\omega) \defeq \{ i' \in \sigma^2_I \setminus  \sigma^{1}_I \ | \ \varphi^c(\theta)(i') \neq \omega(i')\}$. By strategy-proofness and Pareto efficiency of $\varphi^c$, it is clear that for each $i' \in I(\omega)$, we must have $\varphi^c(\theta)(i') \ \theta_{i'} \ \omega(i')$. Take an arbitrary individual in $i \in I(\omega)$.

We show that there exists $k \in \mathcal{I}$, such that $\hat{I}\defeq \{i,k\}$ detects $\omega$. First, it is clear that $\varphi^c(\theta)(i) \in \bar{O}_{\sigma_{t-1}(\theta)}$. Moreover, we know that for each $l \in \sigma^1_I$, $\varphi^c(\theta)(l)=\omega(l)$. Hence, there exists $j\neq i$ unmatched under $\sigma^1$, such $\omega(j)=\varphi^c(\theta)(i) \defeq o$. Consider the following case distinction based on Definition \ref{def:viceownership}. 

\paragraph{Case 1:} Let $i$ be a principal owner. First, suppose that $i$ owns $o$ at $\sigma^1$. It is clear that if $i$ owns $o$ at $\sigma_{\emptyset}$, then $i$ must detect $\omega$. Thus, $|\sigma_I^1|\geq 1$. Hence there are two principal owners $i,k \in I_p$. Furthermore, by condition Definition \ref{def:viceownership} (4), we know that $|\sigma_I^1|\leq 2$. 
Next, if $j$ is a regular or residual owner, then, $\hat{I}=\{i,j\}$ detects $\omega$ by Definition \ref{def:viceownership} (5).  Finally, if $j$ is a vice owner, then either $\hat{I}=\{i,k\}$ or $\hat{I}=\{i,j\}$ detect $\omega$ by Definition \ref{def:viceownership} (4) i. or (4) ii., respectively.

Second, suppose $i$ does not own $o$ at $\sigma^1$. Thus, there are two principal owners $i,k \in I_p$. Again, by Definition \ref{def:viceownership} (5), if $j$ would be a regular or residual owner, then $\hat{I}=\{i,j\}$ detects $\omega$. 
Next, since $i$ must own $o$ if $|\sigma^1_I|\geq 2$ by Definition \ref{def:viceownership} (4), $0\leq|\sigma_I^1|\leq 1$. Suppose that $|\sigma_I^1|=0$. Then $\varphi^c(\theta)(k)=o'$ for some $o'$ that $i$ owns at $\sigma^1=\sigma_{\emptyset}$. Moreover, $o$ is owned by $k$ at $\sigma^1$. Thus, $i$ and $k$ must receive their first ranked objects on $\theta_i$ and $\theta_k$, respectively. Hence $\hat{I}=\{i,k\}$ detects $\omega$. Suppose that $|\sigma_I^1|=1$. Then $\varphi^c(\theta)(k)=\omega(k)$. Otherwise, $\sigma_1$ cannot be of minimal size. Hence, it remains the scenario where $j \in I_v$ and $j$ owns $o$ at $\sigma^1$, while $\varphi^c(\theta)(j)=o'$ and $i$ owns $o'$ at $\sigma^1$. In this scenario, if $\omega(j) \neq o$, then there is $l  \notin I_p$ with $\omega(l)= o$. If $l \in I_v$, then by Definition \ref{def:viceownership} (4) ii., $\hat{I}=\{i,l\}$ detects the deviation. If $l$ is regular or a residual owner, then $\hat{I}=\{i,l\}$ detects the deviation by Definition \ref{def:viceownership} (5). 
Thus, $\omega(j) = o$. In this case, if $\omega(i) = o'$, then $\hat{I}=\{i,j\}$ would detect $\omega$, because $\omega$ is not Pareto efficient for $i$ and $j$ (given their type reports and outcomes). Hence, $\omega(i) \neq o'$ and there must be $l \notin I_p$ with $\omega(l)= o'$: If $l$ is a regular or residual owner, then $\hat{I}=\{j,l\}$ detect the deviation by Definition \ref{def:viceownership} (5). If $l$ is a vice owner, then if $i$ owns $o'$ at $\sigma_{\emptyset}$, $\hat{I}=\{i,j\}$ would detect $\omega$. In fact, in this scenario, given their information as a group, $i$ and $j$ would know that they should have been in a cycle to exchange $o$ and $o'$ at step $t$  with input $\theta$ . Finally, if $i$ does not own $o'$ at $\sigma_{\emptyset}$, then Definition \ref{def:viceownership} (4) ii. requires that if $j$ and $o'$ are unmatched, then $j$ owns $o'$, given that $i$ and $k$ are matched. However, then $\hat{I}=\{j,l\}$ detects $\omega$. 
 
\paragraph{Case 2:} Let $i$ be a vice owner. First, suppose that $i$ owns $o$ at $\sigma^1$: If $j$ is a regular or residual owner, then again $\hat{I}=\{i,j\}$ detects $\omega$ by Definition \ref{def:viceownership} (5). If $j$ is a vice owner, then $\hat{I}=\{i,j\}$ detects $\omega$ by Definition \ref{def:viceownership} (4) i. 

Second, suppose that $i$ does not own $o$ at $\sigma^1$. Again, if  $j$ is a regular or residual owner, then, $\hat{I}=\{i,j\}$ detects $\omega$ by Definition \ref{def:viceownership} (5). If $j$ is a principal owner, then the minimal size selection of $\sigma^1$ ensures that we are in the same scenario as in Case 1.2. Likewise, we can use arguments as those in Case 1.2, if $j$ and $i$ are both vice owners.  It remains the scenario where $j \in I_v$. Then $j$ must own $o$ at $\sigma^1$ and we must have $\varphi^c(\theta)(j)=o'$ for $o'$ that $i$ owns at $\sigma^1$. 
If $\omega(j) \neq o$, then there is $l \neq i$ with $\omega(l)= o$. However, this means that $l$ is regular, since $\omega(k) = \varphi(\theta)(k)$  and $\omega(k) = \varphi(\theta)(k)$ for both principal owners $k,k' \in I_p$ by the minimum size selection of $\sigma^1$. Thus, $\hat{I}=\{i,l\}$ detects the deviation by Definition \ref{def:viceownership} (5). 
Hence it remains the scenario where, $\omega(j) = o$. If $\omega(i) = o'$, then $\hat{I}=\{i,j\}$ would detect $\omega$ since $\omega$ is not Pareto efficient for $i$ and $j$ (given their type reports and outcomes). It remains the case, where $\omega(i) \neq o'$. Here, we can use symmetric arguments as in the last paragraph, since $j$ is also a vice owner. That is, there is $l'$ with $\omega(l') = o'$ that is either a regular or a residual owner. Thus, $\hat{I}=\{j,l\}$ detects the deviation by Definition \ref{def:viceownership} (5). 

\paragraph{Case 3:} Let $i$ be a regular owner. By Definition \ref{def:viceownership} (5), for all individuals $k$ that are in a strictly lower level than $i$, we have $\omega(k) = \varphi(\theta)(k)$. Thus, $j$ is a regular or residual owner. Then, if $j$ is a residual owner or in a strictly lower level than $i$, $\hat{I}=\{i,j\}$ detects $\omega$ by \ref{def:viceownership} (5).

Next, suppose that $i$ owns $o$ at $\sigma^1$. If then if $j$ is also in level $n$,  then $\hat{I}=\{i,j\}$ detects $\omega$ by Definition \ref{def:viceownership} (4) i..

Second, suppose that $i$ does not own $o$ at $\sigma^1$. If $j$ is in level $n$. Thus, $j$ must own $o$ at $\sigma^1$ and $\varphi^c(\theta)(j)=o'$ for some $o'$ that is unmatched at $\sigma^1$. If $j$ owns $o'$ at $\sigma^1$ and we are back in the case of the last paragraph with $j$ taking the role of $i$. Thus, $i$ must owns $o'$ at $\sigma^1$ and either, $o'$ is assigned to a lower level individual $l$ or assigned to $i$ under $\omega$. In the former scenario,  $\hat{I}=\{j,l\}$ detects $\omega$ by Definition \ref{def:viceownership} (3). In the latter scenario, since  $\omega(j') = o$ and $\omega(i) = o'$, $\hat{I}=\{i,j'\}$ would detect $\omega$ since $\omega$ is not Pareto efficient for $i$ and $j$.  

\paragraph{Case 4:} Let $i$ be a residual individual. First note that by our minimal size selection of $\sigma^1$, $i$ cannot be a level-$N$ owner. We have two subcases to consider: First, suppose that $i$ owns $o$ at $\sigma^1$. Then $j$ is in a weakly lower level and thus $\hat{I}=\{i,j\}$ detects the deviation by Definition \ref{def:viceownership} (2) and (5). Second, suppose that $i$ does not own $o$ at $\sigma^1$. Then $j$ owns $o$ at $\sigma^1$ and $\varphi^c(\theta)(j)=o'$, where $i$ owns $o'$ at $\sigma^1$. Hence $\omega(i) = o'$ and $\omega(j) = o$ and as such $\hat{I}=\{i,j\}$ would detect $\omega$ since $\omega$ is not Pareto efficient for $i$ and $j$ (given their type reports and outcomes).   

This completes the proof of the `if' direction

\subsection{Proof of Proposition \ref{proposition:seq}}
\label{appendix:seq-dict-N-1}

We start with some additional notation. Given any $\theta \in \Theta$, and any $n \in \{1,\dots, N\}$, let $I^*_n(\theta)$ be the set of individuals that are owners of an object for a suballocation of size $n-1$ under the TTC algorithm with input $\theta$ for some step $t$. If no such suballocation exists, then this set is empty. Let $I^*_n \defeq \cup_{\theta \in \Theta} \ I^*_n(\theta)$. Note that $I_p=I^*_1$.

Consider the sequential dictatorship $\varphi^c$ specified as follows: For all $n \in \{1,\dots,N-2\} $, let $|I^*_n|=1$. Denote $\{i,j\}\defeq I^*_{N-1} \cup I^*_N$ and let $I^*_{N-1} \cap I^*_N \neq \emptyset$. Thus, there exists a problem $\theta'$ such that $I_{N-1}(\theta')=i$. Let $\Theta'\defeq \{ \theta'' \in \Theta | \ \sigma_{N-2}(\theta'')= \sigma_{N-2}(\theta')\}$ and assume that for all $\theta^*\notin \Theta'$, we have  $I^*_{N-1}(\theta^*)=j$. In the following, consider the two objects $\{o_1,o_2\}\defeq \bar{O}(\sigma_{N-2}(\theta'))$.  
  
Consider a problem $\theta$ such that
\begin{enumerate}

    \item $o_1 \ \theta_i \ o_2$ and $o_1 \ \theta_j \ o_2$ and for all $\hat{o} \notin \{o_1,o_2\}$, we have $\hat{o} \ \theta_i \  o_1$ and $\hat{o} \ \theta_j \  o_1$,
    \item  $\varphi^{c}(\theta')(i')$ ranks first on $\theta_{i'}$ for all $i' \in \mathcal{I} \setminus \{i,j\}$.  
\end{enumerate}

A quick inspection reveals that for all $i' \in \mathcal{I} \setminus \{i,j\}$, it must be  $\varphi^{c}(\theta')(i') = \varphi^{c}(\theta)(i')$. Hence, $\theta \in \Theta'$, which implies that $\varphi^{c}(\theta)(i)=o_1$ and $\varphi^{c}(\theta)(j)=o_2$. Let $\omega \neq \varphi^{c}(\theta)$ be a deviation at problem $\theta$, where

\begin{itemize}
    \item $\omega(i)= \varphi^{c}(\theta)(j)$ and $\omega(j)= \varphi^{c}(\theta)(i)$,
    \item $\omega(i')= \varphi^{c}(\theta)(i')$, for all $i'\in \mathcal{I} \setminus \{i,j\}$.
\end{itemize}   

Consider an arbitrary $\hat{I} \subseteq \mathcal I$ with $|\hat{I}|=N-2$. It suffices to show that $\hat{I}$ does not detect the deviation. Since $N\geq 6$, we must be in one of the following four cases:

In the first case, suppose that $\hat{I} \cap \{i,j\}= \emptyset$. It is clear that since $\omega(i')= \varphi^{c}(\theta')(i')$ for all $i'\in \mathcal{I} \setminus \{i,j\}$, individuals $\hat{I}$ cannot detect $\omega$. 
In the second case, suppose that $j \notin \hat{I}$ and $i \in \hat{I}$. Thus, there exists $k \notin \hat{I}$ with $k \neq j$. Consider a problem $\hat{\theta}= (\hat{\theta}_{\{j,k\}},\theta_{-\{j,k\}})$ where $k$ ranks $o_1$ first on $\hat{\theta}_{k}$ and $j$ ranks $\varphi^{c}(\theta)(k)$ first on $\hat{\theta}_{j}$. This leads to $\omega(k')= \varphi^{c}(\hat{\theta})(k')$ for all $k'\in \hat{I}$ and hence $\hat{I}$ does not detect $\omega$. 
In the third case, suppose that $i \notin \hat{I}$ and $j \in \hat{I}$.
Consider a problem $\tilde{\theta}= (\tilde{\theta}_{i},\theta_{-i})$, where $i$ ranks $o_2$ first on $\tilde{\theta}_{i}$. Then it must be $\varphi^{c}(\tilde{\theta})= \omega$, which means that $\hat{I}$ does not detect $\omega$. 
In the fourth case, let $\{i,j\} \subsetneq \hat{I}$. Since $N\geq 6$, we can select two individuals $l,l' \notin \hat{I}$. Consider a problem $\bar{\theta}= (\bar{\theta}_{\{l,l'\}},\theta_{-\{l,l'\}})$ such that $l$ ranks $\varphi^{c}(\theta)(l')$ first on $\bar{\theta}_{l}$ and $l'$ ranks $\varphi^{c}(\theta)(l)$ first on $\bar{\theta}_{l'}$. Then $\varphi^{c}(\bar{\theta})(k')= \omega(k')$ for all $k'\in \hat{I}$. Thus, $\hat{I}$ does not detect $\omega$. We thus conclude that $ \# \varphi \geq N-1$.

\end{document}